\documentclass{eas}
\usepackage{rotating}
\usepackage{natbib}
\usepackage{bm}
\usepackage{color}
\usepackage{graphicx}
\usepackage{const}
\usepackage{cours_sismo_eas}
\usepackage{amsmath,amssymb,amsfonts}
\def\diff{{\rm d}}
\def\ind#1{_{\mathrm{#1}}}

\def\Ylm{Y_\ell^m}

\renewcommand{\contentsname}{Summary}
\renewcommand{\listfigurename}{Figures}
\renewcommand{\listtablename}{Tables}
\def\aap{A\&A}

\def\apjs{ApJS}\def\apjl{ApJL}

\begin{document}

%%-----------------------------
%%      the top matter
%%-----------------------------
\title{Stellar oscillations - the adiabatic case}
\author{B. Mosser}\address{LESIA, CNRS, Universit\'e Pierre et Marie Curie,
 Universit\'e Denis Diderot, Observatoire de Paris, 92195 Meudon
 cedex, France}
\begin{abstract}
This lecture on adiabatic oscillations is intended to present the
basis of asteroseismology and to serve as an introduction for
other lectures of the EES 2014. It also exposes the
state-of-the-art of solar-like oscillation analysis, as revealed
by the space missions CoRoT and \emph{Kepler}. A large part of the
lecture is devoted to the interpretation of the modes with a mixed
character that reveal the properties of the radiative cores of
subgiants and red giants.
\end{abstract}
\maketitle
%%-----------------------------
%%      your text
%%-----------------------------

\tableofcontents
\section{Introduction}

\subsection{Aim of this lecture}

The favorite activity of a seismologist consists in seeking to
look much below the surface of his preferred object, as everyone
who prefers a deep to a superficial analysis, to get to the bottom
of things, to look an issue in more depth, to delve into
something, to get to the root of the matter, to seek out the
underlying issue... This was the program proposed by Athanasius
Kircher, a German Jesuit (1601-1680), in his model of the Earth's
interior: volcanoes were seen as outlets of the Earth's internal
fires (Fig.~\refer{fig-kircher}). This is the program proposed by
the 2014 \'Evry Schatzman School: Ast\'erosismologie et nouvelles
contraintes sur les mod\`eles stellaires / Asteroseismology and
new constraints for stellar modelling.\\

Attendees of the EES were all supposed to have a background in
physics at the Master level, including a background in
hydrodynamics. Nevertheless, the first chapters are intended to
smoothly introduce the physics of the stellar (or solar)
oscillations. This introduction chapter presents relevant
definitions and terms on which we have to agree. Chapter
\ref{chap2} deals with the propagation of waves, in the general
case applicable to the interior of the stars, but limited to the
adiabatic case. So, any energy exchanged between the wave and the
star interior is here neglected and omitted; non-adiabatic
oscillations are considered in the contribution to the EES 2014 by
Samadi \& Belkacem.

Chapter \ref{chap3} shows how the adiabatic oscillation waves
become modes when boundary conditions are used. It also shows the
acute power of asteroseismology for probing efficiently stellar
interiors.

Chapter \ref{chap4} presents the wealth of information to be
gained from oscillation frequencies. This information first relies
on the identification of the pattern, then on the link between
\emph{global}\footnote{Global seismic parameters are integrated
quantities. They are also known as seismic indices since they
present the double properties of being observable and being
related to the stellar mass and radius.} seismic parameters and
interior structure information; the
\emph{asymptotic}\footnote{Here, asymptotic means, in practice,
valid in the high-frequency domain.} expansion helps providing
both information. Then, global seismic parameters, frequency
differences, or frequencies are analyzed for making the best of
the seismic information.

Chapter \ref{ensemble} is devoted to \emph{ensemble}
asteroseismology: even if restricted to global seismic parameters,
seismology already delivers key information on the stellar
interior structure and can provide valuable information on large
ensemble of stars, in a way similar to the Hertzsprung-Russell
diagram (see Fig.~\ref{fig-HR-sismo}).

The Appendix deals with more technical issues. Spherical harmonics
are briefly presented in Section \ref{ylm}. Section
\ref{justi-asymp} presents how asymptotic development operates. In
Section \ref{variationalprinciple}, we see how the variational
principle is used to make the best of observed oscillations
spectra and address the inverse problem. Section
\ref{perturbationrotation} deals with rotation, and Section
\ref{surface} with surface effects.

\begin{figure}[!t]
\fichier{10.6}{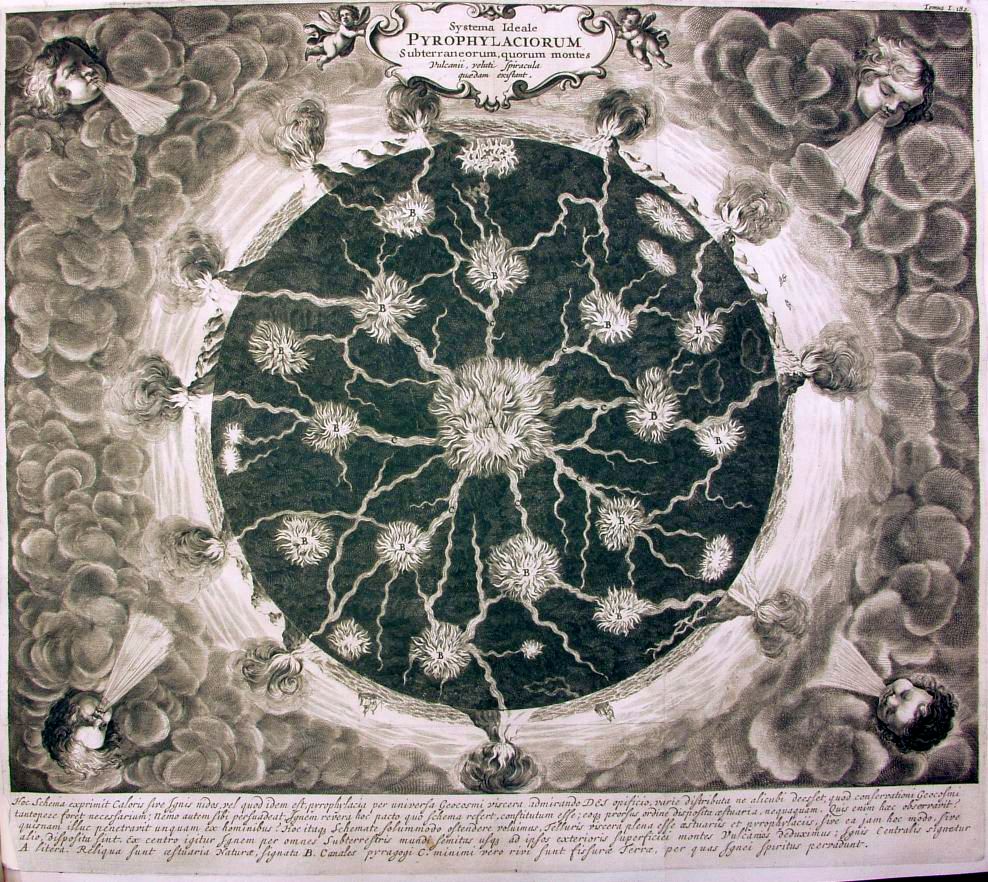} \legende{Earth's internal
fires}{Kircher's model of the Earth's internal fires, from Mundus
Subterraneus. Courtesy History of Science Collections, University
of Oklahoma Libraries. \labell{fig-kircher}}
\end{figure}

\begin{figure}[!t]
  \fichier{8.6}{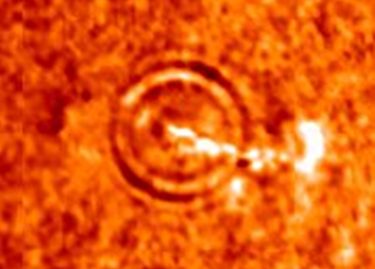}
  \legende{Sun quakes}{Flare generated Sun quakes
  observed by the Solar \& Heliospheric Observatory (ESA/NASA).
  \labell{fig-quakes}}
\end{figure}

\subsection{Tell me how you oscillate, I will tell you who you
are}

A few examples may highlight the physical concept behind
seismology and illustrate the title of the section, freely adapted
from Miguel de Cervantes, Sainte-Beuve, Antoine de
Saint-Exup\'ery, Marcel Pagnol, and the folk wisdom.\\

The railway worker who hits the axletree of a wagon listens to the
produced sound; any disequilibrium or crack in the wheels will
induce a specific signature. If you wish to hang out the picture
of your mother-in-law on a wall covered with wall paper, a flick
of the fingers will indicate which type of bit has to be used to
drill the wall. Similarly, {\sl music}\footnote{Hereafter denoted
as {\sl global oscillations}} from the stars provides us with a
unique view on their interior structure, as music carries a lot of
information (Bach is not Eminem, and conversely). If you listen to
me, you rapidly get an idea of my skills: poor English, French
accent\footnote{Still present in the written version of the
lecture...}, addict to red giant seismology ;-).

\subsection{Definitions, vocabulary, and tricks}

\subsubsection{Seismology}

Greek etymology helps understanding the word seismology:
``seiein'' [shake] + ``logos'' [study]. Then, we can develop
seismology in helioseismology (no Earth quakes on the Sun surface,
but Sun quakes, as seen in Fig.~\ref{fig-quakes}),
asteroseismology, dioseismology
\citep[e.g.,][]{1993A&A...267..604M,2011A&A...531A.104G},
selenoseismology
\citep[e.g.,][]{1974JGR....79.4351D,1993SGeo...14..239L}...\\

Asteroseismology encompasses two notions, which are physically
quite different: \\
- solar-like oscillations, namely stochastically excited and
damped oscillations; \\
- stellar variability, which corresponds to thermal unstable
oscillations (Fig.~\ref{fig-oscil}).\\
This lecture is focussed on solar-like oscillations; the formalism
is however suited to analyze oscillations in any class of
pulsating stars. Solar-like oscillations have much smaller
amplitudes than intrinsically unstable oscillations. Their spectra
are however directly intelligible, what is not necessarily the
case for unstable pulsations. With the space missions CoRoT and
\emph{Kepler} \citep{2006cosp...36.3749B,2010Sci...327..977B},
solar-like oscillations are observed in a large variety of
low-mass stars, all along their evolution from the main sequence
to the upper red and asymptotic giant branches
\citep{2008Sci...322..558M,2009Natur.459..398D,2010ApJ...713L.176B,2011Sci...332..213C},
as shown in Fig.~\refer{fig-ensemble-intro}.

\begin{figure}[!t]
\fichier{10.1}{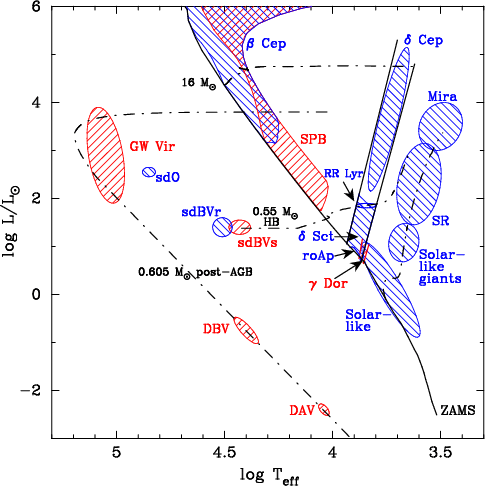} \legende{Oscillations in the HR
diagram}{Hertzsprung-Russel diagram with the different types of
oscillations seen in stars. This lecture focusses on solar-like
oscillation in solar-like stars, with lower effective temperature
than the $\delta$-Scuti, RR Lyrae and cepheid instability strip
indicated by two parallel black lines
\credit{2010aste.book.....A}. \labell{fig-oscil}}
\end{figure}

\begin{figure}[!t]
 \fichier{11.}{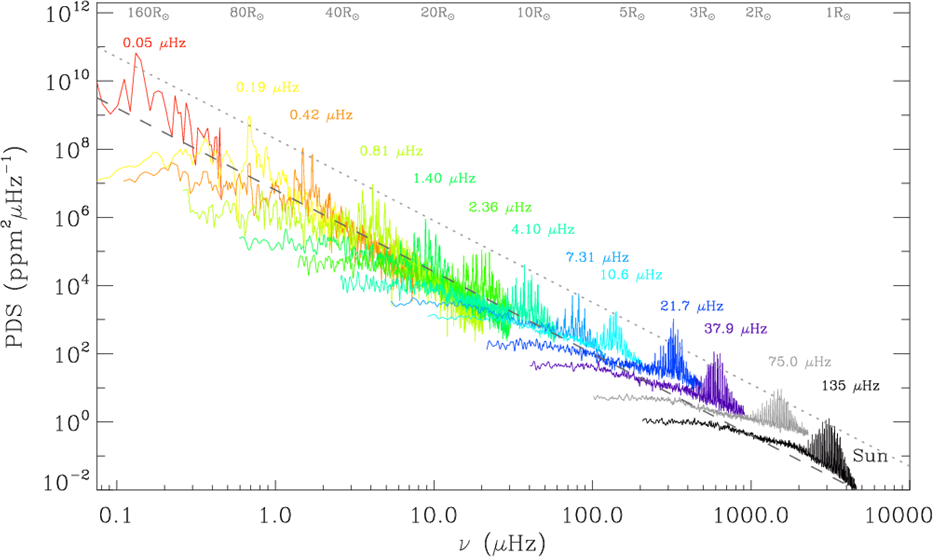}
 \legende{Ensemble asteroseismology (1)} {Solar-like oscillations in low-mass stars from
 the main sequence to semi-regular variables
 \credit{2013sf2a.conf...25M}.
 \labell{fig-ensemble-intro}}
\end{figure}

\subsubsection{Waves \emph{versus} modes}

Waves and modes are different physical concepts. A wave is
propagating; a mode is a standing wave. As such, a mode does not
propagate as it is stationary. Mathematically, a phase term
varying as $\omega t - {\bf k}.{\bf r}$ corresponds to a wave
whereas a decoupled time and space variation as $\cos\omega t
.\cos{\bf k}.{\bf r}$ corresponds to a mode. Anyway, any mode can
be obtained by a sum of waves, and any wave can be obtained by a
sum of modes as well.

Angular frequency ($\omega = 2\pi \nu$) and cyclic frequency ($\nu
= \omega / 2 \pi$) carry similar information but are used in
different contexts. When studying waves, it is more convenient to
use the angular frequency in order to have a compact expression of
the phase of the wave, especially with an harmonic form $\exp i
\omega t$. When boundary conditions are introduced, the concept of
normal modes supersedes the concept of waves. Modes are preferably
characterized by their cyclic frequency.

%%%%%%%%%%%%%%%%%%%% chapitre 2 %%%%%%%%%%%%%%%%%%%%%%%%%%%%%%
%\clearpage
\section{Waves and propagation\labell{chap2}}

This chapiter aims at presenting how waves propagate in a star.
For more information, the reader can refer to
\cite{1989nos..book.....U} and to \cite{1980tsp..book.....C}
\citep[see also][]{2010aste.book.....A}).

\subsection{Equations describing the fluid\label{equation-fluide}}

%\subsubsection{Mass continuity}

The temporal variation of a mass element of a fluid, inside a
volume $V$ surrounded by a surface $S$, is related to the
conservation of mass
\begin{equation}
{\diff \over \diff t } \int\!\!\!\!\!\int\!\!\!\!\!\int_V \rho\
\diff V
 \ = \
- \int\!\!\!\!\!\int_S \rho\ \vv . {\bf n}\ \diff S,
\end{equation}
where $\rho$ is the density and $\vv$ is the particle velocity.
The negative sign comes from the orientation of the normal vector
${\bf n}$ towards the exterior, which is opposite to the usual
sign convention that counts positively what is entering the
system. Using Ostrogradski's theorem\footnote{for any vector
$\mathbf{A}$,  $\int\!\!\!\int\!\!\!\int_V  {\bm\nabla} . {\bf A}\
\diff V \ =\ \int\!\!\!\int_S {\bf A} . {\bf n}\ \diff S $}, the
previous equation, known as the equation of mass continuity, can
be rewritten
\begin{equation}
{\diff \rho\over \diff t} + \rho\  {\bm\nabla} . {\bf v} \ = \ 0 .
\label{eqt-mass-conti}
\end{equation}

%\subsubsection{Conservation of momentum}

When rotation, viscosity, and electromagnetic forces are
neglected, the equation of motion writes
\begin{equation}
{\partial \vv \over \partial t} + \bigl[ \vv . {\bm\nabla} \bigr]
\vv \ = \ -{{\bm\nabla} p\over \rho} + {\bf g},
  \label{eqt-moment}
\end{equation}
where $\bf g$ is the gravity field, which derives from the gravity
potential $\psi$ and depends on the density $\rho$:
\begin{equation}
 {\bf g} \ = \ -  {\bm\nabla} \psi \ \ \hbox{ and } \ \ \nabla^2 \psi \ = \
 4\pi {\cal G} \rho .
 \label{eqt-poisson}
\end{equation}

%\subsubsection{Energy equation}

The energy equation is quite a complex issue, treated in Samadi \&
Belkacem's lecture. For the following, we suppose that
adiabaticity is met: over a cycle, there is no energy exchanged
between the wave and the fluid. This is justified by the fact that
the temporal evolution of the wave is much more rapid than the
characteristic time of any mechanism transferring energy as heat
between the different fluid elements.

If, as in the following part, we note $\rho$ and $p$ the Eulerian
perturbations associated to the oscillations (and with a subscript
$\tiny 0$ for the variables at equilibrium), the energy equation
is then
\begin{equation}
 \left({\diff \ln p\over \diff \ln \rho}\right)\ind{ad}
  \egaldef\ \Gamma_1 .
  \labell{eqt-justison}
\end{equation}
It helps defining the first adiabatic exponent $\Gamma_1$. As one
can imagine, this relation if of prime importance for this lecture
on adiabatic oscillations. There is no phase lag between the
pressure perturbation and the density perturbation, contrary to
the non-adiabatic case.

\subsection{Linearization}

\subsubsection{Euler versus Lagrange}

To develop the characteristics of the fluid dynamics, we have to
choose a way to treat the variables. On the one side, the
\emph{Eulerian} (fixed-position) approach describes the physics of
a fluid in terms of fields. On the other side, the
\emph{Lagrangian} description corresponds to the case where a
fluid element is individually followed. The position of the
particle varies with time. Mathematically, these different points
of view are reconciled with the operator relation
\begin{equation}
  {\diff \over \diff t}
  \ = \
  {\partial \over \partial t} + {\diff {\bf r}\over \diff t}.{\nabla}
  \ = \
  {\partial \over \partial t} + {\bf v}.{\nabla}
  \label{eqt-def-eul-lag}
\end{equation}
or
\begin{equation}
  \delta x \ = \ x + {\bf \delta r}. \nabla x_0 ,
\end{equation}
with the following notations:
\begin{equation}
\left\{%
\begin{array}{rl}
  \hbox{Variable at equilibrium: }          & x_0 \\
  \hbox{Eulerian perturbation: }   & x \\
  \hbox{Lagrangian perturbation: } & \delta x \\
\end{array}%
\right.
\end{equation}
In the literature, Eulerian perturbations are often denoted with a
prime, so that there is no confusion between a variable $x$, its
value at equilibrium $x_0$, and its Eulerian perturbation $x'$.
Since the next Section introduces the linearization of the wave,
we prefer to lighten the equations and to avoid introducing
primes.

In the following, equations are written according to the Eulerian
specification of the flow field.  However, in many cases it is
necessary to go back to the Lagrangian form, for situations where
localization matters. For instance, referring to a boundary
condition requires the use of the Lagrangian approach.

\subsubsection{Notations}

Interior structure variables are marked with an index  0 to
distinguish them from the Eulerian perturbations. So, pressure,
density, velocity field, and gravity field (or gravity potential)
are denoted:
\begin{eqnarray}
\left\{%
\begin{array}{rccccl}
  \hbox{At equilibrium:}       &p_0, &\rho_0, & {\bf 0}, &{\bf g_0}& (\hbox{or\ }
  \psi_0)\\
  \hbox{Eulerian perturbation:}&p, &\rho, &\vv, &{\bf g} & (\hbox{or\ } \psi)\\
\end{array}%
\right.
\end{eqnarray}
Perturbations, if small, can be treated at first order. This
provides a linearized approach, which is evidently much simpler
than the complete approach since non-linearity is discarded. So,
as soon as the linear approach is chosen, some tricky problems
cannot be considered any more, as those concerning the amplitude
of the waves. In other words, the eigenvalue problem can be solved
to the first order, but the absolute amplitude of eigenfunctions
will remain undetermined.\\

We also consider that the equilibrium structure obeys the
hydrostatic equilibrium. This relies on the assumption of
stationarity and of absence of any movement at equilibrium. For
any interior structure parameter $x_0$, this implies
\begin{equation}
{\partial x_0\over \partial t} \ = \ 0 .
\end{equation}
Accordingly velocity field is zero. This also means that we
consider non-rotating stars. Including rotation would require
either a specific study or a perturbation treatment. Rotation is
briefly presented, as a perturbation only, in Appendix
\ref{perturbationrotation}. Rapidly rotating stars requires a
specific non-pertubative study \citep[e.g.,][and references
therein]{2009A&A...506..189R,2012A&A...542A..99O}.
\\

We can thus rewrite the equations describing the fluid (Section
\ref{equation-fluide}). With the preceeding hypothesis, the
advection term $\vv .\nabla \vv$ in the equation of motion
(Eq.~\ref{eqt-moment}) is a second-order term and can be
neglected. So, the equation of motion writes
\begin{equation}
 (\rho_0 + \rho) {\partial \vv \over \partial t} \ = \
 -{\bm\nabla}
 (p_0+p) - (\rho_0 + \rho) {\bm\nabla} (\psi_0 + \psi) .
\end{equation}
To go further, we use the hydrostatic equilibrium
\begin{equation}
 {\bm\nabla} p_0 \
 =
 \ - \rho_0 \ {\bm\nabla} \psi_0 ,
\end{equation}
so that, to first order, one gets
\begin{equation}
  {\partial \vv\over \partial t}
  \ = \
  -{1\over\rho_0}\ {\bm\nabla} p - {{\bm\nabla} \psi_0\over \rho_0} \ \rho -
  {\bm\nabla}\ \psi
  .
  \labell{eqt-mvt1}
\end{equation}
On the right-hand term, it is easy to recognize three
contributions to the restoring force: pressure gradient; buoyancy;
perturbation of the gravitational potential.\\

To first order, the mass conservation (Eq.~\ref{eqt-mass-conti}),
rewritten as
\begin{equation}
  {\partial (\rho_0 + \rho) \over \partial t} +  {\bm\nabla}. ((\rho_0+\rho) {\bf v}) \ =
  \ 0 ,
  \label{eqt-masse}
\end{equation}
becomes
\begin{equation}
 {\partial \rho \over \partial t} +  {\bm\nabla} . (\rho_0 \vv) \ = \ 0 .
 \labell{eqt-masse1}
\end{equation}

%\subsubsection{Poisson equation}

The previous form of the Poisson equation (\ref{eqt-poisson})
being linear, it simply rewrites
\begin{equation}
 \nabla^2 \psi \ = \ 4\pi {\cal G} \rho .
 \labell{eqt-poisson1}
\end{equation}

%\subsubsection{Adiabatic propagation}

Eq. (\refer{eqt-justison}) that expresses the adiabatic
propagation becomes
\begin{equation}
{\diff p\over \diff t} \ = \ \Gamma_1 {p_0\over\rho_0}\ {\diff
\rho\over \diff t}. \labell{eqt-son1}
\end{equation}
This is a Lagrangian form, corresponding to the Eulerian form
\begin{equation}
 {\partial p\over \partial t} + \vv . {\bm\nabla} p_0
 \ = \
 \Gamma_1
 {p_0\over \rho_0} \ \left({\partial \rho \over \partial t} + \vv .
 {\bm\nabla} \rho_0 \right)
 .
 \labell{eqt-adiab1}
\end{equation}
Combinations with the hydrostatic equilibrium and the mass
conservation then provide
\begin{equation}
   {\partial p \over \partial t}
   \ =\ -\Gamma_1 p_0\ {\bm\nabla}. \vv - \rho_0\ {\bf g_0} . \vv
   .
   \labell{eqt-dpdt}
\end{equation}

\begin{figure}[!t]
 \fichier{9.231}{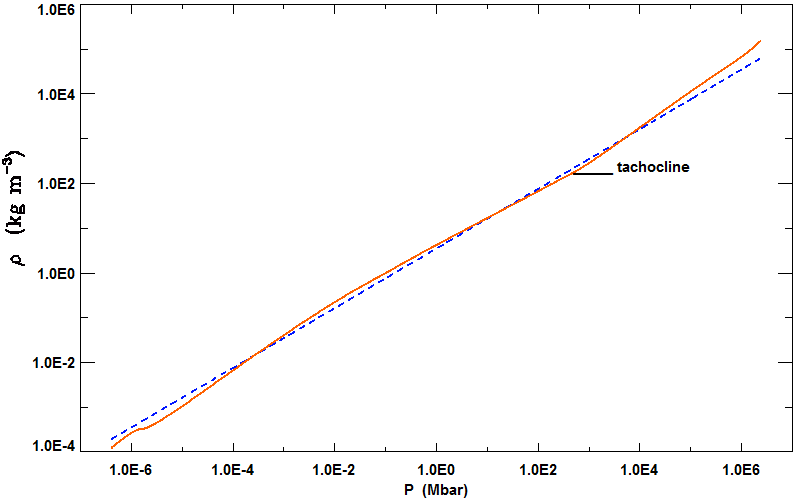}
 \legende{Solar pressure-density profile}{Solar pressure-density profile (red continuous line).
  This profile is close to a pure adiabat with an adiabatic index 5/3 (blue dashed line).
The horizontal tick shows the location of the tachocline, namely
the transition between the outer convective envelope and the inner
radiative region where a large gradient in the rotation profile is
observed.
  \labell{fig-prho-sol}}
\end{figure}

\begin{figure}[!t]
 \fichier{9.231}{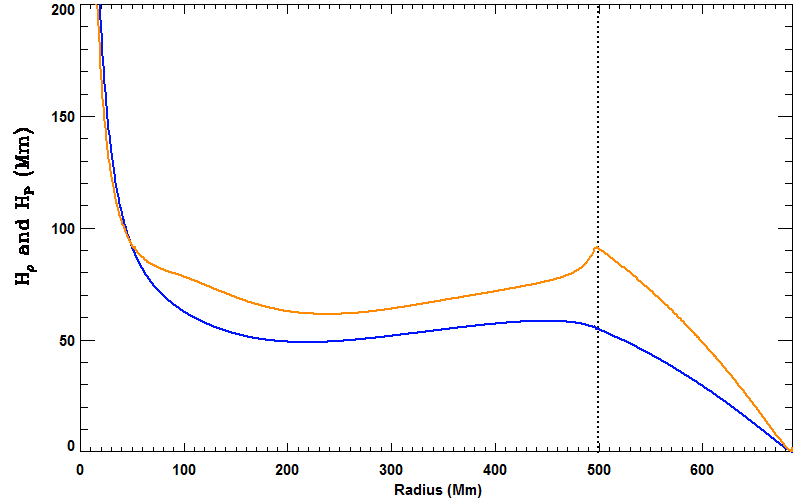}
 \legende{Pressure and density scale heights}{Pressure (blue line) and density (red line) scale
 heights in the Sun. We note three domains: in the outer
 convective envelope, the scale heights vary approximately
 linearly  with the depth, and the ratio of the scale heights is
 of about 5/3 (see Fig.~\ref{fig-prho-sol}); in the radiative region, the scale heights are
 approximately uniform; in the core, their values are
 approximately equal since the fluid is nearly isothermal, and diverge at the center,
 where pressure and density show limited gradients.
 The dotted vertical line shows the transition between the radiative interior
 and the convective envelope.
 \labell{fig-height-sol}}
\end{figure}

\subsection{Sound waves}

%A sound wave is a pressure perturbation. The perturbation is longitudinal, since there is no shear in a fluid.

Here, we consider a highly simplified case for introducing useful
concepts, methods and parameters. We consider the case of a
plane-parallel atmosphere and no gravity gradient. Moreover, all
gradients in the interior structure are neglected, so that all
scale heights are much larger than the wavelength $\lambda$. The
scale height $H\ind{X}$ of any equilibrium variable $X$ (such as
temperature, pressure, density, sound speed, adiabatic
gradient...) is defined by $H\ind{X} = (-\diff\ln X / \diff
r)^{-1}$.

Pressure and density inside the Sun are plotted in
Fig.~\refer{fig-prho-sol}; their scale heights are given in
Fig.~\refer{fig-height-sol}. This shows that the assumptions as
presented are not adapted to study solar pressure waves; this
paragraph intends to present a case study and not a realistic
description of sound waves inside the Sun. We also suppose the
Cowling approximation \citep{1941MNRAS.101..367C}: the Eulerian
perturbation of the gravitational potential is neglected in the
equation of motion
(Eq.~\refer{eqt-mvt1}).\\

\begin{figure}[!t]
 \fichier{9.231}{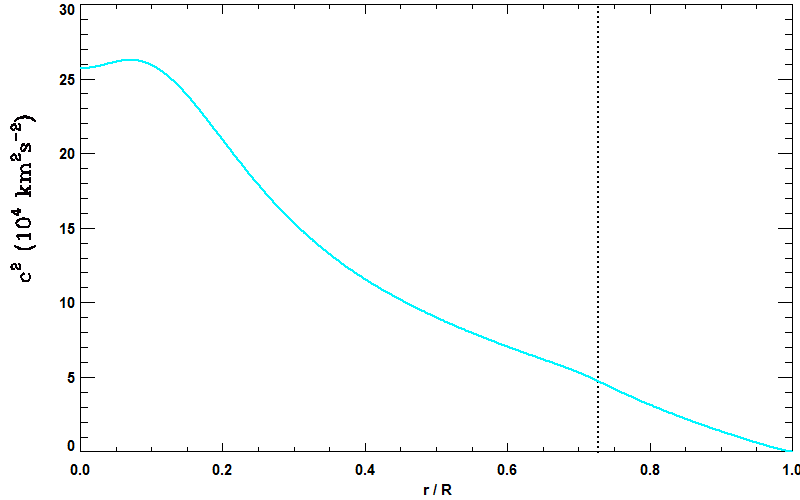}
  \fichier{9.231}{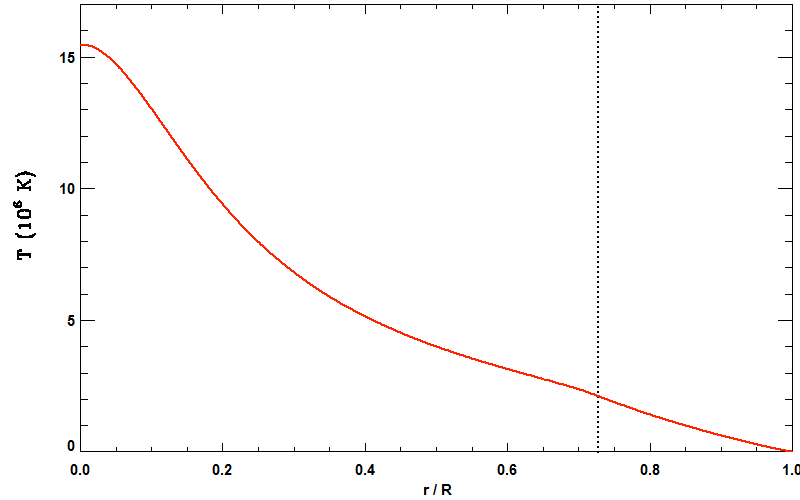}
 \legende{Sound-speed profile}{Solar sound-speed profile
 (squared) and temperature profile. The main difference between
 the profiles is seen in the core, where the mean particle mass $\mu$ is
 increased by the product of the nuclear reactions. The dotted vertical line
 shows the transition between the convective envelope and the radiative interior.
 \labell{fig-intc-sol}}
\end{figure}

%\subsubsection{Sound speed}

We introduce the harmonic description
\begin{equation}
{\partial \over
\partial t} = i \omega ,
\end{equation}
taking into account the temporal variation of all perturbed terms
varying as $\exp i\omega t$.  So, from the adiabatic relation
(Eq.~\refer{eqt-son1})
\begin{equation}
i\omega p \ =\ \Gamma_1{p_0\over \rho_0} i\omega \rho ,
\end{equation}
we get
\begin{equation}
  p\ =\ \Gamma_1\disp{p_0\over\rho_0} \rho\ =\ c_0^2 \rho .
\labell{eqt-son2}
\end{equation}
The sound speed profile (Fig.~\refer{fig-intc-sol}) is defined by
\begin{equation}
  c_0^2
  \ \egaldef \
  \left( {\partial p \over \partial \rho}
 \right)\ind{S} \ \egaldef\ \Gamma_1\ \disp{p_0\over\rho_0}  .
  \labell{eqt-son3}
\end{equation}
Note that this definition is independent on the
approximations used in the simplified case.\\

In an isothermal atmosphere, at temperature $T_0$, where the
classical non-degenerate perfect gas law is relevant,
Eq.~(\ref{eqt-son3}) reduces to
 \begin{equation}
 c_0^2\ = \ \disp{\Gamma_1 \kb T_0 \over \mu}
 , \labell{eqt-def-son}
 \end{equation}
where $\mu$ is the mean particle mass, so that the temperature and
sound-speed profiles are closely linked (Fig.~\ref{fig-intc-sol}).\\

%\subsubsection{Dispersion relation}
According to the set of hypothesis previously defined,
perturbations are described by:
\begin{eqnarray}
\left\{
\begin{array}{rcl}
i \omega\ \rho_0  \vv  &= & - {\bm\nabla} p \\
i \omega\ \rho + \rho_0\  {\bm\nabla} . \vv &=& 0 \\
p &=& c_0^2 \ \rho\\
\end{array}
\right.
\end{eqnarray}
In the equation of motion, the restoring force reduces to the
pressure gradient owing to hypothesis on the scale heights. The
expressions of $p$ and $\vv$ can be eliminated for obtaining an
equation for $\rho$
\begin{equation}
- \omega^2 \rho  = - i \omega \ \rho_0\  {\bm\nabla} .
 \vv  = +  {\bm\nabla} .  {\bm\nabla} p = \Delta \ p = c_0^2\ \Delta \rho .
\end{equation}
Under the hypothesis that the wave is a plane wave, the phase of
the wave varies as
$$
\exp i\ \bigl[w t - {\bf k}.{\bf r}\bigr]
$$
so that $\mathbf{\nabla} \equiv - i \mathbf{k}$ and $\Delta \equiv
- k^2$. Therefore, we obtain the dispersion relation
\begin{equation}
\omega^2 \ = \ k^2 c_0^2. \labell{eqt-disperson}
\end{equation}
Owing to all simplifications already made, this equation is in
fact not dispersive:  the phase velocity $\omega / k$ and group
velocity $\diff \omega / \diff k$ are the same for all
wavelengths.

\subsection{Gravity waves}

The main restoring force of a gravity wave is not due to pressure
gradient but to buoyancy: a particle moved away from its
equilibrium position has to find it again due to the
disequilibrium in density. This happens when the motion is slow
enough for ensuring that the pressure equilibrium is achieved.
Indeed, gravity waves have a cyclic frequency smaller than the
dynamic frequency $2\pi/t\ind{dyn}$ associated to the dynamic
time. However, the motion is rapid enough for ensuring that the
movement is adiabatic. As a result, the density of the
perturbation differs from the local density in the fluid.
Necessarily, gravity waves cannot exist in convective regions
since the Schwarzschild criterion expresses that buoyancy cannot
provide a restoring force in a convective region since it is the
disturbing force that moves convective cells.

As for the sound waves, we address a simplified case of gravity
waves that can be expressed in terms of plane waves. The vertical
and horizontal components of the wave vector $\bf k$ are noted
$\kv$ and $\kh$, respectively. For sake of simplicity, we do not
consider the vector behavior of the horizontal component of $\bf
k$. As for sound waves, we apply the Cowling approximation
\citep{1941MNRAS.101..367C}: in the equation of motion
(Eq.~\refer{eqt-mvt1}), the Eulerian perturbation of the
gravitational potential is neglected.

\subsubsection{The dispersion relation} % vérifié : "dispersion relation"

The motion and mass equations (Eq.~\refer{eqt-mvt1} and
\refer{eqt-masse1}) express then
\begin{eqnarray}
\left\{
\begin{array}{rcl}
 \iw \rho_0 \vh &=& \ikh p \\
 \iw \rho_0 \vr &=& \ikv p - g_0\ \rho \\
 \iw \rho       &=& \rho_0 \left( \ikh \vh + \ikv \vr \right)\\
\end{array}
\right.
\end{eqnarray}
Hence, the horizontal component of the velocity is
\begin{equation}
\vh \ = \ {\kh \over \rho_0 \omega} \ p .
\end{equation}
Using the mass equation, we have
 \begin{equation}
 \vh = {\omega\over \kh} \ {\rho\over \rho_0} - {\kv \over \kh}
 \vr . \labell{eqt-grav-p}
 \end{equation}
This result is used in the equation of the vertical movement
\begin{equation}
\iw \rho_0 \vr = \left( \ikv {\omega^2\over \kh^2} - g_0 \right)\
\rho - \rho_0 \iw {\kv^2 \over \kh^2}\ \vr .
\end{equation}
The right-hand term of this expression is composed of terms in
$\rho$ and $\vr$. Since we only deal with long-period waves, the
coefficient proportional to $\omega^2$ is negligible compared to
the other terms, so that:
 \begin{equation}
 \iw \rho_0\ \left( 1 + {\kv^2\over \kh^2} \right) \ \vr \ = \ - g_0\
 \rho . \labell{eqt-grav1}
 \end{equation}
This equation shows that the vertical motion is mainly due to the
buoyancy term. The inertia of the displacement increases for low
values of the horizontal wavenumber. The coupling between the
vertical and horizontal motions results from the conservation of
the mass.\\

Finally, we have to introduce the adiabatic evolution of the
system (Eq.~\refer{eqt-adiab1}):
\begin{equation}
\iw p + \vr {\diff p_0\over\diff r} \ = \ \Gamma_1 {p_0\over
\rho_0} \ \left( \iw \rho + \vr {\diff \rho_0\over\diff r} \right)
.
\end{equation}
In the left-hand term of this equation, the contribution of $p$
can be neglected with respect to the contribution of $\vr$ since
it is proportional to $\omega^2$ (Eq.~\refer{eqt-grav-p}). So,
adiabaticity gives
\begin{equation}
\vr \left( {\diff p_0\over\diff r} - \Gamma_1 {p_0\over\rho_0} \
{\diff \rho_0\over\diff r}\right) \ = \ \iw \Gamma_1
{p_0\over\rho_0}\ \rho .
\end{equation}
Hence, with the introduction of the density scale height,
\begin{equation}
\vr \left(1- \disp{1\over \Gamma_1} \disp{\diff\log p_0\over\diff
\log \rho_0}\right) \ {\rho_0\over H_\rho} \ = \ \iw \rho .
\end{equation}
Combination with Eq.~(\refer{eqt-grav1}) provides
\begin{equation}
\omega^2 \rho_0\ \left( 1 + {\kv^2\over \kh^2} \right) \ \vr \ = \
\left(1- \disp{1\over \Gamma_1} \disp{\diff\log p_0\over\diff \log
\rho_0} \right) \ {g_0\rho_0\over H_\rho} \ \vr
\end{equation}
This gives the dispersion relation
\begin{equation}
\omega^2 \ \left( 1 + {\kv^2\over \kh^2} \right) \ = \ {g_0\over
H_\rho}\ \left(1- \disp{1\over \Gamma_1}\disp{\diff\log
p_0\over\diff \log \rho_0}\right) \ =\ \NBV^2 .
\labell{eqt-dispergrav}
\end{equation}
So, with the introduction of the \BV\ frequency discussed in the
next paragraph, the relation between the horizontal and vertical
components of the wavevector is
\begin{equation}
 \kv^2 \ =\
 \left({\NBV^2 \over \omega^2} -1\right) \kh^2 .
 \labell{eqt-dispergrav-kvkh}
\end{equation}

\subsubsection{\BV\ frequency \label{section-BV}}

The dispersion relation (Eq.~\refer{eqt-dispergrav}) makes use of
the \BV\ frequency $\NBV$. This frequency helps comparing the real
gradient $\Gamma$ in the fluid to the adiabatic gradient
$\Gamma_1$:
\begin{equation}
  \NBV^2 \ \egaldef\ {g_0\over H_\rho}\ \left(1- \disp{
  \left.\disp{\diff\log p\over\diff \log \rho}\right|\ind{fluid}
  \over \left.\disp{\diff\log p\over\diff \log
  \rho}\right|\ind{adiab} } \right)
  = {g_0\over H_\rho}\ \left(1- {\Gamma \over \Gamma_1}\right).
  \labell{eqt-BV}
\end{equation}

\begin{figure}[!t]
 \fichier{9.231}{solar_Gamma1}
 \legende{Real and adiabatic gradients in the Sun}{Real gradient
 $\Gamma = (\diff\log p / \diff\log\rho)\ind{fluid}$ (blue line) and adiabatic gradient
 $\Gamma_1$ (red line)
  in the solar interior. The horizontal dashed line
  indicates the constant value 5/3: $\Gamma_1$ is close to it except in the outer region where ionization processes occur.  $\Gamma$ is less than
  $\Gamma_1$ in the radiative region and close to it in the
  convective outer envelope. The vertical dotted line indicates
  the transition between the radiative and convective regions.
  \labell{fig-gamma1-soleil}}
\end{figure}

\begin{figure}[!t]
 \fichier{9.231}{solar_cavity}
 \legende{\BV\ frequency in the Sun}{\BV\ frequency (blue curve) and $S_\ell$ profiles in
 the solar interior. The $S_\ell$ function is defined later (Eq.~\ref{eqt-lamb}); the value of the degree
 $\ell$ is indicated for each curve. The vertical dotted line indicates
  the transition between the radiative and convective regions. The
  horizontal dashed lines indicate the approximate lower and upper frequency limits of the observed
  pressure modes in the Sun, significantly above the maximum value
  of $\NBV$.
  \labell{fig-BV-SL-soleil}}
\end{figure}

According to the definition, $\NBV=0$ in a convective region
(Fig.~\refer{fig-gamma1-soleil}, \refer{fig-BV-SL-soleil}). In
fact, the small overadiabaticity necessary to evacuate the stellar
energy makes that $\NBV^2<0$. This confirms that gravity waves
cannot propagate in a convective region. In the Sun, gravity waves
are trapped in the inner radiative region. Their detection is
highly difficult due to their evanescent evolution in the
convective envelope
\citep{2009A&A...494..191B,2010A&ARv..18..197A}.

Different forms for expressing $\NBV$ can be found in the
literature.  Eq.~(\refer{eqt-BV}) can write
\begin{equation}\labell{eqt-def-general-NBV}
  \NBV^2
  \ =\
  g_0\ \left( {1\over \Gamma_1 p_0}  {\diff p_0 \over \diff r}
    -{1\over \rho_0}  {\diff \rho_0 \over \diff r}  \right) .
\end{equation}
Under the assumption of the ideal gas law for a fully ionized gas,
and taking the composition gradient into account, the expression
becomes \citep[Section 13.4 of ][]{1989nos..book.....U}
\begin{equation}
  \NBV^2\ =\ {g_0^2 \rho_0 \over p_0}\
  \left( \nabla\ind{ad} - \nabla + \nabla_\mu \right),
\end{equation}
with
\begin{equation}
  \nabla\ind{ad} = \left( {\partial\ln T_0 \over \partial
  \ln p_0}\right)\ind{ad}
  ,\
  \nabla= {\diff\ln T_0 \over \diff\ln p_0}
  , \hbox{ and }
  \nabla_\mu = {\diff\ln \mu \over \diff\ln p_0}.
\end{equation}
In a plane-parallel atmosphere with a uniform gravity field, it
simplifies into.
\begin{equation}
  \NBV^2\
  =\ g_0\ \left( {1\over H_\rho}  - {g_0\over c_0^2} \right)
  \label{eqt-NBV-simplip}
  .
\end{equation}
The contribution of the gravity term in the \BV\ frequency yields
the increase of $\NBV$ in the stellar interior when the star
evolves, as a result of the contracting core
(Fig.~\ref{fig-BV-RGB0}).

\begin{figure}[!t]
 \fichier{9.0}{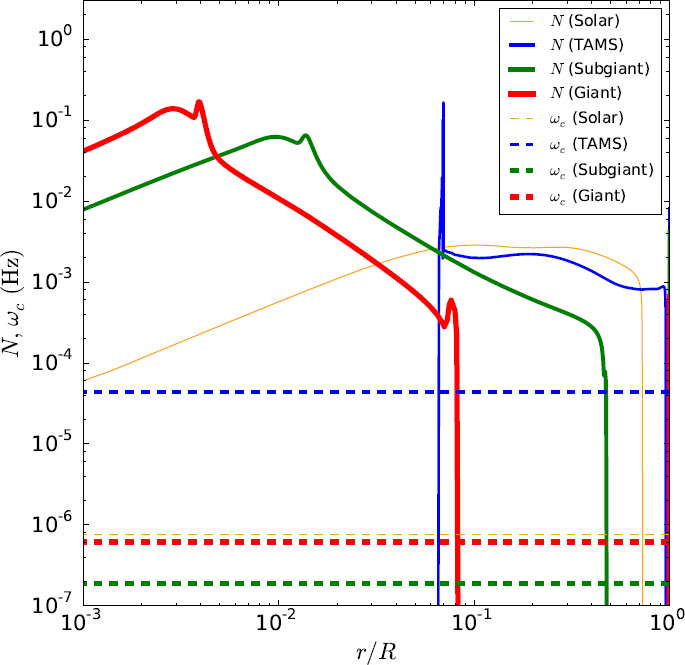} \legende{\BV\ frequency
 }{\BV\ frequency at different evolutionary stages
 \credit{2014ApJ...796...17F}.
 \labell{fig-BV-RGB0}}
\end{figure}

According to Eq.~(\refer{eqt-dispergrav}), the frequency of a
gravity wave is necessarily below the \BV\ frequency $\NBV$ since
one has
\begin{equation}\label{eqt-grav-1}
    {\kh \over k}  = {\omega \over \NBV}.
\end{equation}
In the low-frequency regime, with $\omega \ll \NBV$, we have $\kh
\ll kv$. Otherwise when $\omega$ is close to $\NBV$, we have  have
$\kh \approx k$.

\subsection{Surface gravity waves}

We may use the previous equations to derive the dispersion
equation at the surface of a fluid object, or at the boundary
between two immiscible fluids. We also suppose the plane-parallel
case, and the fluid to be incompressible.

From incompressibility, the mass equation (Eq. \refer{eqt-masse1})
gives
\begin{equation}
  {\bm\nabla} .\vv \ = \ 0.
\end{equation}
If the gradient and the perturbation of the gravitational
potential are neglected, the equation of motion is
\begin{equation}
  \label{eqt-motion-simple}
  \rho_0 {\partial \vv \over \partial t} \ = -  {\bm\nabla} p .
\end{equation}
The perturbed pressure then obeys to
 \begin{equation}
 \Delta \ p \ = \ 0 . \labell{eqt-laplace-p}
 \end{equation}
We aim to search for a periodic solution, moving parallel to the
surface or to the interface. With $z$ the variable of the axis
perpendicular to the surface and $x$ along this surface, we have
\begin{equation}
  p \ = \ P(z) \exp i (\omega t - kx ).
  \labell{eqt-laplace-p-prop}
\end{equation}
Using Eq.~(\refer{eqt-laplace-p-prop}) in
Eq.~(\refer{eqt-laplace-p}) gives
\begin{equation}
{\diff^2 P\over \diff z^2} - k^2 P \ = \ 0 .
\end{equation}
Hence $P(z)$ writes:
\begin{equation}
P(z) \ = \ P_- \exp( -k z) + P_+ \exp(+ k z).
\end{equation}
The diverging solution must be excluded if the width of the layer
is large. At the surface, pressure equilibrium implies that the
Lagrangian perturbation is null. This is typically a case where,
despite the solution found with Eulerian variables, it is
necessary to use the Lagrangian term to express the properties of
a boundary:
 \begin{equation}
 {\diff p \over \diff t} \  = \ {\partial p \over \partial t} + \vv
 .  {\bm\nabla} {p_0} \ = \ {\partial p \over \partial t} + \rho_0 \vv .
 {\bf g_0} \ = \ 0. \label{eqt-boundary}
 \end{equation}
Without information on the surface (presumably it just has small
variations around $z=0$), we can get the dispersion equation when
combining Eq. (\refer{eqt-boundary}) with the equation of vertical
motion (Eq. \refer{eqt-motion-simple})
\begin{eqnarray}
\left\{
\begin{array}{rll}
\disp{\partial p \over \partial t} &+ \rho_0 g_0\ v_z &=  0, \\
\rho_0\disp{\partial v_z \over \partial t} &+ \disp{\partial
p\over
\partial z} &=  0 .\\
\end{array}
\right.
\end{eqnarray}
This gives
\begin{equation}
\disp{\partial^2 p \over \partial t^2} \ = \ g_0 \disp{\partial
p\over
\partial z},
\end{equation}
hence the dispersion equation
\begin{equation}
\omega^2 = \ g_0 k .
\end{equation}
The equation of the free surface, with no Lagrangian perturbation,
is then
\begin{equation}
\delta p = p + \depl .  {\bm\nabla} p_0 = 0 ,
\end{equation}
where $\depl$ is the wave displacement. Hence
\begin{equation}
\xi\ind{v}  = {1\over \rho_0 g_0}\ p .
\end{equation}

In such a case, the group velocity $\diff \omega / \diff k$ is
half the phase velocity $\omega / k$. This can be easily seen when
throwing a small stone in a pond. Such waves resembles the
fundamental waves, at the limit between pressure and gravity waves
(Section \ref{classifi}).

\subsection{Differences between pressure waves and gravity
waves\labell{difference-p-g}}

This paragraph aims at emphasizing the difference between pressure
and gravity waves. It also shows the relevance of the Cowling
approximation. Heavy assumptions are used for hand-made physics
based on Eqs.~(\ref{eqt-mvt1}) and (\ref{eqt-poisson1}).
\begin{eqnarray}
\left\{
\begin{array}{rrll}
\hbox{Motion equation:} &
 \rho_0\ \disp{\partial \vv\over \partial t} &= &
- \nabla p +\disp{\nabla p_0\over \rho_0} \ \rho
- \rho_0\ \nabla\ \psi  \labell{eqt-diffpg}\\
\hbox{Poisson equation:} &
\nabla^2\psi  & = & 4\pi \G \rho\\
\end{array}
\right.
\end{eqnarray}

\subsubsection{Pressure waves}

For high-frequency waves, the operation of partial derivative with
respect to time,  $\partial /
\partial t$, provides much bigger terms than the gradient
operator. Adiabatic evolution implies, as a consequence of the
definition of $\Gamma_1$ (Eq.~\ref{eqt-son3}):
\begin{equation}
p\ = \ \Gamma_1{p_0\over \rho_0} \ \rho \ = \ c_0^2 \ \rho .
\end{equation}
In the inner regions, sound speed and gravity verify
\begin{equation}
c_0^2 \propto {\cal G}M/R \propto g_0\ R .
\end{equation}
For a crude estimate of the restoring forces, one can use the
following approximations for the gradients
\begin{eqnarray}
\left\{
\begin{array}{rlcr}
\hbox{equilibrium:} & \nabla X_0 &\simeq&   X_0 / R ,\\
\hbox{perturbation:}& \nabla X   &\simeq& n  X  / R ,\\
\end{array}
\right.
\end{eqnarray}
where $n$ is the number of wavelengths along the stellar radius.\\

From the Poisson equation, one derives  $\nabla^2 \psi \simeq
n^2\, \psi / R^2$, hence $\psi \simeq 4\pi \G R^2 \rho / n^2$.
Accordingly, the three components of the restoring force can be
estimated:
\begin{eqnarray}
\left\{
\begin{array}{rrllllll}
 \hbox{pressure:} &\disp{\nabla p}
                  &\simeq n\ \disp{p\over R}
                  &\simeq n\ \disp{c_0^2\over R}\ \rho
                  &\propto      & n& g_0\ \tvb\rho,\\
 \hbox{buoyancy:} &\disp{\nabla p_0\over \rho_0}\ \rho
                  &%\simeq  \disp{1\over R} \disp{p_0\over\rho_0}\ \rho
                  &%\simeq \disp{c_0^2\over\Gamma_1 R}\ \rho
                  &\propto      &  & g_0\ \rho,\\
 \hbox{potential:}&\rho_0\ \nabla\ \psi
                  &\simeq 4\pi \G R \rho_0 \disp{\rho \over n}
                  &\propto \disp{1 \over n} \disp{c_0^2\over R}\
                  \rho
                  &\propto & \disp{1\over n}& g_0\ \rho.\\
\end{array}
\right.
\end{eqnarray}
The pressure term evidently dominates the restoring force when the
number $n$ is large: this is a pressure wave. Furthermore, the
perturbation of the gravitational potential is the smallest term.
This justifies the Cowling approximation (see
paragraph~\ref{section-cowling}).

\subsubsection{Gravity waves}

Gravity waves evolves slowly with time. In that case, spatial
gradient terms are dominating the terms obtained from the operator
$\partial / \partial t$, so that Eq.~(\refer{eqt-diffpg}) gives,
when we assume that the perturbation of the gravitational
potential can be neglected:
\begin{eqnarray}
\begin{array}{ccclc}
 &\nabla p &\simeq& \disp{\nabla p_0\over \rho_0}& \rho ,\\
n&\disp{p\over R} &\simeq& \disp{p_0\over R \rho_0} & \rho , \\
 n& p &\simeq& c_0^2 &\rho .\\
\end{array}
\end{eqnarray}
This means that the pressure gradient and the buoyancy term are
equilibrated. Compared to a pressure wave for which the pressure
and density perturbations are linked with $p\ind{p} = c_0^2 \rho$,
here we have $p\ind{g} = c_0^2 \rho / n$. Hence, the pressure
perturbation $p\ind{g}$ verifies $p\ind{g} \ll c_0^2 \rho$. This
is a gravity wave.

%\clearpage

\section{Global oscillations\labell{chap3}}

As seen in the previous chapter, four variables are used for
describing the oscillations in a non-rotating object: pressure
$p$, density  $\rho$, velocity field $\vv$, and gravity field $\bf
g$ (or gravitational potential $\psi$). These terms, defined as
small perturbations, are bound with a set of four equations:
equation of motion, continuity equation, Poisson equation, and
energy equation. \\

A few hypotheses for studying global oscillations in a star are
implicitly present: taking  small perturbations into account
yields linear equations; rotation being neglected, spherical
symmetry is used. Different treatments can be made, depending on
the set of hypotheses. In this chapter, we do not intend to redo
all calculations from the physical equations to the dispersion
equation. These calculations are available in many references
\citep[e.g.,][]{2010aste.book.....A}. We just point out a few
important steps in the understanding of the physics. The aim is to
obtain a propagation diagram (Section~\ref{propagation}). A
solution of the eigenvalue problem, the asymptotic expansion,
is presented in the next chapter.\\

The choice of the hypothesis used for simplifying and reducing the
size the differential equation system plays a significant role, as
one can imagine. Among the main hypothesis (and apart from the
non-rotating spherical case already mentioned), we can
find the following cases:\\
- Plane-parallel atmosphere, corresponding in practice to neglect
the spherical case; this can be useful for phenomena in the upper
stellar envelope
but cannot fit a real case of a spherically symmetric star; \\
- Cowling approximation: the perturbation of the gravitational
field plays a negligible role in the equations. With the Cowling
approximation, this term is omitted; this interestingly
simplifies the computation \citep{1941MNRAS.101..367C};\\
- Ray-tracing: the properties of the eigenfunctions are omitted
when the dispersion equation is obtained, and the wave is
described by the properties of the wave vector only.\\

The most fertile approach is the JWKB approximation, used in
Section \ref{chap4} for delivering an asymptotic solution of the
eigenfrequency pattern. Here, we use it first in a very simple
manner, following \cite{2007AN....328..273G} who presents `an
elementary introduction to the JWKB approximation': an adequate
change of variable helps simplifying the dispersion equation.

\subsection{Simplified case\labell{simple-case}}

A simple case can help catching most of the physics of seismic
waves. Therefore, we neglect the spherical symmetry, consider
propagation in a plane-parallel atmosphere, and use the Cowling
approximation.\\

Variables are:
\begin{eqnarray}
\left\{%
\begin{array}{rcccc}
  \hbox{At equilibrium:}        &p_0, &\rho_0, & \vv_0 = {\bf 0}, &{\bf g_0} = g_0\, {\bf u}\ind{r}\\
  \hbox{Eulerian perturbation:} &p,    &\rho,   &\vv,     &{\bf 0} \hbox{ (Cowling approx.)}\\
\end{array}%
\right.
\end{eqnarray}
As a consequence of the simplification, only three equations are
needed:
\begin{eqnarray}
\left\{%
\begin{array}{rcccccl}
\rho_0\displaystyle{\partial \vv \over \partial t}
&=&-&{\bm\nabla} p &+& {\bf g_0}\ \rho \hfill
&\quad\hbox{Motion}\\
\displaystyle{\partial \rho \over
\partial t} &=&-&{\bm\nabla}.(\rho_0 \vv)& &\hfill
&\quad\hbox{Continuity}\\
%
% \noalign{\medskip} \
\displaystyle{\partial p \over \partial t} &=&-&\Gamma_1 p_0
{\bm\nabla}. \vv &-& \rho_0\ {\bf g_0} . \vv \hfill
&\quad\hbox{Adiabaticity}\\
\end{array}%
\right.
 \labell{eqt-set-equadif}
\end{eqnarray}

\subsubsection{Change of variables}

With a dedicated change of variables (again, reading
\cite{2007AN....328..273G} is useful to understand how it works),
we can describe the propagation  in a form as close as possible to
a plane wave. We use
\begin{eqnarray}\labell{eqt-cas-simple}
  \left\{%
\begin{array}{rcl}
  \eta & = & v \sqrt{\disp{p_0 \over c}}  , \labell{eqt-eta}\\
  \diff \tau & = & \disp{\diff r \over c} . \\
\end{array}%
\right. ,
\end{eqnarray}
where we have omitted the subscript 0 for the sound speed $c$
since there is no ambiguity. If we further assume that the
adiabatic exponent $\Gamma_1$ is almost uniform, we get from
Eq.~(\refer{eqt-eta}) that \citep[e.g.,][]{1995A&A...293..586M}
\begin{equation}\label{eqt-eta2}
    \eta^2 \propto \rho_0\  c \ v^2,
\end{equation}
which means that $\eta^2$ varies as the kinetic energy flux. The
variable $\tau$ is the acoustic radius (which is a time and not a
radius). Its use helps counterbalancing the rapid
variation of the sound speed.\\

The resolution of the set of differential equations
(Eq.~\ref{eqt-set-equadif}), in the simplified considered case and
with the ad hoc change of variables, implies that the new function
$\eta (\tau)$ follows the differential equation
\citep[e.g.,][]{1994MNRAS.268..880R,1995A&A...293..586M}
\begin{equation}\labell{eqt-dispers-simple}
    {\partial^2 \eta \over \partial\tau^2} + (\omega^2 - \wc^2) \
    \eta = 0 ,
\end{equation}
where we have introduced the harmonic time dependence $\partial /
\partial t \equiv i\omega$ and the
cutoff frequency $\wc$. Due to the change of variable, the
expression of $\wc$ derived from Eq.~(\ref{eqt-dispers-simple}) is
more complex than the original term found \cite{Lamb1908}, but is
in fact very close to it. Here, we only introduce the leading
term:
\begin{equation}
  \omega\ind{c} \ =\ {c\over 2 H_\rho}, \labell{eqt-cutoff}
\end{equation}
which corresponds to the term found by Lamb (1908). This frequency
is small everywhere except near the surface
(Fig.~\refer{fig-wc-sol}). In a convective region, the cutoff
frequency $\wc$ defined Eq.~(\refer{eqt-cutoff}) can also be
written
\begin{equation}\label{eqt-wc-g}
    \wc \ =\ { g \over 2c},
\end{equation}
as is clear from the expression of sound-speed
(Eq.~\ref{eqt-def-son}) and density scale height ($H_\rho =
\Gamma_1 \,H\ind{p} = \Gamma_1 \kB T / \mu g = c^2/g$).

\begin{figure}[!t]
 \fichier{9.231}{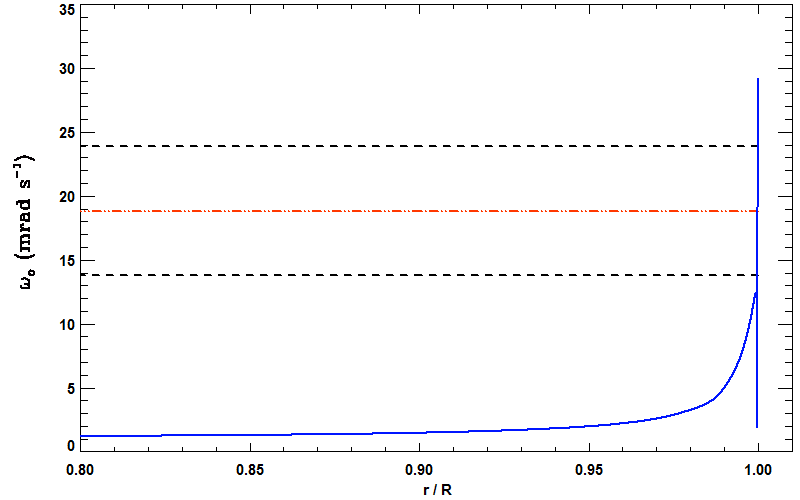}
 \legende{Cutoff frequency}{Solar cutoff frequency $\wc$ as a function of the normalized radius.
  The dashed lines indicate the approximate lower and higher limits
  of the frequency range where solar-like oscillations are
  observed. The red dashed line shows the frequency where
  solar oscillations have maximum amplitudes. The ratio between
  this frequency and the maximum of $\wc$ near the photosphere is constant from star to
  star \citep{2011A&A...530A.142B}, as discussed in paragraph
  \ref{numax}.
 \labell{fig-wc-sol}}
\end{figure}

\begin{figure}[!t]
 \fichier{11.}{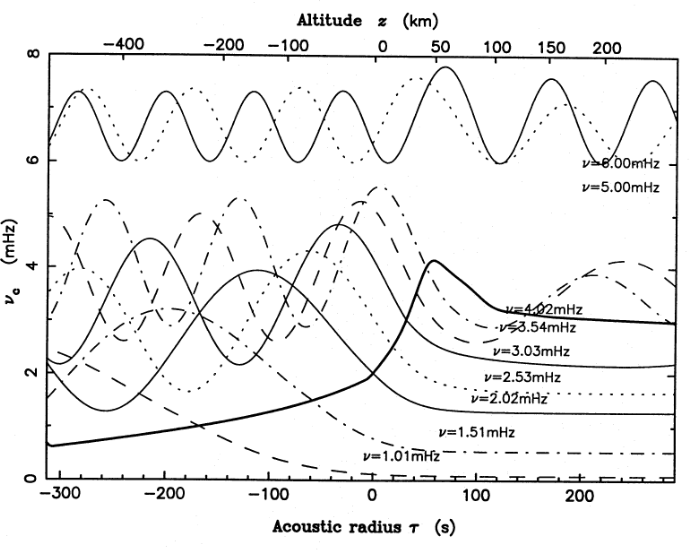}
 \legende{Cutoff frequency}{Cutoff cyclic frequency (full bold line) and wave reflection in the Jovian
 upper envelope. Standing waves are represented by $\eta^2\propto \rho_0 c v^2$
 (Eq.~\refer{eqt-eta2})
 \credit{1995A&A...293..586M}.
 \labell{fig-coupure-jup}}
\end{figure}

\subsubsection{Dispersion equation}

From Eq.~(\refer{eqt-dispers-simple}), we can deduce the
dispersion equation
\begin{equation}\labell{eqt-dispersion-simple}
 \omega^2 = \kappa^2 + \wc^2,
\end{equation}
if we assume a plane-wave form for $\eta(\tau)$ with an $\exp i
\kappa \tau$ dependence. Note that, with the acoustic variable
$\tau$, $\kappa$ behaves as a radial wavevector but is homogeneous
to an angular frequency. As shown in Fig.~\ref{fig-wc-sol}, $\wc$
has small values everywhere except near the surface, so that the
dispersion equation reduces to $\omega = \kappa$. The wave behaves
as a plane wave; the modulus  of the kinetic energy flux  $\eta$
is supposed to be nearly uniform. This is largely
verified (e.g., Fig.~\ref{fig-coupure-jup}, \ref{fig-pot-sol}).\\

With this dispersion equation (Eq.~\ref{eqt-dispersion-simple}),
we derive the reflection of the waves with a frequency lower than
the maximum frequency reached in the upper stellar atmosphere. We
also retrieve the propagation of pressure waves, when $\wc$ can be
neglected in most regions of the stars. In fact, the expression of
$\wc$ is close to the expression of $\NBV$\footnote{In an
isothermal atmosphere, $\wc$ and $\NBV$ have similar variation in
$\mu g^2/\kB T$,as derived from Eqs.~(\ref{eqt-NBV-simplip}) and
(\ref{eqt-cutoff}).}, so that gravity waves can be also derived
from this expression.

From the study of the simplified case, we retain that the acoustic
radius is better adapted than the radius for describing a pressure
wave. Moreover, a change of variable is able to provide new
variables showing reduced amplitude gradient
(Fig.~\refer{fig-coupure-jup}).

\subsection{Differential equations\label{complet}}

In this section, we examine the main steps for achieving a
solution in the general case. The methodology for entering more
complex solutions based on less strong hypotheses than those in
the previous simple case is in fact basically the same: an
appropriate change of variables, as done in
Eq.~(\refer{eqt-cas-simple}), provides a new system of
differential equations derived from Eq.~(\refer{eqt-set-equadif}).
This system has then to be organized in a convenient way for a
numerical resolution. Boundary conditions are used to solve the
eigenvalue problem.

\subsubsection{Chosen set of variables}

The equations explaining the propagation of the waves are
expressed with the  following variables:
\begin{eqnarray}
  \labell{eqt-system-set}
 \left\{%
 \begin{array}{rll}
   \hbox{vertical displacement:} &\xir  & \left(\hbox{so that } v\ind{r} =
\disp{\diff \xir\over \diff t}\right) \\
   \hbox{pressure:}              & p \\
   \hbox{gravitational potential:} &\psi \\
  \end{array}%
  \right.
\end{eqnarray}

The time dependence, for all terms, is harmonically expressed by
$\exp i\omega t$. Similarly, the angular dependence is expressed
by the spherical harmonics $\Ylm$ (Appendix \refer{ylm}). Strictly
speaking, the notations should change, in order to take this
factorization into account. For simplicity, we keep it the same,
but have in mind that we should move to notations $x \to \tilde x$
such that
\begin{equation}\label{eqt-nota}
   x (\mathbf{r}, t)\ \equiv  \ \tilde x(r) \ \Ylm (\theta, \varphi)  \ \exp i\omega t.
\end{equation}
So, slowly but surely we leave the real space for the Fourier
space; the angles $\theta$ and $\phi$, and the time $t$ are
superseded with the quantum numbers $\ell$ and $m$, and the
frequency $\omega$, respectively. As a consequence of
Eq.~(\refer{eqt-nota}), radial partial derivatives are replaced by
simple derivatives in the following equations, since $\partial x /
\partial r = \diff \tilde x / \diff r$.

\subsubsection{Density and horizontal displacement}

Compared to the set of variables used for describing the fluid
(Section \ref{equation-fluide}), with now the velocity $\bm v$
replaced by the displacement $\bm \xi$, we have left the density
and the horizontal component of the displacement in the new set of
variables (Eq.~\ref{eqt-system-set}).

It is worthwhile to express the Eulerian density perturbation as a
function of the pressure perturbation and the radial displacement
(Eq.~\ref{eqt-def-eul-lag}) in order to emphasize the difference
between pressure and gravity waves. From the Lagrangian form of
the adiabatic propagation (Eq.~\ref{eqt-adiab1}) and from the
definition of the sound speed profile (Eq.~\ref{eqt-son3}), we
have
\begin{equation}\labell{eqt-diff-rho0}
    \rho
    = {p \over c^2}
    + \rho_0 \ \left( {1\over c^2}  {\bm\nabla} p_0 -  {\bm\nabla} \rho_0 \right)\ {\bm
    \xi}
    .
\end{equation}
Due the absence of horizontal gradient of structure parameters in
the spherical symmetry, this simplifies into
\begin{equation}\labell{eqt-diff-rho}
    \rho
    = {p \over c^2}
    - \rho_0 \ \left( {1\over \Gamma_1 \Hp} - {1\over H_\rho} \right)\ \xir
    .
\end{equation}
If we introduce the \BV\ frequency, we get
\begin{equation}\labell{eqt-diff-rho1}
    \rho
    = {p \over c^2}
    + {\rho_0\, \NBV^2\over g_0}\ \xir
    .
\end{equation}
This demonstrates that the Eulerian density perturbation is always
related to the Eulerian pressure perturbation and to the radial
displacement. The importance of this latter term
increases with increasing $\NBV$ values, and decreases with increasing $c$ values.\\

Similarly, the horizontal displacement can express as a function
of the pressure perturbation and the gradient of the potential
\citep[e.g., section 3.3 of][]{2010aste.book.....A}. This derives
from the horizontal component of the equation of motion. From
Eq.~(\ref{eqt-mvt1}) and from the fact that interior structure
parameters have no horizontal variation, one gets
\begin{equation}\labell{eqt-diff-xih0}
    \omega^2 \,\xih
    =
    \nabla\ind{h}\left( {p\over \rho_0} + \psi \right)
    .
\end{equation}
The horizontal gradient and the vectorial dependence of the
horizontal displacement (Eq.~\ref{eqt-ylm-disp}) match, so that
\begin{equation}\labell{eqt-diff-xih}
    \xih
    =
    {1 \over r \omega^2}  \
    \left( {p\over \rho_0} + \psi \right)
    .
\end{equation}

\subsubsection{System of differential equations}

We go back to the equations with the three selected variables.
They obey the set of differential equations:
\begin{eqnarray}
\scriptsize{\left\{
\begin{array}{rrrr}
 \disp{\diff \xir\over \diff r} = &
 \left(\disp{1\over\Gamma_1 \, \Hp} -\disp{{2\over r}}\right)\ \xir &
 + \disp{1\over \rho_0 c^2} \left(\disp{{S_\ell^2\over \omega^2}-1} \right)\ p &
 + \disp{\ell (\ell + 1) \over \omega^2 r^2} \ \psi \\
 \disp{\diff p\over \diff r} = &
 \rho_0 (\omega^2 - \NBV^2) \ \xir &
 - \disp{1\over \Gamma_1 \,\Hp}\ p &
 - \rho_0\ \disp{\diff \psi\over \diff r} \\
 \disp{1\over r^2}
 \disp{\diff\over \diff r} \left( r^2 \disp{\diff \psi\over \diff r}
 \right) = &
   \disp{4\pi{\cal G} \rho_0\over g_0} \NBV^2 \ \xir &
 + \disp{4\pi{\cal G} \over c^2} \ p &
 + \disp{\ell (\ell + 1) \over r^2}\ \psi \\
 \end{array}
 \right.
 \labell{eqt-syst-base}
 }
\end{eqnarray}
Where does this come from? The aim of this lecture is not to redo
calculations that are published in many places, e.g., section 14.1
of \cite{1989nos..book.....U} or section 3.3 of
\cite{2010aste.book.....A}, but to help understanding how things work:\\
 - we have here an apparent set of three differential equations for three variables;
 in fact, due to the double derivative of $\psi$, this is a fourth-order system of differential equations
 with the four (dependent) variables $\xir$, $p$, $\psi$ and $\diff \psi /
\diff r$;\\
- the frequency $\omega$ has replaced all time derivatives;\\
- the terms $1/r$ and $1/r^2$ were introduced by the operators
divergence ($\mathbf{\nabla}$) et Laplace operator ($\nabla^2$) in
spherical coordinates;\\
- the term $\ell (\ell+1)$ comes from the horizontal gradient of
the spherical
harmonics (Eqs.~\refer{eqt-ylm-ell}, \refer{eqt-kh});\\
- all coefficient are real, so that the eigenvalues $\omega^2$ are
real too. We can get either a sinusoidal variation or an
exponential decay, but not decaying or growing oscillations.\\

Unsurprisingly, seismic parameters introduced in the previously
studied simplified cases are used in these equations: sound speed
$c$, \BV\ frequency $\NBV$, pressure scale height (Section
\ref{section-BV}). As a consequence of the supposed absence of any
horizontal gradient and of the properties of the spherical
harmonics (Eqs.~\refer{eqt-ylm-ell}, \refer{eqt-kh}), the
horizontal wavevector verifies
\begin{equation}
   \labell{eqt-diff-kh}
   \kh^2
   =
   {\ell (\ell+1) \over r^2}
   .
\end{equation}
This property is similarly expressed by the Lamb frequency
$S_\ell$,
\begin{equation}\label{eqt-lamb}
    S_\ell^2
    \egaldef
    \ell (\ell + 1)\ {c^2\over  r^2}
    =
    c^2 \kh^2
    .
\end{equation}
The variations of the functions $S_\ell$ and $\NBV$ are shown in
Fig.~\refer{fig-BV-SL-soleil} for the Sun and in
Fig.~\refer{fig-BV-RGB} in the red giant case. We notice that the
azimuthal order $m$ is absent in the previous system of equations
(\refer{eqt-syst-base}) since the problem is fully spherically
symmetric.

We stress that the system of equations (\ref{eqt-syst-base})
is not the only way to address the solution. With the variables
$\xir$, $\xih$ et $\psi$, the equations of hydrodynamics should
have been expressed as
\begin{eqnarray}
 \scriptsize{
 \left\{\begin{array}{rrrr}
 \disp{\diff \xir\over \diff r}
 =&
 \left[ \disp{1\over\Gamma_1\,\Hp}
- \disp{2\over r} +
 \right]
 \xir &
 +\disp{r\omega^2\over c^2} \left[ \disp{S_\ell^2\over
 \omega^2}-1\right]  \xih &
 -\disp{1\over c^2} \psi
 \\
 \disp{\diff \xih\over \diff r}  =&
 \disp{1\over r}\left[1- \disp{\NBV^2\over\omega^2} \right] \xir &
 + \left[ \disp{\NBV^2\over g_0} - \disp{1\over r} \right]  \xih &
 - \disp{\NBV^2 \over r g_0 \omega^2} \psi
 \\
 \disp{1\over r^2}
 \disp{\diff\over \diff r} \left( r^2 \disp{\diff \psi\over \diff
 r} \right) = &
 \disp{4\pi{\cal G} \rho_0\over g_0}\ \NBV^2\, \xir&
 +
 \disp{4\pi{\cal G} \rho_0 r\over c^2}\ \omega^2\, \xih &
 + \left[ \disp{L^2 \over r^2} - \disp{4\pi{\cal G}\rho_0 \over c^2} \right]
 \psi
 \\
 \end{array}\right.
 .
 }
 \end{eqnarray}

Regardless of the choice of the variables, the next steps, not
presented here, consist in the conditioning of the equations
\citep[e.g., section 3.3 of][]{2010aste.book.....A}. With an
appropriate change of variables, the numerical resolution is
derived from a system in a matricial form
\begin{equation}\label{eqt-sys-resol}
    {\diff\over \diff r} X \ = [M] \ X ,
\end{equation}
where $X$ is a vector of 4 elements according to the set of
differential equations (Eq.~\refer{eqt-syst-base}), completed with
the boundary conditions presented hereafter.

\subsubsection{Boundary conditions\label{section-boundcond}}

Boundary conditions must now be introduced to solve the previous
system of differential equations (Eq.~\ref{eqt-syst-base}).\\

- At the center, due to the spherical symmetry, vectorial
perturbations should cancel. Diverging terms must also be avoided.
From the  properties of the Legendre polynomials, we get
\begin{eqnarray}
\left\{
\begin{array}{rcl}
\xir &=& \ell\ \xih\cr \disp{\diff \psi\over \diff r} &=& \ell\
\disp{\psi \over r}\cr
\end{array}
\right.
\end{eqnarray}
These conditions come from the properties of the Legendre
development, with terms proportional to $r^\ell$.\\

- At the surface, the gravitational potential has to fulfil a
continuity condition with the external potential. Out of the star,
it expresses
\begin{equation}
 \nabla^2 \psi = 0 ,
\end{equation}
then
\begin{equation}
 {\partial \psi \over \partial r} + {\ell + 1\over r} \psi \ = \ 0
 .
\end{equation}
The boundary condition on the pressure depends on the atmospheric
model. If the surface is free of any constraint, the Lagrangian
pressure perturbation is necessarily zero
\begin{equation}
  \delta p
  =
  p + \depl .  {\bm\nabla} p_0
  = 0 .
\end{equation}
The hypotheses used for simplifying the equations are not valid in
the uppermost layers where the density and pressure scale heights
are small and where the radiative time scale is short. This
complicates the seismic analysis (see Section~\refer{surface}).

\subsubsection{Cowling approximation\labell{section-cowling}}

The Cowling approximation consists in neglecting the restoring
force due to the Eulerian perturbation of the gravitational
potential \citep{1941MNRAS.101..367C}. As seen in Section
\refer{difference-p-g}, this is valid where the number of angular
and radial nodes is high enough for blurring the influence of the
perturbation of the potential. In this case, the system of
differential equations (Eq.~\refer{eqt-syst-base}) reduces to a
much more tractable second-order differential system of equations
\citep[e.g., Section 15.1 of ][]{1989nos..book.....U}.

\begin{eqnarray}
\left\{
\begin{array}{rcccccc}
  \disp{\diff \xir\over \diff r}
  &=& - \left[\disp{2\over r} - \disp{1\over\Gamma_1 \Hp}\right] &\xir
  &+&\disp{1\over \rho_0 c^2} \left[ \disp{S_\ell^2\over \omega^2}-1\right] &p \\
  \disp{\diff p\over \diff r}
  &=&
  \rho_0\left[\omega^2- \NBV^2 \right] &\xir
   &-&\disp{1\over\Gamma_1 \Hp} &p
   .\\
\end{array}
\right. \label{eqt-system-cowling}
\end{eqnarray}
For low degrees and low radial orders, encountered for instance in
the red giant oscillation spectrum, the Cowling approximation may
be too crude. For dipole modes ($\ell = 1$),
\cite{2006PASJ...58..893T} has found a way for reducing the system
of differential equations to second order without using the
Cowling approximation.

\begin{figure}[!t]
 \fichier{9.5}{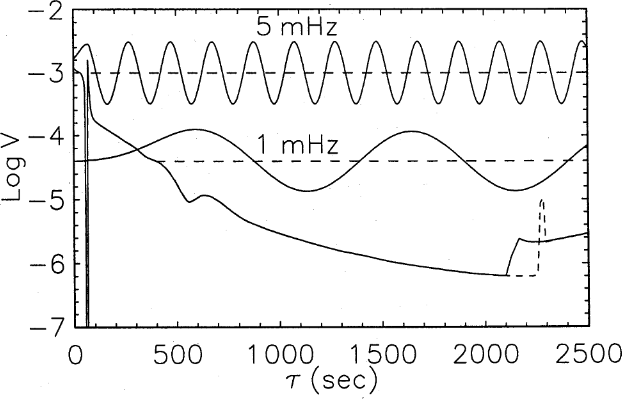}
 \legende{Acoustic potential}{Acoustic potential versus acoustic depth
 in a standard solar model. Also shown are the wavefunctions $\zeta(\tau)$ for the wave frequencies 1 and 5\,mHz
 \credit{1994MNRAS.268..880R}.
 \labell{fig-pot-sol}}
\end{figure}

\subsubsection{Dispersion equation}

If spatial variations of the interior structure parameters are
neglected compared to the variations of the perturbed
terms\footnote{This is relevant in most regions of the stars, but
not near the surface, where the pressure scale height has very low
values.}, the system (Eq.~\ref{eqt-system-cowling}) reduces to
\begin{eqnarray}
\left\{
\begin{array}{rccc}
 \disp{\diff \xir\over \diff r}
 &=&
 \disp{1\over \rho_0 c^2} \left[\disp{S_\ell^2\over \omega^2}-1\right] &p, \\
 \disp{\diff p\over\diff r}
 &=&
 \rho_0 \left[\omega^2- \NBV^2 \tvi\right] &\xir .\\
\end{array}
\right.
\end{eqnarray}
This gives, when assuming that the derivatives of the equilibrium
quantities can be neglected too,
\begin{equation}
 \disp{\diff^2 \xir\over \diff r^2}
 \ =\ - {\omega^2\over c^2}
 \left[1- \disp{ \NBV^2 \over \omega^2}\right]
 \left[1- \disp{S_\ell^2\over \omega^2}\right]\ \xir .
 \labell{eqt-dif-cow}
\end{equation}
So, we can derive the dispersion equation
\begin{equation}
 c^2 \kr^2 \ = \ \omega^2 \left[1- \disp{\NBV^2\over \omega^2}\right]
 \left[1- \disp{S_\ell^2\over \omega^2}\right]
 ,
 \labell{eqt-dispersion}
\end{equation}
from which it is possible to infer the main properties of the
waves.\\

There are many different ways to reach a dispersion equation
similar to Eq.~(\refer{eqt-dispersion}).
\cite{1989nos..book.....U} propose the change of variables (see
their chapter 16):
\begin{eqnarray}
\left\{
\begin{array}{rclll}
 u
 &=&
 \rho_0^{1/2}\ c r\ \omega & \left|\omega^2 - S_\ell^2\right|^{-1/2}  &
 \xir ,
 \\
 v
 &=&
 \rho_0^{1/2}\ r^2\ \omega^2 & \left|\NBV^2 - \omega^2\right|^{-1/2}  &
 \xih .
 \\
\end{array}
\right. \labell{eqt-changes}
\end{eqnarray}
These new variables are both solutions of the second-order
differential equation (Eq.~\refer{eqt-dif-cow}).

\begin{figure}[!t]
 \fichier{10}{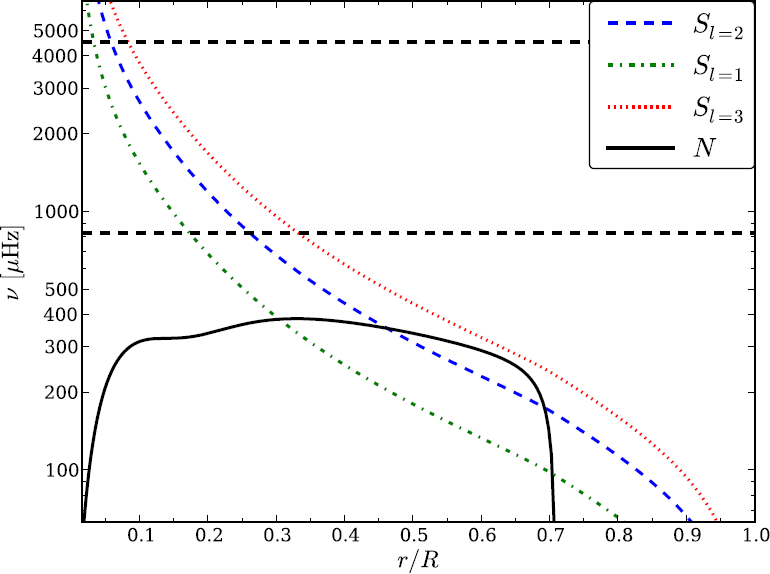}
 \legende{Propagation diagram in the solar interior} {Propagation
 diagram: pressure and gravity waves oscillate in different
 cavities. Solar pressure waves oscillate around 5\,min, or 3\,mHz,
 or 20\,mrad\,s$^{-1}$, much above the \BV\ cavity. The horizontal
 dashed lines show the frequency range  where solar pressure modes are observed
 \credit{2014ApJ...790..121L}.
 \labell{fig-propagation}}
\end{figure}

\begin{figure}[!t]
 \fichier{10.0}{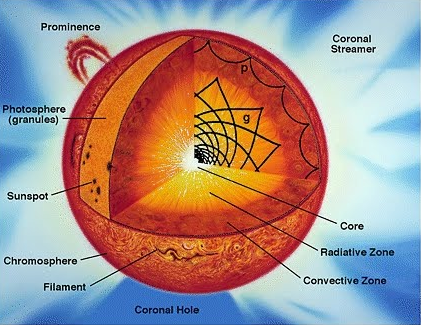}
 \legende{Oscillation modes} {Gravity and
 pressure modes in the Sun \textsl{(NASA/GONG)}. \labell{fig-modespg}}
\end{figure}

\subsubsection{Acoustic potential}

As an alternative, solutions to the system of differential
equation can be found with a methodology based on the acoustic
potential $V$ \citep[e.g.,][]{1994MNRAS.268..880R}. Similarly to
the development proposed in Section \refer{simple-case}, a change
of variables provides an equation similar to
Eq.~(\refer{eqt-dispers-simple})
\begin{equation}\labell{eqt-potentiel}
    {\partial^2 \zeta \over \partial\tau^2} + \left[\omega^2 - V(\tau)\right] \
    \zeta
    =
    0 ,
\end{equation}
where $\tau$ is the acoustic radius (or depth) and $\zeta$ is a
function close to the kinetic energy density
(Fig.~\refer{fig-pot-sol}). In the spherical case, the expression
of $\zeta$ cannot correspond to $\eta$ previously defined by
Eq.~(\refer{eqt-eta2}) since it has to account for the sphericity.
Then, the potential $V$ (Fig.~\refer{fig-pot-sol}) captures the
physics of the wave.

\subsection{Propagation diagram\labell{propagation}}

The propagation diagram is important to define the region where a
wave propagates. This region depends on the type of the wave. The
dispersion equation (\refer{eqt-dispersion}) shows that the
propagation condition (ie. $\kr^2 > 0$) implies
\begin{eqnarray}
 \left[
 \begin{array}{ccc}
  \omega^2 > S_\ell^2 &\hbox{ and }& \omega^2 > \NBV^2 \cr
                &\hbox{ or }&                \cr
  \omega^2 < S_\ell^2 &\hbox{ and }& \omega^2 < \NBV^2 . \cr
  \end{array}
  \right.
\end{eqnarray}
Since we have $\NBV<S_\ell$ nearly everywhere (cf.
Fig.~\refer{fig-propagation}), the condition most often reduces to
\begin{equation}\label{eqt-modes-p-ou-g}
    \omega  >  S_{\ell}
     \hbox{ \ or  \ }
 \omega  <  \NBV.
\end{equation}
So, the  propagation diagram emphasizes waves of two types
(Fig.~\refer{fig-propagation} and \refer{fig-modespg}) :

- Pressure waves with $\omega\ > \ S_\ell$. \\
- Gravity waves with $\omega \ < \ \NBV$.

\subsection{Radial quantization}

\subsubsection{Radial trapping}

The frequencies $S_\ell$ et $\NBV$ are crucial for determining the
region where the waves propagate. Here, we study the simple cases
where a wave has either a very low or a very high frequency:
$\omega \ll \NBV$ or $\omega \gg S_\ell$, respectively.

If we note $\ra$ and $\rb$ the internal and external turning
points of the waves, the resonance condition writes
\begin{equation}
 \int_{\ra}^{\rb} \kr\  \diff r \ = \  (n+\alpha_\ell)\ \pi
 ,
 \labell{eqt-reson-n}
\end{equation}
where $\alpha_\ell$ is a small constant term that expresses the
fact that complex phenomena occurring mainly in the upper
atmosphere were neglected. In these regions, non-adiabaticity
cannot be neglected; any shift due to such an effect has a large
influence since then the low value of the sound speed may add
non-negligible offsets.

\subsubsection{High-frequency pressure modes}

From the dispersion equation (Eq.~\ref{eqt-dispersion}), we derive
a simple form at high frequency ($\omega \gg \SL$)
\begin{equation}
c^2 \kr^2 \ = \ \omega^2 .
\end{equation}

We further simplify the integration with the assumptions that
$\ra$ is close to the center of the star, which is valid for
radial modes with $\ell=0$,  and that $\rb$ is close to the
surface. Then, we get from the boundary conditions
\begin{equation}
\int_0^{R} \kr\ \diff r \ \simeq \  (n+\alpha_\ell)\ \pi .
\end{equation}
So, with the dispersion equation we have
\begin{equation}
  \int_0^{R} {\diff r \over c} \ \simeq \  {(n+\alpha_\ell)\ \pi \over
  \omega}
  .
\end{equation}
This result, expressed in cyclic frequency instead of angular
frequency, gives
\begin{equation}
  \nu_{n,\ell} \simeq (n+\alpha_\ell)\ \Dnuas
  ,
  \labell{eqt-rad-as}
\end{equation}
where we identify the asymptotic frequency spacing
\begin{equation}
  \Dnuas = \left( 2 \ \int_0^{R} {\diff r \over c} \right)^{-1} ,
  \labell{eqt-dnu-as}
\end{equation}
which corresponds to the inverse of the stellar acoustic diameter.

\subsubsection{Low-frequency non-radial gravity modes}

From the dispersion equation (Eq.~\ref{eqt-dispersion}) and the
definition of the Lamb frequency, we derive a simple form of the
radial wavevector at very low frequency ($\omega \ll \NBV$)
\begin{equation}
\kr \ = \ { \sqrt{\ell (\ell+1)} \over r}\ {\NBV \over \omega} .
\end{equation}
The integration of the boundary equation (Eq.~\ref{eqt-reson-n})
then gives, with $\mathcal{R}$ denoting the radiative region,
\begin{equation}
 { \sqrt{\ell (\ell+1)} \over   \omega} \int_\mathcal{R} \NBV {\diff r \over
 r}\ \simeq \  (n+\alpha')\ \pi .
\end{equation}
Introducing the period of the mode, we get
\begin{equation}
   P_{n,\ell\not=0} \simeq (n+\alpha')\ \Delta\Pi_\ell ,
\end{equation}
where we identify the asymptotic period spacing
\begin{equation}
  \Delta\Pi_\ell =  { 2\pi^2 \over \sqrt{\ell (\ell+1)}} \ \left(
  \int_\mathcal{R} \NBV {\diff r \over r} \right)^{-1} .
\end{equation}

\subsection{Reflection and refraction}

\subsubsection{The cutoff frequency}

We go back to the dispersion equation
(Eq.~\ref{eqt-dispersion-simple}) introduced by the simplified
approach and note the role of the cutoff frequency
(Eq.~\ref{eqt-cutoff}). Near the surface, the dispersion equation
is
\begin{equation}
   \omega^2 \ =\ k^2 c^2\ + \omega^2_c
   .
   \labell{eqt-reflex}
\end{equation}
The formal difference with Eq.~(\ref{eqt-dispersion-simple})
derives from the fact that the wavevector is here associated with
the radius and not the acoustic radius.

One can find multiple expressions of the cutoff pulsation in the
literature, since the way to derive it depends on the variable
used for the calculation: $\depl$, $\vv$, $p_0^{-1/2} \vv$
\citep{1974ahl..book.....B}, $\rho_0^{1/2} c^2 \nabla.\vv$
\citep{1986hmps.conf..117G}, or even  more complex variables as
seen in Eq.~\refer{eqt-changes}. All expressions of the cutoff
frequency have the same first-order, and all present similar
variations (Fig.~\refer{fig-wc-V0})
\begin{equation}
\omega_c\ \egaldef\  {c\over 2H_\rho} .\labell{eqt-coup}
\end{equation}
Since the density variation with altitude is much more important
than the sound-speed variation, especially near the \emph{surface}
of the star\footnote{The definition of the surface of an object is
an issue, even more for a fully fluid object.}, the highest value
of $\wc$ is reached at the level where the density scale height is
minimum, close to the level where the temperature is minimum
(Fig.~\refer{fig-wc-V0}). This define the `seismic surface', and
implies that waves are reflected there.

The cutoff frequency shows large values and a steep profile at the
surface. This ensures that all waves are approximately reflected
at the same location. As already stated (Section
\ref{section-boundcond}), things are more complex in this region:
a phenomenological analysis of the conditions of the reflection is
presented in Section.~\refer{surface}. In the upper regions, the
dispersion equation writes
\begin{equation}
  \omega^2 \
  = \ k^2 c^2 + \omega^2_c \ \simeq \ \kv^2 c^2 + L^2\
  {c^2\over R^2} + {c^2\over 4 H_\rho^2}
  ,
\end{equation}
according to Eqs.~(\ref{eqt-diff-kh}) and (\ref{eqt-coup}), and
with $L^2 = \ell (\ell +1)$. In the ray-tracing approach, $\kv=0$
corresponds to the upper level where the wave (here, the ray) is
reflected. The leading order of the expression of $\wc$ indicates
that, except for very high angular degrees, the reflection of the
wave indeed occurs when $\kv =0$, so that $\omega =
\omega\ind{c}$. In the layers just below this level, $\kv \gg
\kh$, except for very high values of $\ell$ since the condition
$\kv \gg \kh$ is equivalent to:
\begin{equation}
\omega\ind{c} \gg c \kh .
\end{equation}
From Eq.~(\ref{eqt-lamb}), this implies that the wave displacement
at the surface is nearly vertical when
\begin{equation}
\ell \ll \ell\ind{max} = {R\over 2 H_\rho} .
\end{equation}
In the solar photosphere, $H_\rho = \Hp=\kB T/\mu g \simeq
200$~km, so that $\ell\ind{max} \simeq 700 $.\\

\begin{figure}[!t]
 \fichier{8.0}{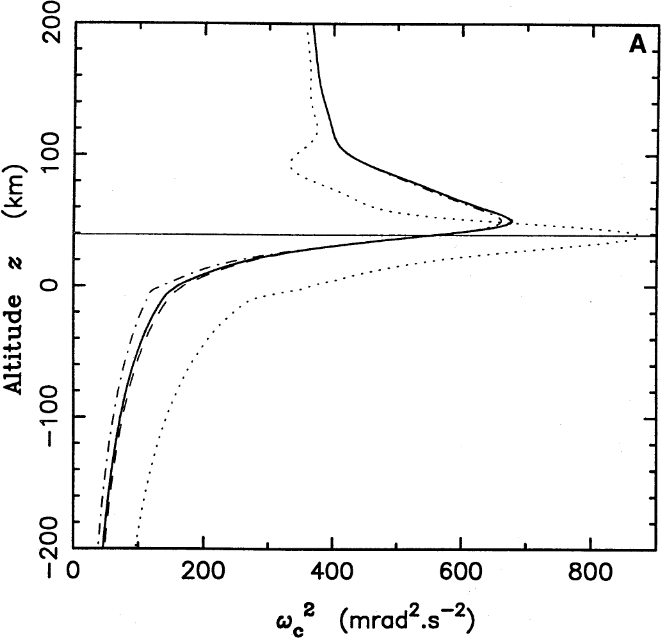}
 \fichier{8.0}{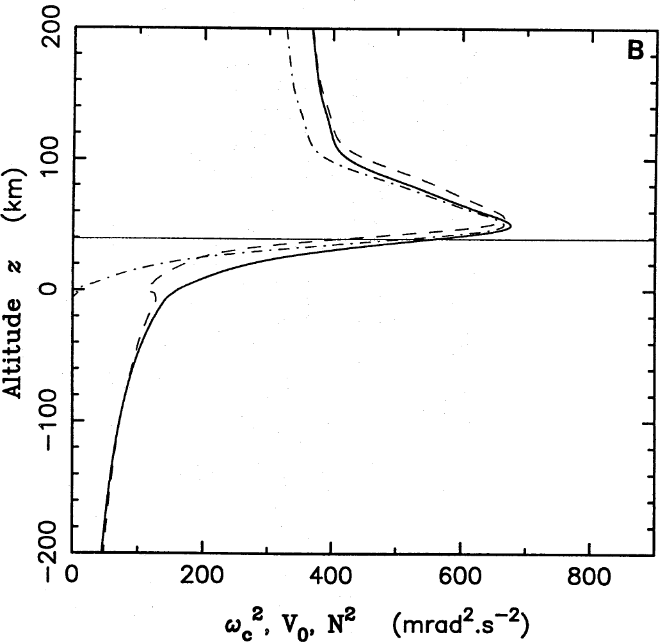}
 \legende{Cutoff
 frequency}{\emph{Top:} Different expressions of the cutoff
 frequency
 \citep{1974ahl..book.....B,1986hmps.conf..117G,1995A&A...293..586M}
 in the Jovian atmosphere. Their maximum value is reached either at
 the altitude with minimum temperature (horizontal line) or at the
 altitude corresponding to the minimum density scale height. $\wc$
 corresponding to Eq.~(\refer{eqt-coup}) is plotted with the full
 line. \emph{Bottom:} comparison with the acoustic potential
 (dashed line) and the \BV\ frequency (dot-dashed line). Full
 expressions are given in
 \cite{1995A&A...293..586M}.
 \labell{fig-wc-V0}}
\end{figure}

To conclude with the cutoff frequency, it is useful to recall that
the frequencies introduced for expressing the physics of the
oscillations are all correlated. In the upper stellar troposphere,
where the atmosphere can be considered as isothermal, the cutoff
and \BV\ frequencies have similar expressions, as seen in
Fig.~\refer{fig-wc-V0}. In fact, the hypotheses imply $H_\rho =
\Hp = \kB T/\mu g$ and $c^2= \Gamma_1 \kB T/\mu$, so that both
$\NBV^2$ and $\wc^2$ vary as $g_0/H_\rho$.

\begin{figure}[!t]
\fichier{7.1}{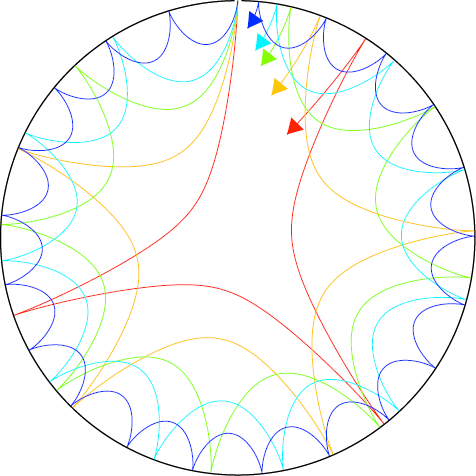} \legende{Propagation and turning
points}{Propagation and turning points: when $\ell$ increases, the
propagation is restrained to shallower regions.
\labell{fig-prop-tourne}}
\end{figure}

\begin{table}[t]
\begin{tabular}{rllllllll}
\hline
$\ell$ &  0&   5&  15&  40& 100&  500&  1000\\
$\rt/R$&  0& 0.2& 0.4& 0.7& 0.9& 0.99& 0.999\\
\hline
\end{tabular}
\legende{Turning points}{Position of the turning point $\rt$ in
the Sun, for a wave with a frequency close to 3~mHz, as a function
of the degree $\ell$; $\rt$ is limited to the upper envelope when
$\ell$ is high. \labell{tab-tourne}}
\end{table}

\subsubsection{The degree-dependent Lamb frequency $S_\ell$}

In most parts of the objects, the pressure mode frequency is much
larger than any characteristic frequency as $\NBV$ or
$\omega\ind{c}$. The dispersion equation is then
\begin{equation}
  \omega^2\ =\ k^2 c^2 \ = \ \kv^2 c^2 + S_\ell^2 , \labell{eqt-refrac}
\end{equation}
with, as introduced earlier (Eq.~\refer{eqt-lamb}),
\begin{equation}
  S_\ell\ \egaldef\ L\ {c(r) \over r}\ \ \hbox{ \ and \ } L \ =\
  \sqrt{\ell (\ell +1)} . \labell{eqt-Sl}
\end{equation}
The $\SL$ term corresponds to a conservation law related to the
spherical symmetry and the absence of any horizontal gradient in
the equilibrium parameters. As already shown
(Eq.~\refer{eqt-diff-kh}), the horizontal component of the wave
vector writes
\begin{equation}
\kh = {L \over r} .
\end{equation}
In other words, the term $L$ expresses a boundary condition. At
each level $r$ where $\kh$ is defined by
\begin{equation}
  \int\ind{r\ = \ cst} \kh\ \diff x\ =\ 2\pi L .
  \labell{eqt-conL}
\end{equation}
This equation requires $\kh$ to be defined, hence, from
Eq.~(\refer{eqt-diff-kh}), $\kv \le k$.

In the ray-tracing approach, $\kv=0$ corresponds to the refraction
of the wave at the turning point with $r=\rt$. From
Eq.~(\ref{eqt-refrac}), it is easy to show that $\kv=0$ when
\begin{equation}
  \omega \ = \ S_\ell (\rt)\ = \ L \ { c ( \rt)\over \rt } .
\labell{eqt-tourne}
\end{equation}
Accordingly, at fixed frequency, the higher $L$, the shallower
$\rt$ (Fig.~\refer{fig-prop-tourne}, Table~\refer{tab-tourne});
similarly at fixed degree, the higher $\omega$, the deeper the
penetration. Only radial waves, with $\ell=0$, visit the center of
the star.

\begin{figure}[!t]
 \fichier{10.0}{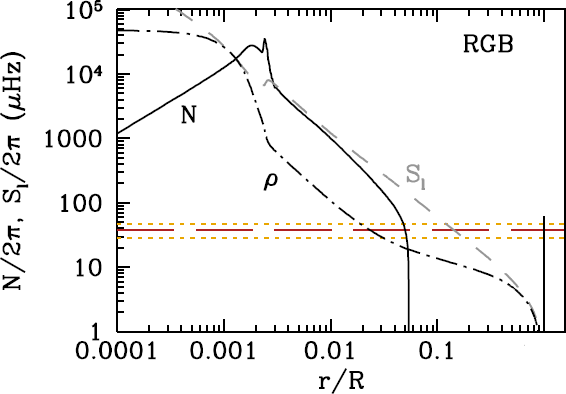}
 \legende{\BV\ frequency
 in red giant stars}{\BV\ frequency in a red giant star, compared
 to $S_\ell$. Both frequencies show similar variation in the
 envelope surrounding the hydrogen burning shell (located for this
 model at $r\ind{H}/R \simeq 0.0025$)
 \credit{2013ASPC..479..435N}.
 \labell{fig-BV-RGB}}
\end{figure}

Figure~\refer{fig-BV-RGB} shows the frequencies $\NBV$ and
$S_\ell$ in the radiative core of a red giant; $\NBV$ and $S_\ell$
have clearly close variations in the region above the
hydrogen-burning shell in a red giant star. This can be explained
by the contrast between the high-density core and the envelope.

Assuming a polytropic function in the envelope ($p_0 \propto
\rho_0^{\Gamma})$, the perfect gas law ($T_0\propto p_0 /\rho_0$),
and a so high density contrast between the core and the
surrounding envelope that $g_0 \propto r^{-2}$ in the envelope, we
derive that both $\NBV$ and $S_\ell$ vary as $r^{-3/2}$,
independent of $\Gamma_1$. With $\Gamma_1 \simeq 5/3$, the density
profile in $r^{1/(1-\Gamma_1)}$ varies also as $r^{-3/2}$. This
explains the parallel variations of $\NBV$, $S_\ell$ and $\rho_0$
just above the hydrogen-burning shell in Fig.~\refer{fig-BV-RGB}.
This property of $S_\ell$ and $\NBV$, generic in a broad
evolutionary range on the red giant branch (RGB) and in the red
clump, has important consequences for the mixed-mode pattern
observed on the RGB (Section \ref{asymp-mixte}).

\subsection{The power of seismology: a differential view}

It is time to explain why seismology is powerful for probing the
interior structure of stars... and of the Earth, the Moon, and
maybe soon the comet Choryumov-Gerasimenko with a radar onboard
the lander Philae launched by the ESA interplanetary probe
Rosetta, Mars with the NASA Discovery Program mission InSight
\citep[Interior Exploration using Seismic Investigations, Geodesy
and Heat Transport,][]{2012LPI....43.1983L}, Jupiter and the giant
planets with dedicated instruments.

Equations (\refer{eqt-refrac}) and (\refer{eqt-tourne}) are enough
to describe the main characteristics of modes, depending on the
radial order $n$ and degree. Modes with cyclic frequencies
$\nu\nl$ and $\nu_{n, \ell\pm 1}$ (or $\nu\nl$ and $\nu_{n\pm 1,
\ell}$) have close turning points $\rt$ and $\rt '$. According to
the ray tracing approach, the frequency differences test the
region between $\rt$ and $\rt '$ (Fig.~\refer{fig-differentiel}).
Precise asteroseismic measurements provide then information about
these regions. The way this is treated in practice in presented in
Appendix \ref{variationalprinciple}.

\begin{figure}[!t]
 \fichier{9.231}{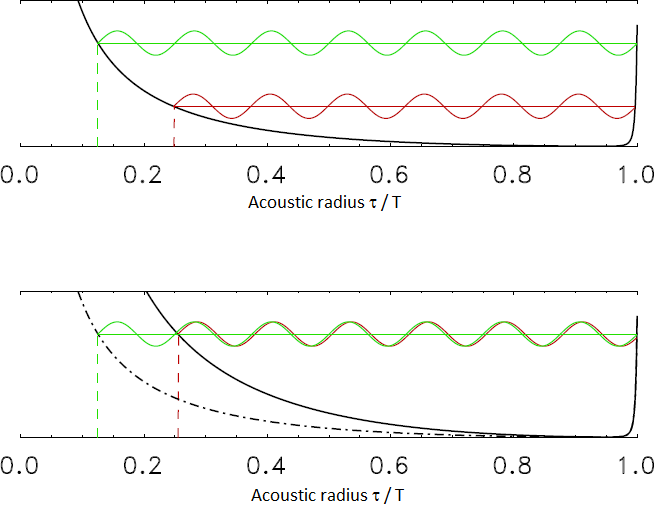}
 \legende{Differential view} {Modes with
 same degree and different frequencies \emph(top),
  or  same frequency and different degrees \emph(bottom),
  probe the stellar interior
  in a different but close way. The comparison of their properties helps
  explaining the properties of the inner region. \labell{fig-differentiel}}
\end{figure}

%%%%%%%%%%%%%%%%%%%%%%%%%%%%%%%%%%%%%%%%%%%%%%%%%%%%%%%%%%%%%%%%%%%%%%%%%%%%%%%%%%
%\clearpage
\section{Normal mode properties\labell{chap4}}

In the previous chapter, the analysis of the wave propagation has
provided us with the definition of global oscillation modes. With
more or less severe assumptions, we could derive simplified
expression for radial pressure modes (evenly spaced in frequency)
and non-radial gravity waves (evenly spaced in period). In this
chapter, we aim at deriving more precise calculations. The lecture
in fact only intends to present the main properties of the
asymptotic expansions. Extended equations can be found in Appendix
\refer{justi-asymp}, which largely refer to the method and
notations introduced by \cite{1980ApJS...43..469T}.

In practice, for asteroseismology there is no general asymptotic
solution for the determination of adiabatic eigenfrequencies
involving a fourth-order system of differential equations as
Eq.~\refer{eqt-syst-base} \citep[e.g.,][]{1986A&A...165..218P}.
Considering the Cowling approximation allows us to reduce the
problem to a second-order differential equation. Then, the
asymptotic method developed by \cite{1954RSPTA.247..307O} applies:
eigensolutions are approximated by Bessel functions near singular
points, and fitted together.

Asymptotic expansions are based on the JWKB method (Jeffreys,
Wentzel, Kramers \& Brillouin) developed in quantum mechanics and
used for finding approximate solutions to linear differential
equations with spatially varying coefficients
\citep{2007AN....328..273G}. The method consists in expressing a
variable with a quasi-harmonic form whose amplitude is smoothly
modulated:
\begin{equation}
  x(r) \ = \ A(r) \exp i {\bf k.r}
  .
\end{equation}
The amplitude gradient fulfills the condition
\begin{equation}
  \left| {\diff A\over \diff r}\right| \ \ll \ | {\bf k}| .
\end{equation}
Equations are manipulated in order to get a second-order
differential equation with the form
\begin{equation}
  {\diff^2 x\over \diff r^2} + K^2 \ x \ = \ 0 , \labell{eqt-asymp-un}
\end{equation}
where $x(r)$ is a function of the wave perturbed terms. With this
development, the factor $K$ has a more or less complex form. For
pressure modes $K \simeq \omega / c$, for gravity modes $K \simeq
\sqrt{\ell (\ell +1)} \NBV / r\omega$, according to
Eq.~(\refer{eqt-dispersion}).\\

\begin{figure}[!t]
\fichier{11.0}{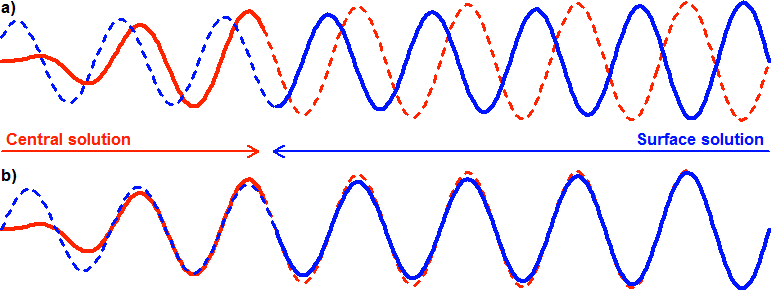} \legende{Asymptotic expansion}{Principle
of the asymptotic expansion: a solution is developed from the
center, another from the surface. They converge in case {\bf b)},
but not in case {\bf a)}. \labell{fig-asymp}}
\end{figure}

\subsection{Low-degree pressure modes}

As a first approximation, we can consider that $K$ in
Eq.~(\ref{eqt-asymp-un}) is equal to $\omega / c$, as for a plane
wave. However, this cannot fit with the solution at the center
of the spherically symmetric star. So, the asymptotic expansion deals with two solutions:\\
- an inner solution is developed in the center; it accounts for
the singularity in $r=0$. \\
- the outer solution takes the outer boundary conditions into account.\\
The solution is valid if independent of the location of the
coupling between the two solutions (Fig.~\refer{fig-asymp}).\\

The development of this solution is proposed in Appendix
\refer{justi-asymp}. The variables, here the Lagrangian pressure
perturbation and the radial displacement, are expressed in terms
of Bessel functions, for both the inner and outer solutions
(Eqs.~\refer{A.3}-\refer{A.6} in Appendix \ref{justi-asymp}). At
first order in frequency, the inner and outer solutions have the
respective phases
\begin{eqnarray}
  \phi_i  &=& \omega \tau_i (r) - (\ell +1/2){\pi\over 2} - {\pi\over 4}\\
  \phi_o  &=& \omega \tau_o (r) - n_o {\pi\over 2} - {\pi\over 4}
  \labell{eqt-asymp-no}
\end{eqnarray}
where $n_o$ is a small offset relating the phase offset in the
outer region, and $\tau_{i}(r)$ and $\tau_{o}(r)$ are integrals of
the acoustic radius
\begin{equation}
  \tau_i (r) = \int_0^r {\diff r \over c}
  \hbox{ \ and \ }
  \tau_o (r) = \int_r^R {\diff r \over c}.
\end{equation}
Continuity of the variables implies that the phases verify
\begin{equation}\label{eqt-raccord-as}
   \phi_i + \phi_o = p\ \pi  \pm {\pi \over 2} ,
\end{equation}
where $p$ is an integer. As a consequence, the first-order cyclic
frequency follows the relation
\begin{equation}\label{eqt-raccord-1}
   \nu \ = \ \left(
   p + {\ell \over 2} + {1 \over 4} + {n_o \over 2}
   \right)
   \ \Dnuas
   ,
\end{equation}
with $\Dnuas = (2 \int_0^R {\diff r / c})^{-1}$, as already
introduced (Eq.~\ref{eqt-dnu-as}). The integer $p$ derives from
the resonance condition; the terms $\ell+1/2$ and $n_o$, both
divided by 2, derive from the Bessel functions used for the
development of the wavefunction.

\begin{figure}[!t]
 \fichier{9.231}{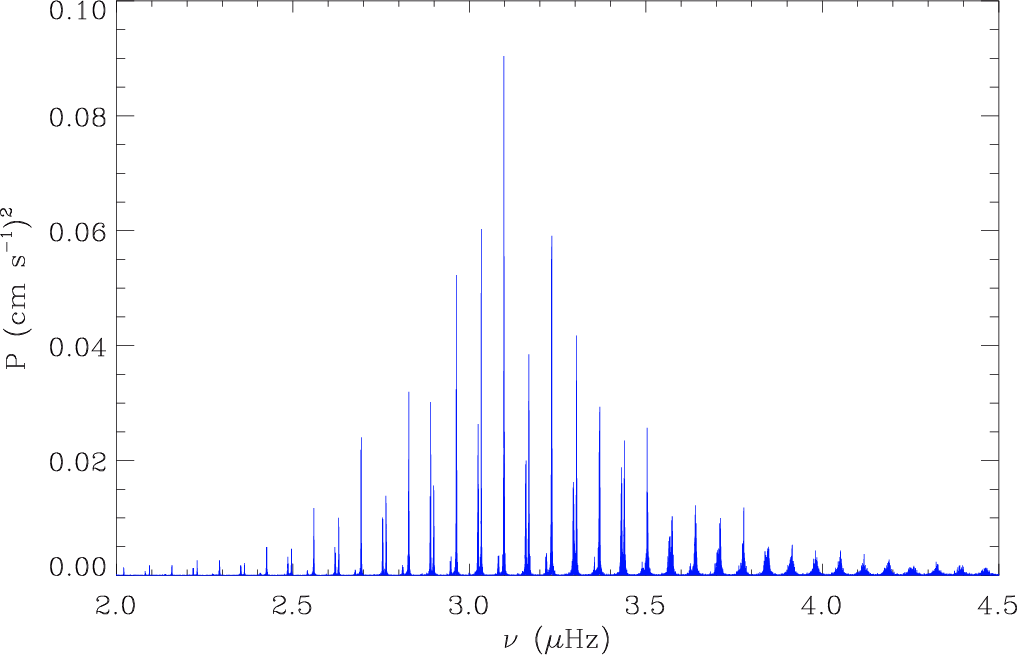}
 \legende{Low-degree oscillation spectrum} {The Sun as a star: low-degree oscillation spectrum
 (SoHO/Golf) exhibit the comb-like pattern corresponding to the asymptotic expansion.\labell{fig-iphir}}
\end{figure}

\subsubsection{Asymptotic oscillation pattern}

Low-degree pressure modes $(\ell \ll n$) follow the asymptotic
relation \citep{1980ApJS...43..469T}, which defines the well-known
comb-like structure of the solar-like oscillation pattern here
expressed as in \cite{1986hmps.conf..117G}
(Fig.~\refer{fig-iphir}):
\begin{equation}
  \nu\nl \ = \ \left( n+ {\ell\over 2} + \varepsilon \right)\ \Dnu \
  -\ {\ell (\ell+1) +\delta \over n+ \displaystyle{\ell\over 2} +
  \varepsilon} \ A ,\labell{eqt-asymp}
\end{equation}
with
\begin{eqnarray}
  \Dnu &=& \displaystyle{\left( 2 \int_0^{R}
     {\diff r\over c} \right)^{-1}} \labell{eqt-def-Dnu},
     \\
   A &=&
   \displaystyle{{1\over 4\pi^2} \left( {c(R)\over R} -
   \  \int_0^{R} {\diff c\over \diff r} {\diff r\over r} \right)},
   \\
    \varepsilon &=& \disp{1\over 4} + {n_o\over 2} \labell{eqt-eps}
   .
\end{eqnarray}

The term $n_o$ is introduced as a small surface correction
(Eq.~\refer{eqt-asymp-no}) but was often affected too big a value,
resulting from a mismatch between the observed and asymptotic
large frequency $\Dnu$ (Eq.~\refer{eqt-def-Dnu}), as explained in
Section \ref{comp-obs}.

The second-order term $A$ accounts for the fact that non-radial
modes do not probe the inner regions, contrary to radial modes.
Since the corresponding waves do not propagate in these regions
where the sound speed is high, the correcting term is negative.

Last but not least, the large separation provides an estimate of
the inverse of the acoustic stellar diameter (Section
\ref{scaling-relations}). This requires that the condition $n\gg
\ell$ is strictly fulfilled, so that the second-order term is
negligible. In fact, such a condition is hardly met (Section
\ref{comp-obs}).

\begin{figure}[!t]
  \fichier{10.0}{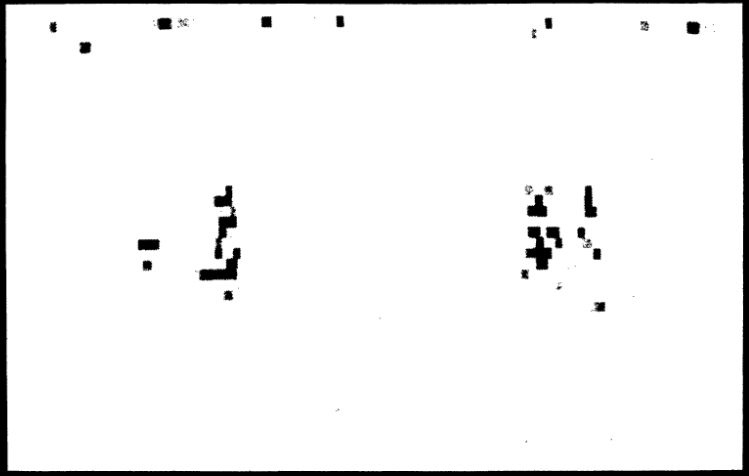}
  \legende{First \'echelle diagram }{First \'echelle diagram of full-disk
  solar observations \citep{1983SoPh...82...55G}. No axes were plotted,
  according to the state-of-the-art of plotting device in 1983. \labell{fig-grec}}
\end{figure}

\begin{figure}[!t]
 \fichier{10.9}{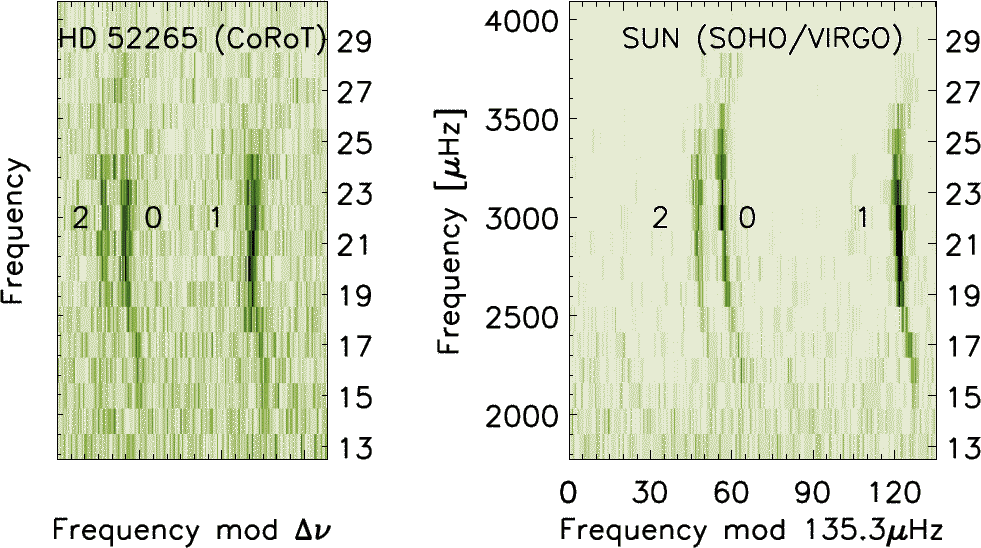}
  \legende{\'Echelle diagrams}{\'Echelle diagram of HD 52265 observed by
  CoRoT \citep{2011A&A...530A..97B}: useful plot for showing the evenly spaced
 frequencies and for enhancing the similarities between solar-like oscillation spectra;
 departure to alignments are due to the second-order term of Eq.~(\refer{eqt-asymp}).
 \labell{fig-echelle52265}}
\end{figure}

\subsubsection{\'Echelle diagram}

The \'echelle diagram\footnote{\'Echelle = ladder in French; the
concept was introduced by \cite{1983SoPh...82...55G}, who reported
full-disk observations of solar oscillations from the geographic
South Pole.} is a very useful representation. It makes profit from
the nearly evenly spaced frequencies. The spectrum being cut in
$\Dnu$-wide slices and these slices being then superposed
vertically, the \'echelle diagram (Fig.~\refer{fig-grec},
\refer{fig-echelle52265}) helps emphasizing the structure of the
oscillation spectra and showing the second-order terms.

As ladders have horizontal bars, the y-axis of the \'echelle
diagram should be the radial order, or the radial order multiplied
by $\Dnu$, but not the frequency.

\begin{figure}[!t]
  \fichier{11}{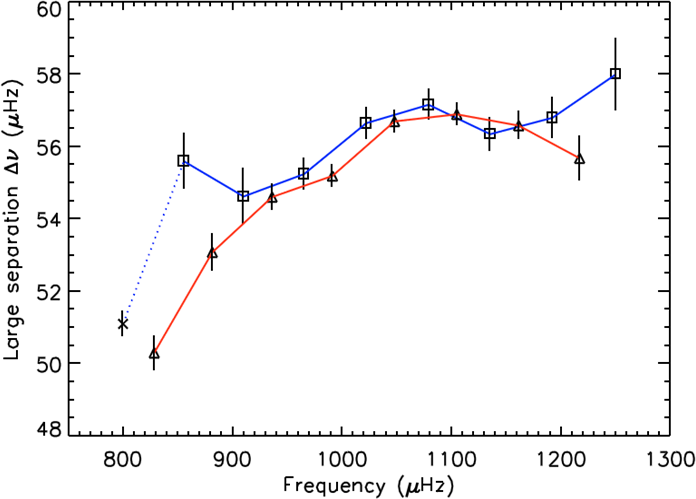}
  \legende{Frequency spacings}
  {Frequency spacings of radial and dipole modes in a
  subgiant, showing a gradient of the frequency difference between
  consecutive radial modes and a significant modulation
  \credit{2011A&A...535A..91D}.
  \labell{fig-sub-seb}}
\end{figure}

\begin{figure}[!t]
  \fichier{10}{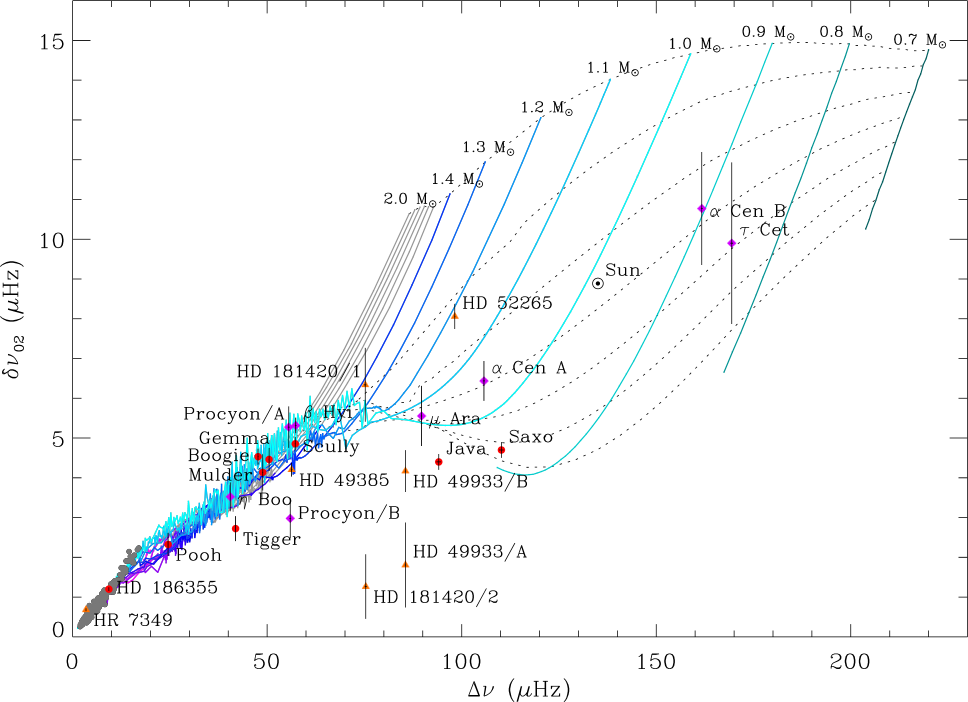}
  \fichier{10}{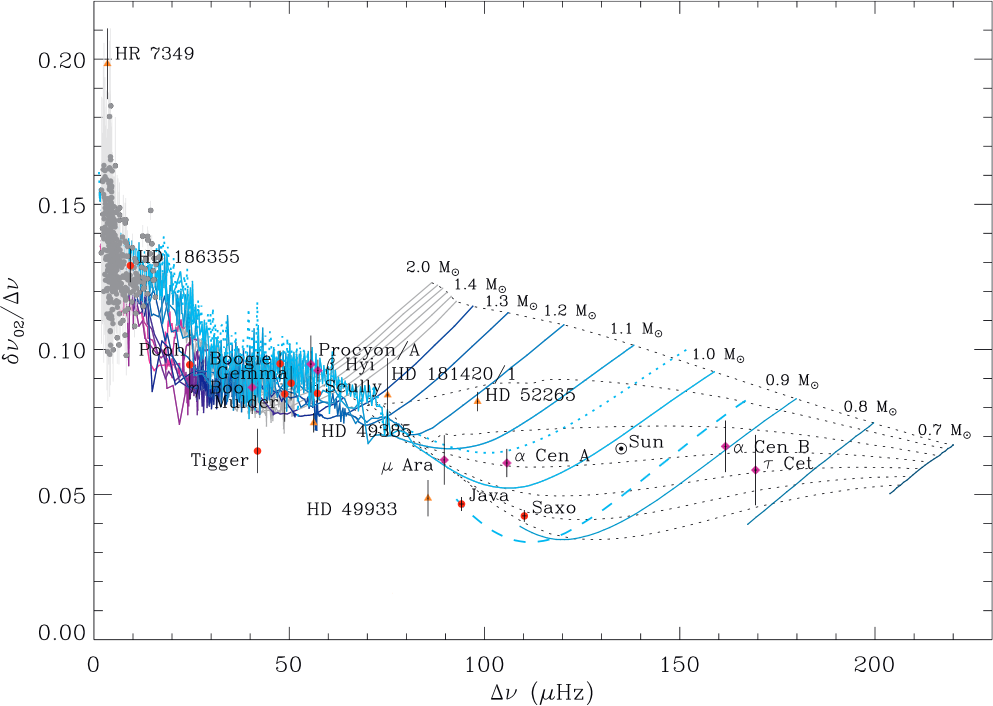}
  \legende{Asteroseismic diagram}
  {Asteroseismic diagram for the small separation between radial and quadrupole modes
  $\delta\nu_{02} \propto\ D_0$ (Eq.~\ref{eqt-def-DO}),
  with isomass levels (full lines) and iso-compositions levels
  (metal-poor dotted lines and metal-rich dashed lines).
  Stars shown were observed by either
  CoRoT (orange triangles), \Kepler\ (red circles), or from the ground
  (purple diamonds)
  \credit{2011ApJ...743..161W}. \labell{fig-HR-D0}}
\end{figure}

\begin{figure}[!t]
  \fichier{10.0}{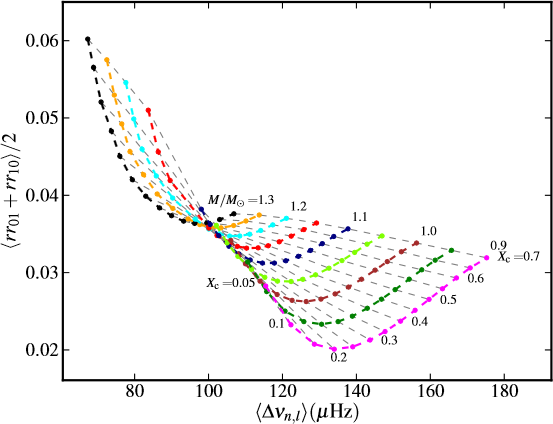}
  \legende{Asteroseismic diagram}
  {Asteroseismic diagram showing the evolution of the mean ratios
  $\langle r_{01} + r_{10}\rangle /2$ as a function of the large
  separation for low-mass main-sequence stars and different hydrogen abundances
  \credit{2013EAS....63..123L}. \labell{fig-ratios-leb}}
\end{figure}

\begin{figure}[!t]
  \fichier{10.0}{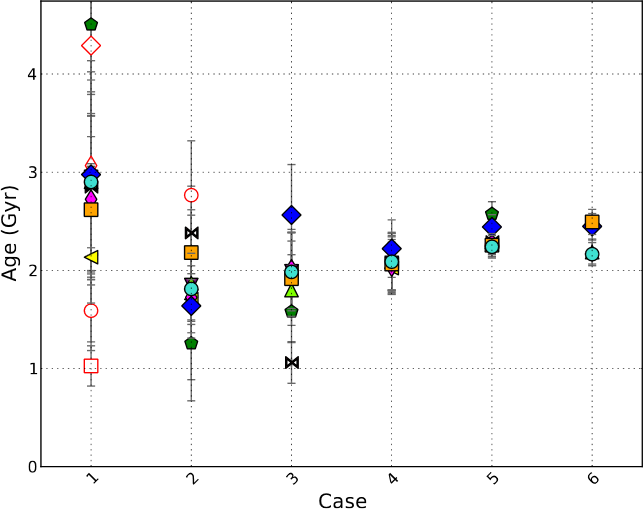}
  \legende{Determination of the stellar age with seismology} {Determination of the age of the main-sequence star
  HD 52265 by different modelling approaches
  based on different sets of constraints. Case 1 corresponds to the classical constraints $\Teff$, $L$ and [Fe/H];
  all other cases includes this constraints plus:
  2) $\langle \Dnu \rangle$;
  3) $\langle \Dnu \rangle$ and $\numax$;
  4) $\langle \Dnu \rangle$ and $\langle d_{02} \rangle$;
  5) $r_{02} (n)$ and $rr_{01/10} (n)$;
  6) $\nu\nl$
  \credit{2013EAS....63..123L}.
  \labell{fig-age-leb}}
\end{figure}

\subsubsection{Large and small frequency spacings}

The low-degree oscillation spectrum provides a relevant
description for solar-like oscillations at various evolutionary
stages. Even if the asymptotic conditions is not met, comb-like
oscillation patterns are observed, so that interesting
equidistances can be derived from the second-order asymptotic
equation.\\

The large frequency separation is closely related to the frequency
spacings between consecutive orders (Fig.~\refer{fig-sub-seb}):
\begin{equation}
\Dnu_{n,\ell} = \nu_{n+1,\ell}- \nu\nl\ \simeq\ \Dnu .
\end{equation}
Small separations are derived from frequency differences between
pairs of frequency with the same $(n+\ell/2)$ value:
\begin{equation}
  \nu_{n, \ell} - \nu_{n-1, \ell+2}
  = {2(2\ell + 3)\ A \over n+\ell/2 + \varepsilon} = 2 (2\ell + 3)\
  D_\ell (\nu)
\end{equation}
This helps defining the term $D_0$ (Fig.~\refer{fig-HR-D0})
\begin{equation}
  \label{eqt-def-DO}
  \delta\nu_{02}
  =  \langle \nu_{n, 0} - \nu_{n-1, 2} \rangle
 \simeq 6\ D_0
  ,
\end{equation}
which provides a useful diagnostic of the evolutionary stage in
combination with the large separation \citep{1988IAUS..123..295C},
as seen in Fig.~\ref{fig-HR-D0}. This seismic diagram shows the
evolution of $\Dnu$ and $\delta\nu_{02}$ for stellar models with a
metallicity close to solar and various masses. Stars in the
main-sequence are more spread than subgiants.
Isochrones are also shown.\\

Weighted small separations, defined by
\begin{eqnarray}
  d^{(2)}_{01} (n)  &=& {1\over 8}
  \left(\nu_{n-1,0} - 4 \nu_{n-1,1} + 6 \nu_{n,0} - 4 \nu_{n,1} + \nu_{n+1,0} \right)
  ,
  \\
  d^{(2)}_{10} (n)  &=& -{1\over 8}
  \left(\nu_{n-1,1} - 4 \nu_{n,0} + 6 \nu_{n,1} - 4 \nu_{n+1,0} + \nu_{n+1,1}
  \right)
  ,
\end{eqnarray}
were proposed by \cite{2003A&A...411..215R} to obtain
dimensionless ratios $r_{01}= d^{(2)}_{01}/\langle\Dnu\rangle$ and
$r_{10}= d^{(2)}_{10}/\langle\Dnu\rangle$ insensitive to gradients
in the frequency differences $\nu_{n,\ell+1} - \nu\nl$
(Fig.~\refer{fig-ratios-leb}).

These ratio are much less perturbed by poorly modelled features
(as the surface effect) than eigenfrequencies. Furthermore, they
allow a more rapid analysis than that conducted with the full set
of frequencies. As an example, the determination of the age of the
main-sequence star HD 52265
\citep{2012A&A...544L..13L,2014A&A...569A..21L} shows how the
precision in the stellar age is increased when seismic parameters
are taken into account (Fig.~\refer{fig-age-leb}). Considering the
small separation (cases 4 and 5) is almost as precise as
considering the full set of eigenfrequencies $\nu\nl$ (case  6).

This justifies the importance of deriving precise eigenfrequency
pattern and emphasizes the useful shortcuts provided by the
different seismic global parameters that make the best of the
seismic information in the different radial orders and degrees.

\begin{table}
  \centering
  \caption{Large frequency spacing measured in Procyon}\label{tab-dnu-procyon}
  \begin{tabular}{rrrl}
 \hline
    Year & $\Dnu$ & Method & Reference \\
 \hline
    1986 & 79.4 & \'Echelle diagram            & (1) resonance cell  \\
    1991 & 71   & Match to asymptotic expansion& (2) FF\'ES\\
    1998 & 53$\pm$3 & Comb response            & (3) FTS \\
    1999 & 55  & CLEAN and asymptotic          & (4) FF\'ES \\
    2004 & 55.5& Mode identification           & (5) FF\'ES \\
    2010 & 56  & 10-telescope network          & (6) FF\'ES\\
  \hline
  \end{tabular}

\small{(1) \cite{1986A&A...164..383G}; (2)
\cite{1991ApJ...368..599B}; (3) \cite{1998A&A...340..457M}; (4)
\cite{1999A&A...351..993M}; (5) \cite{2004A&A...422..247E}; (6)
\cite{2010ApJ...713..935B};

FF\'ES = fiber-fed \'echelle spectrometer

FTS = Fourier transform spectrometer
 }
\end{table}

\begin{figure}[!t]
  \fichier{9.0}{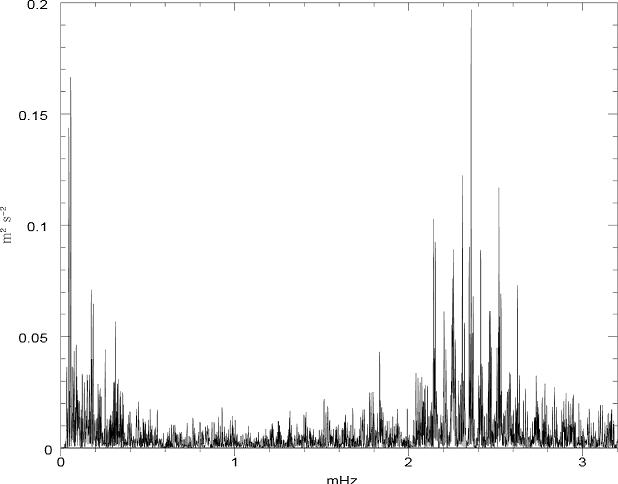}
  \legende{Oscillation spectrum $\alpha$-Cen} {Oscillation pattern in $\alpha$-Cen interpreted
  with the asymptotic expansion of low-degree p modes \citep{2001A&A...374L...5B}.
  This result corresponds to the first unambiguous detection
  of individual solar-like oscillation modes in another star than the Sun
  \labell{fig-acen}}
\end{figure}

\subsubsection{Observations analyzed with the asymptotic expansion}

In the first times of asteroseismology, the identification of a
comb-like spectrum, hence of the large separation, has been the
main tool for identifying solar-like oscillations. Apart from the
\'echelle diagram representation, different methods have been
promoted for the most efficient search of the large separation in
stellar oscillation spectra observed from the ground as the comb
response \citep{1995AJ....109.1313K}. Compared to the current
standards with the space-borne missions with month-long or even
year-long continuous time series, these spectra were derived from
awfully short, non-continuous, and/or noisy time series.\\

A favorite target for searching solar-like oscillations was the F5
star Procyon. This star has however proven to be a complex seismic
target, as all F-type stars
\citep{2008A&A...488..705A,2009A&A...507L..13B}. The quest of its
global oscillations took a long time, as seen by the measurements
of the large separation reported by different observers
(Table~\refer{tab-dnu-procyon}). Poor frequency resolution and low
duty cycle both contributed to explain the difficulty of the task.
Apart from the Sun, the first identification of a full comb-like
stellar oscillation spectrum with duly identified modes was done
with $\alpha$-Cen \citep[Fig.~\refer{fig-acen},
][]{2001A&A...374L...5B}.\\

\begin{figure}[!t]
  \fichier{7.0}{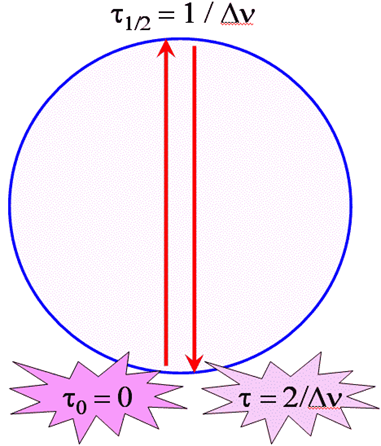}
  \legende{Autocorrelation} {Autocorrelation of the time series.
  The travel time to cross two times the stellar diameter is
  directly related to the large separation.
  \labell{fig-principe-autocor}}
\end{figure}

With space-borne results free of the aliasing effect introduced by
the day-night alternance, efficient methods were introduced to
analyze solar-like oscillations spectra with a comb-like pattern.
A presentation of these methods can be found in, e.g.,
\cite{2011MNRAS.415.3539V} and \cite{2011A&A...525A.131H}. All the
methods work well; comparing them help discriminating real from
false positive detections. However, methods based on the concepts
developed by \cite{2006MNRAS.369.1491R}  are certainly optimized.
They make profit of the equivalence between the autocorrelation of
the oscillating time series and the Fourier spectrum of the
Fourier spectrum. The principle is explained in
Fig.~\ref{fig-principe-autocor}: an event observed in the time
series at $\tau=0$ reappears after a delay corresponding to the
double cross of the stellar diameter. This takes a time equal to
$2 /\Dnu$. The autocorrelation of the time series, performed in
the Fourier space with two consecutive Fourier transforms,
provides the signature of the large separation.
\cite{2009A&A...506..435R} has complemented the method with the
introduction of narrow frequency-windowed autocorrelation to
enhance its diagnostic capability. \cite{2009A&A...508..877M} have
shown how the method can be automated and how its performance can
be estimated with a test based on the null hypothesis.

Oscillation spectra recorded with a very high signal-to-noise
ratio
\citep[e.g.,][]{2010A&A...515A..87D,2012ApJ...748L..10M,2013PNAS..11013267G}
carry much more information than the large separation. However,
for faint stars, only the large separation and the frequency
$\numax$ of maximum oscillation signal can be measured
\citep[e.,g.,][]{2013A&A...558A..79O}.

%\begin{figure}[!t]
\begin{sidewaysfigure}
%  \fichier{11.2}{evolution_ensemble_stello2.png}
%  \fichier{13.5}{evolution_ensemble_stello.png}
  \fichier{14.5}{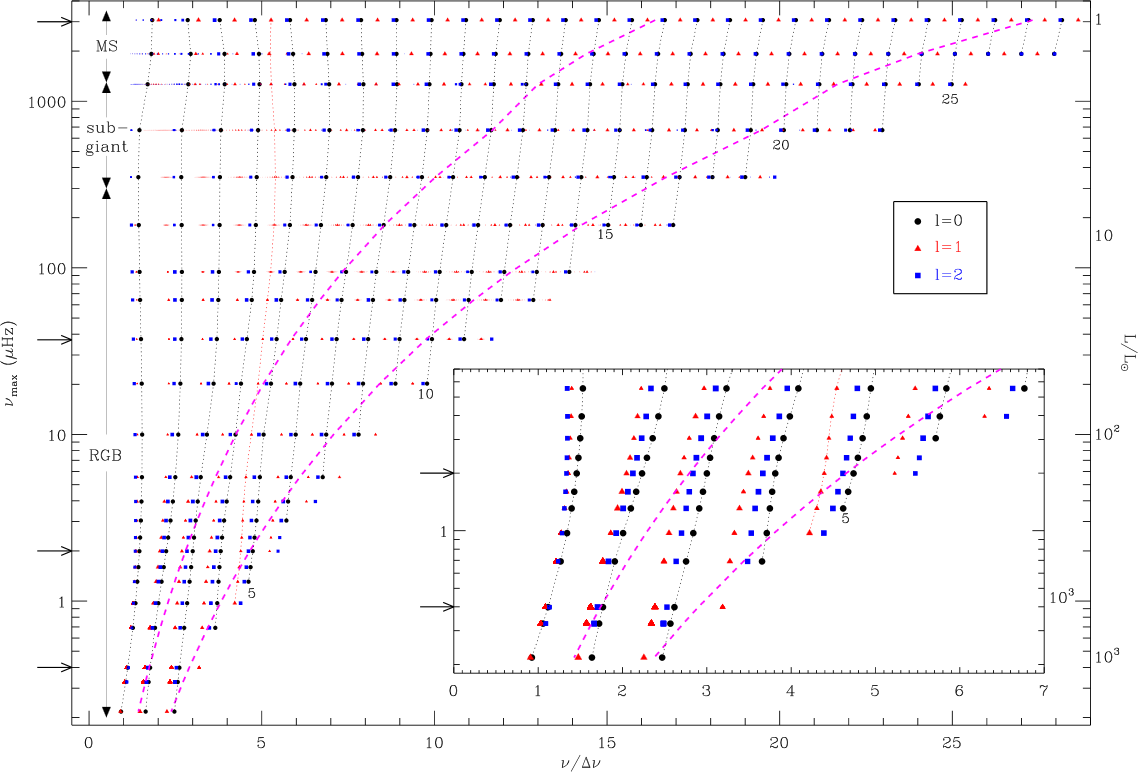}
  \legende{Model frequencies} {Model frequencies,
  in units of the asymptotic large frequency separation, for solar-metallicity models along a
  1-$M_\odot$ track. Each model is plotted according to its   $\numax$
  \credit{2014ApJ...788L..10S}.
  \labell{fig-evolution-ensemble}}
\end{sidewaysfigure}
%\end{figure}

\subsubsection{Numerical computations}

The asymptotic expansions for pressure and gravity modes are
wonderful tools for a rapid analysis of oscillation spectra.
However, the exquisite precision reached with the space missions
CoRoT and \emph{Kepler} now exceeds the precision of the classical
second-order asymptotic expansion. Numerical computations  are
used to enter the details of the oscillation spectra and depict
all the subtle features of a low-degree oscillation spectrum, even
if they remain limited by some features as the surface term
(Section~\refer{surface}).\\

Long-term efforts have provided precise and robust oscillation
codes to compute adiabatic oscillations, as the ADIPLS code
\citep{2008Ap&SS.316..113C,2011ascl.soft09002C}. These codes can
be used for the study of a large set of stars, as for fitting
low-degree modes with $\ell\le 2$ observed at all evolutionary
stages by CoRoT and \Kepler. Fig.~\refer{fig-evolution-ensemble}
shows modelled frequencies, with a close agreement with asymptotic
expansion.

\begin{figure}[!t]
  \fichier{11}{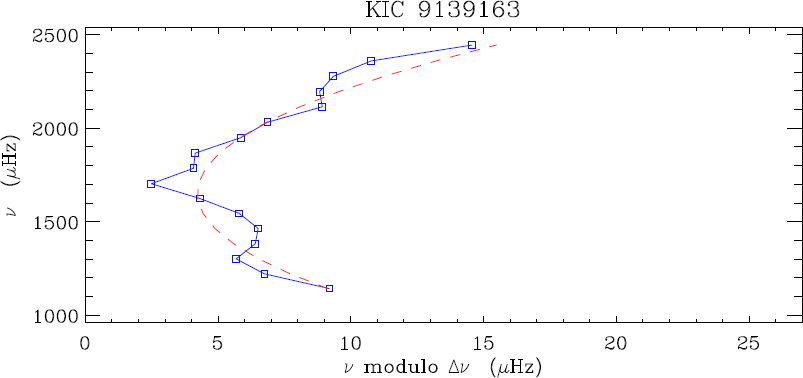}

  \fichier{11}{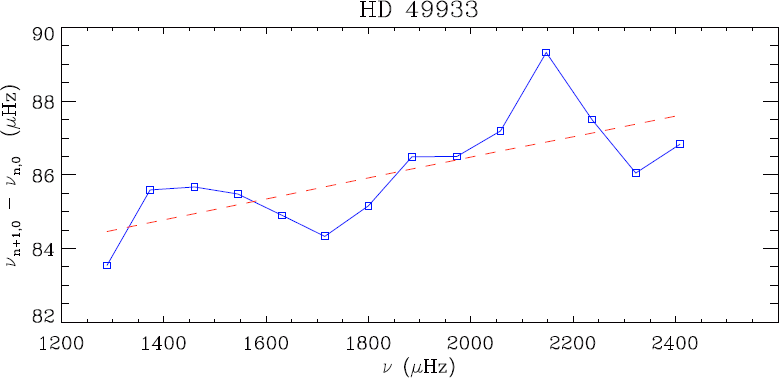}
  \legende{\emph{Top:} radial modes of the star KIC 9139163} {\'Echelle diagram of the radial modes of the star KIC
  9139163 \citep[from][]{2012A&A...543A..54A}. The red dashed line
  indicates the quadratic fit that mimics the curvature.
  \emph{Bottom:} variation of the radial frequency difference $\nu_{n+1,0}
  -\nu_{n,0}$ as a function of $\nu_{n,0}$ for the star HD49933
  \citep[from][]{2009A&A...507L..13B}. The red dashed line indicates a linear
  fit
  \credit{2013A&A...550A.126M}.
  \labell{fig-9139163}}
\end{figure}

\begin{figure}[!t]
  \fichier{10}{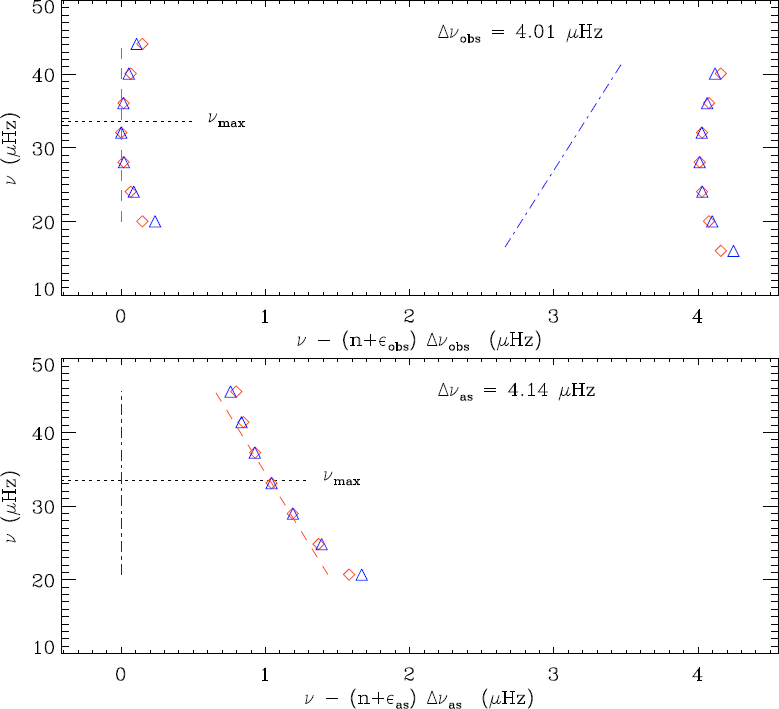}
  \legende{\'Echelle diagram: asymptotic versus observation} {\'Echelle diagrams of the radial modes of a
  typical red-clump giant, comparing the asymptotic expansion (blue
  triangles) and the development describing the curvature (red
  diamonds).
  \emph{Top:} diagram based on $\Dnuobs$ observed at
  $\numax$; the dashed line indicates the vertical asymptotic line
  at $\numax$; the dot-dashed line indicates the asymptotic line at
  high frequency. For clarity, the ridge has been duplicated modulo
  $\Dnuobs$.
  {\sl Bottom:} diagram based on $\Dnuas$; the dot-dashed
  line indicates the vertical asymptotic line at high frequency
  \credit{2013A&A...550A.126M}. \labell{asymp-ech1}}
\end{figure}

\subsection{Comparison with observations\labell{comp-obs}}

The asymptotic expansion of low-degree pressure modes is valid at
large radial orders only. Pressure modes are however not observed
at high frequency. This results from their stochastic excitation
in the upper convective envelope: modes are excited at periods
corresponding to the lowest periods of the breaking of convective
cells.

The frequency of maximum oscillation signal largely depends on the
stellar evolutionary stage (Fig.~\refer{fig-ensemble-intro}). For
the Sun, $\numax \simeq 3100\ \mu$Hz. The highest peaks in the
Solar oscillation spectrum have then a radial order close to 21.
This is not high enough for a strict application of the asymptotic
expansion.

In more evolved stars, the situation is even worse. The radial
orders
$n$ corresponding to $\numax$ are approximately:\\
- 18-22 for main-sequence stars, \\
- 15-18 in subgiants,\\
- below 15 in red giants: typically 8 for a clump star (a star of
the red horizontal branch with core helium burning) and 3 for a
semi-regular variable
\citep{2013A&A...559A.137M,2014ApJ...788L..10S}. \\

\begin{sidewaysfigure}
%\begin{figure}[!t]
%  \fichier{11.1}{epsilon_as_obs2.png}
%  \fichier{13.5}{epsilon_as_obs.png}
  \fichier{14.5}{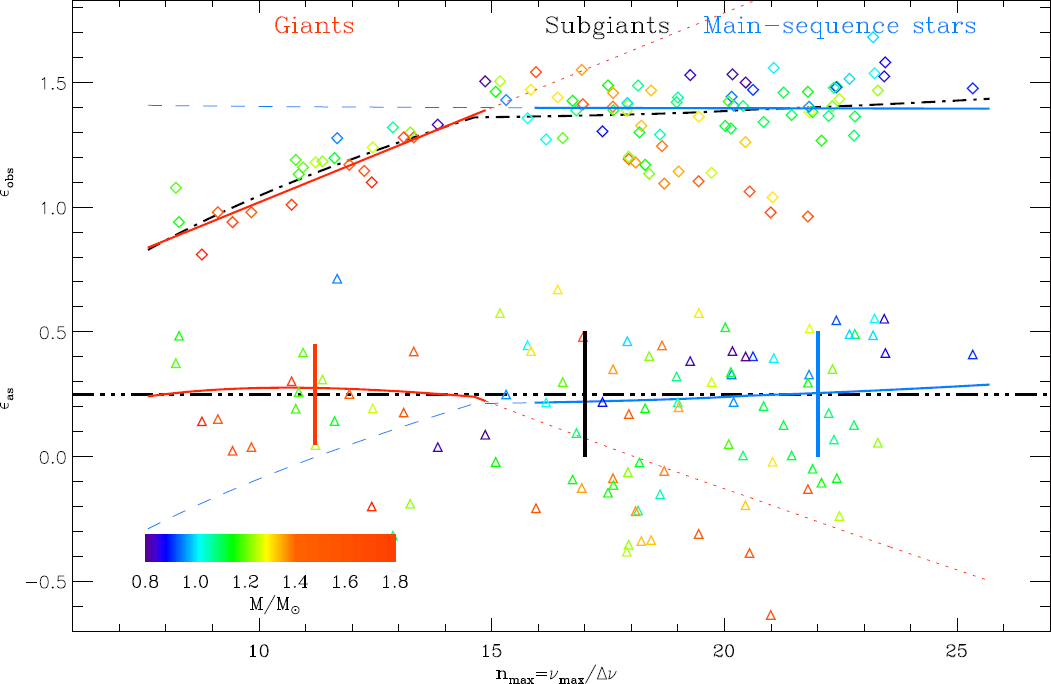}
  \legende{Observed and asymptotic offsets} {Observed and asymptotic offsets $\varepsilon$
   as a function of $\nmax$, fitted with dotted
   lines in the red giant regime and dashed lines in the
   main-sequence regime. Thicker lines indicate the domain of
   validity of the fits. The triple-dot-dashed line represents the
   Tassoul value $\epsas=1/4$, and the dot-dashed line is the model
   of $\epsobs$, varying with $\log\Dnuobs$ in the red giant regime,
   and constant for less-evolved stars
   \credit{2013A&A...550A.126M}. \labell{fig-epsilon}}
%\end{figure}
\end{sidewaysfigure}

So, it is clear that observations do not match the asymptotic
conditions and cannot directly derive asymptotic parameters. In
other words, the frequency spacing derived from alignments of
modes in the \'echelle diagram cannot be asymptotic. To enhance
the quality of the fit of radial modes, many authors have proposed
including the curvature of the radial ridge
\citep{1983SoPh...82..469C,1983SoPh...82...55G,2004ApJ...614..380B,2005ApJ...635.1281K,2008A&A...478..197M}.
This curvature corresponds to the signature in the \'echelle
diagram of the non-negligible second-order asymptotic terms
(Fig.~\refer{fig-9139163}). It induces a significant gradient in
the frequency spacing between consecutive radial modes.\\

In order to fit the radial mode pattern, we use the expression
\begin{equation}\label{tassoul_obs}
\nu_{\np,0} = \left(\np  + \epsobs
 + {\alfa \over 2}\; [ \np - \nmax ]^2 \right) \Dnuobs,
\end{equation}
where $\Dnuobs$ is the observed large separation, measured in a
wide frequency range around the frequency $\numax$ of maximum
oscillation amplitude, $\alfa$ is the curvature term, and
$\epsobs$ is the offset. The index  $\nmax$ equivalent to $\numax$
is
\begin{equation}\label{eqt-def-nmax}
    \nmax= \numax / \Dnuobs - \epsobs .
\end{equation}
With such a definition, $\nmax$ is not an integer. \\

Two different regimes are seen for the curvature:
\begin{eqnarray}
   \alfa\ind{MS} &\simeq& 1.14/ \nmax^2 \hbox{ \ for subgiants and main-sequence stars, }
  \labell{eqt-curv-MS}\\
   \alfa\ind{RG} &\simeq&  0.076/ \nmax   \hbox{ \ for red giants,}
  \labell{eqt-curv-RG}
\end{eqnarray}
where the different exponents emphasizes different properties of
the interior structure, depending on the evolutionary stage. The
link between asymptotic and observed descriptions of the radial
oscillation pattern with a second-order development is expressed
by
\begin{eqnarray}
  \Dnuas &=& \Dnuobs\   \left({ 1 + {\nmax\alfa\over2}} \right),  \label{eqdnu}\\
  \Aas   &=& {\alfa \over 2} {\nmax^3 \over 1  + \nmax \displaystyle{\alfa\over 2}} , \label{eqsecond0}\\
  \epsas &=& {\epsobs - \nmax^2 \alfa \over 1  + \nmax \displaystyle{\alfa\over 2}} . \label{eqeps0}
\end{eqnarray}
According to Eq.~(\refer{eqt-curv-MS}) and (\refer{eqt-curv-RG}),
we may consider that the ridge curvature is small enough for
ensuring $\nmax \alfa/2 \ll 1$, so that $\Aas$ and $\epsas$ become
\begin{eqnarray}
  \Aas   &\simeq& {\alfa \over 2}\ \nmax^3 , \label{eqsecond}\\
  \epsas &\simeq& \epsobs \left({1-{\nmax\alfa\over2}}\right) - \nmax^2 \alfa .\label{eqeps}
\end{eqnarray}
These developments provide a reasonable agreement between the
asymptotic and observed forms. The difference between $\Dnuobs$
and $\Dnuas$ is high enough for providing very different observed
and asymptotic \'echelle diagrams (Fig.~\refer{asymp-ech1}).

The comparison of the observed and asymptotic values of the offset
$\varepsilon$ is enlightening (Fig.~\ref{fig-epsilon}): $\epsas$
is close to 1/4, as predicted by the asymptotic expansion
(Eqs.~\ref{eqt-raccord-1} and \ref{eqt-eps}), whereas $\epsobs$ is
above 1 for subgiants and main-sequence stars and varies
significantly with the large separation in the red giant regime.
According to the holomological property of red giant interior
structure, \cite{2011A&A...525L...9M} have identified that
$\epsobs$ follows a strict relationship with $\Dnu$. Empirically,
one observes
\begin{equation}\label{eqt-eps-RG}
    {\epsobs}\ind{,RG } = 0.60 + 0.52 \log( \Dnu ) + 0.0091 \Dnu ,
\end{equation}
with $\Dnu$ expressed in $\mu$Hz. This relation is only slightly
modified at very low $\nmax$, for depicting solar-like
oscillations in semi-regular variables
\citep{2013A&A...559A.137M}:
\begin{equation}\label{eqt-eps-RGlowDnu}
    {\epsobs}\ind{,RG } = 0.623 + 0.599\log( \Dnu )
\end{equation}
for $\Dnu$ less than 1\,$\mu$Hz.

\begin{figure}[!t]
\fichier{8.8}{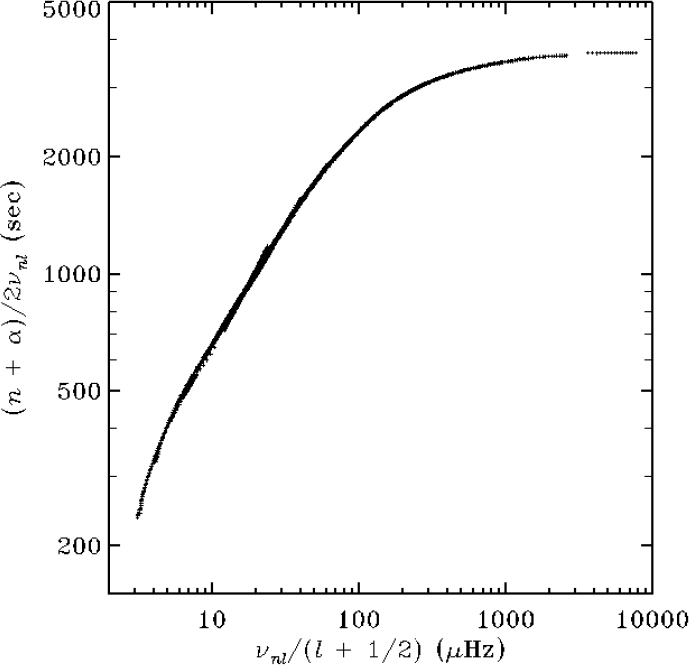} \legende{Duvall diagram} {Duvall diagram
\citep{1982Natur.300..242D}. The angular degree used to derive the
abscissae comes from imaging; the radial order used on the y-axis
is adjusted from the counting of oscillation ridges.
\labell{fig-duvall}}
\end{figure}

\subsection{Pressure modes with medium $\ell$}

Since modes with intermediate degree do not probe the inner
region, surface terms are dominating their evolution. We address
the solar case only,  since high-degree modes cannot be observed
in non-resolved distant stars.

Radial quantization is expressed by
\begin{equation}
\int_{\rt}^{R} \kr\ \diff r \ = \  (n+\alpha)\ \pi .
\labell{eqt-reson}
\end{equation}
 As in
Eq.~(\refer{eqt-rad-as}), $\alpha$ is a small constant term. The
turning point $\rt$ is related to the frequency $\omega$ by
\begin{equation}
\omega \ = \ S_\ell (\rt),
\end{equation}
or conversely:
\begin{equation}
{\rt\over c(\rt ) } \ =\ \disp{L\over \omega}
\end{equation}
with $L\ =\ \sqrt{\ell ( \ell +1)}$.\\

In the convective outer envelope, the role of $\NBV$ is
negligible, so that the radial component of the wave vector can be
approximated by
\begin{equation}
\kr^2\ = \ {\omega^2\over c^2} - {L^2\over r^2} .
\end{equation}

The resonance condition (Eq.~\refer{eqt-reson}) can be written
\begin{equation}
{\pi \ (n+\alpha) \over \omega\nl} \ =\ \int_{\rt}^R \left[ 1-{L^2
c^2\over r^2 \omega\nl^2} \right]^{1/2} {\diff r\over c} .
\end{equation}
This equation corresponds to the relation proposed by
\cite{1982Natur.300..242D}
\begin{equation}
{\pi \bigl( n+\alpha \bigr) \over \omega} \ = \ F\left( {\omega
\over L} \right) , \labell{eqt-duvall}
\end{equation}
which has been verified by the observations in the solar case
(Fig.~\refer{fig-duvall}).

\begin{figure}[!t]
 \fichier{10.6}{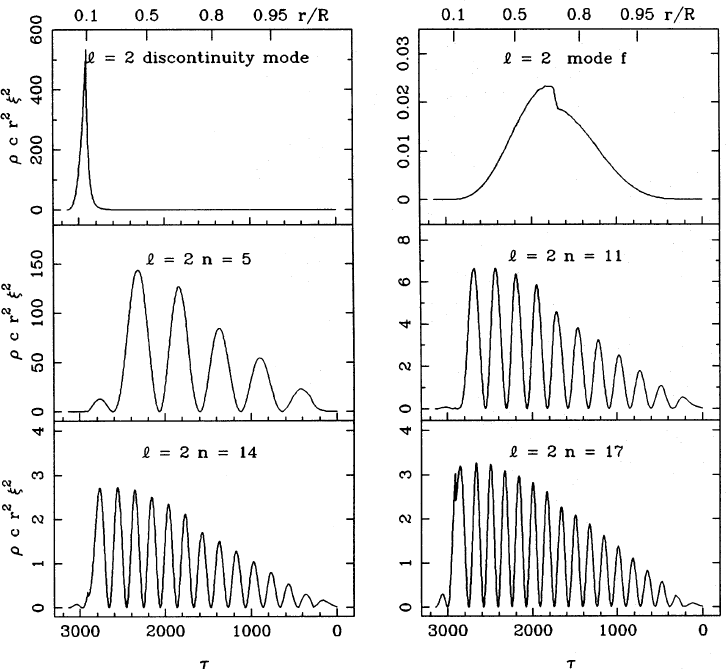}
 \legende{Eigenfunctions} {Eigenfunctions of
 various modes, expressed with a variable related to
 the density of kinetic energy flux
 \credit{1993A&A...274..595P}.
 \labell{fig-amplip}}
\end{figure}

\subsection{Eigenfunctions}

Since now, we have focussed our attention on the eigenfrequencies.
From the eigensolutions, we can derive eigenfunctions. For
pressure modes, the radial displacement can be approximated by
\begin{equation}
\xir \ \propto \ {1\over r} {1\over \sqrt{\rho_0 c}} \cos \left[
\omega \int_{\rt}^R \left( 1- {L^2 c^2\over \omega^2 x^2}
\right)^{1/2} {\diff x\over c} - \varepsilon \pi\right] ,
\end{equation}
according to \cite{1989nos..book.....U}. With the hypothesis
$\omega \gg S_\ell$ and the change of variables from ($r, \xi$) to
($\tau, X$) defined by
\begin{eqnarray}
  X^2 &\propto& r^2 \ \rho_0 c\ \xir^2,\labell{eqt-densi} \\
  \diff \tau &=& \diff r / c , \labell{eqt-rayon-acou}
\end{eqnarray}
one gets
\begin{equation}
X(\tau) \ \propto \cos^2 \Bigl( \omega \bigl[\tau(R) - \tau (r)
\bigr] \Bigr) .\labell{eqt-ampli}
\end{equation}
This looks like a typical stationary plane wave. As already seen,
the acoustic radius $\tau$ is in fact a more natural variable than
the radius for stellar pressure modes. The variable $X$ defined by
Eq.~(\ref{eqt-densi}) is close to the density flux of kinetic
energy $u \propto r^2 \rho_0 c  \xi^2 \omega^2$. According to
Eq.~(\ref{eqt-ampli}), this variable has nearly the same amplitude
all along the star (Fig.~\refer{fig-amplip}).\\

It is useful to introduce the mode mass and mode inertia. They
derive from the mean kinetic energy of the modes, which writes
\begin{equation}\label{eqt-ek}
    E\ind{kin} = \int {1\over 2}\; \rho_0\; {\mathbf{v}}^2 \diff^3
    \mathbf{r}
    .
\end{equation}
Its mean value, where mean means both time-averaged and
space-averaged, is
\begin{equation}\label{eqt-mean}
    \langle E\ind{kin} \rangle
    = {1\over 4}\ \omega^2\ \int_0^R
    \bigl(\xir^2  + \ell(\ell+1) \xih^2 \bigr)
    \ 4\pi \rho_0 r^2 \diff r
    ,
\end{equation}
where we recognize the contribution of the variable $X$ and of its
horizontal counterpart. The extra factor $1/2$ comes from the time
average. This defines the mode inertia $\inertia$
\begin{equation}\label{eqt-def-inertia}
    \inertia\ind{n,\ell} =
    \int_0^R
    (\xir^2  + \ell(\ell+1) \xih^2) \ 4\pi \rho_0 r^2 \diff r ,
\end{equation}
and similarly the mode mass
\begin{equation}\label{eqt-def-mode-mass}
   \mathcal{M}\ind{n,\ell} =
    \int_0^R
    \mathbf{\xi}^2 \  \diff m \bigm/ \mathbf{\xi}^2\ind{R}.
\end{equation}
Mode inertia are used, for instance, for comparing mode
amplitudes: two modes with similar degree and similar frequency
show similar amplitudes only if they have close inertia. This is
not the case for dipole mixed modes in red giants
\citep{2009A&A...506...57D,2014A&A...572A..11G}. The relative
contributions of mode inertia in the core and in the envelope of
the star can also be compared to derive properties such as the
rotational splittings (Appendix \ref{perturbationrotation}).

\begin{figure}[!t]
  \fichier{9.6}{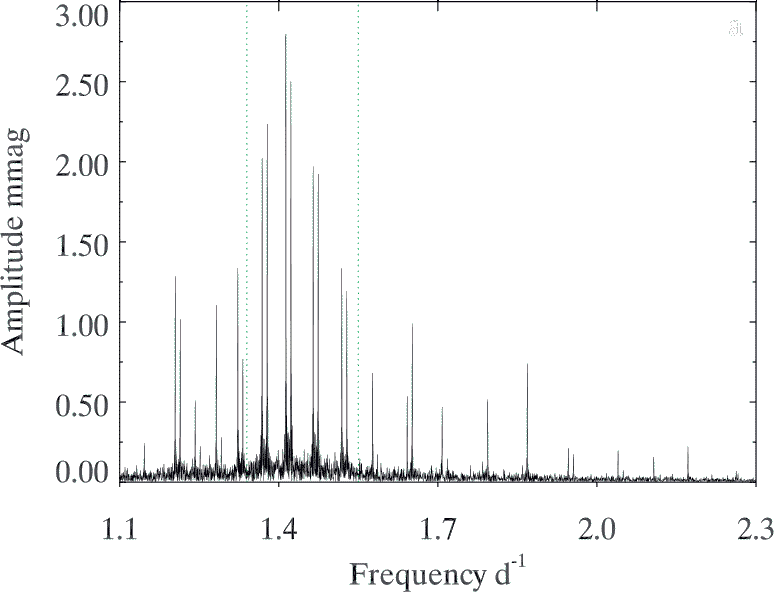}
  \legende{Gravity mode spectrum in an
  A star}{Amplitude spectrum for the gravity dipole modes splitted by rotation
  in the A-type star KIC 11145123 observed by \Kepler\
  \credit{2014MNRAS.444..102K}. \labell{fig-Astar}}
\end{figure}

\subsection{Gravity modes}

The radial component of the wave vector derived from
Eq.~(\refer{eqt-dif-cow}) is
\begin{equation}
  \kr^2
  \  = \ {S_\ell^2\over c^2}\ \left( {\NBV^2\over \omega^2} - 1 \right)
  \  = \ {\ell (\ell+1)\over r^2}\ \left( {\NBV^2\over \omega^2} - 1 \right)
\end{equation}
when $\omega \ll S_\ell$. In the \BV\ cavity $\mathcal{R}$ defined
by $\NBV^2 \ge 0$, the resonance equation (Eq.~\refer{eqt-reson})
writes
\begin{equation}
  {\sqrt{\ell (\ell + 1)} \over \omega}\
  =
  \ \pi (n+\alpha)
  \left[
  \int_\mathcal{R} \left( 1- {\omega^2 \over \NBV^2} \right)^{1/2}
  {\NBV \over r}\, \diff r
  \right]^{-1}
  .
\end{equation}
Similarly to the development that provides
Eq.~(\refer{eqt-asymp}), the second-order asymptotic period
pattern of low-frequency gravity modes ($\omega\ll \NBV$) is,
following \cite{1986A&A...165..218P}:
\begin{equation}
  P\nl \ = \ \left( n+ {\ell\over 2} - {{\ifmmode{1\over 4}\else
  {$1\over 2$}\fi}}- \theta \right)\ \Delta\Pi_\ell \
  +\ \bigl({\ell (\ell+1) V\ind{1g} + V\ind{2g} }\bigr)
  \ {{\Delta\Pi_\ell}^2\over P\nl} ,
\end{equation}
with
\begin{equation}
  \Delta\Pi_\ell\ =\ {2\pi^2\over \sqrt{\ell (\ell+1)}}\  \Bigm/
  \displaystyle{\int_\mathcal{R} {N \over r}\ \diff r}.
\end{equation}
The dimensionless terms $V\ind{1g}$ and $V\ind{2g}$ are functions
of $\NBV(r)$. The phase factor $\theta$ is a complex function
sensitive to the stratification just below the convection zone. At
first order, one gets for dipole modes
\begin{equation}
  P_{n,1} \simeq \left(n + {1\over 4} - \theta \right) \ \Delta\Pi_1
\end{equation}
Gravity modes exist in mostly radiative stars as $\gamma$-Dor type
stars, white dwarves, or SpB stars
\citep[e.g.,][]{2010A&A...516L...6C,2011A&A...525A..23C}
(Fig.~\refer{fig-Astar}).  In the Solar case, strong evanescence
occurs in the outer convective envelope, so that their
observations is highly difficult \citep{2010A&ARv..18..197A}.

\subsection{Mixed modes\labell{asymp-mixte}}

Gravity waves  propagate in the radiative core of evolved low-mass
stars. In red giants, the convective envelope is much too large
for observing pure gravity modes trapped in the core. However,
gravity waves propagating in the stellar core can couple with
pressure waves propagating in the convective envelope. This yields
mixed modes, behaving as gravity modes in the core and as pressure
modes in the envelope. Such pressure-gravity mixed modes are
observed in subgiants
\citep{2007ApJ...663.1315B,2013ApJ...767..158B} and red giants
\citep{2009Natur.459..398D,2010ApJ...713L.176B}.  Their presence
significantly modifies the oscillation spectrum compared to the
case observed in main-sequence stars (Fig.~\refer{echelle-mixte}
and \refer{mixed-pattern}). The coupling of gravity waves trapped
in two different \BV\ cavities of ZZ Ceti stars (pulsating DA
white dwarfs) also constructs mixed modes
\citep{1992ApJS...80..369B}.

\begin{figure}[!t]
 \fichier{9.6}{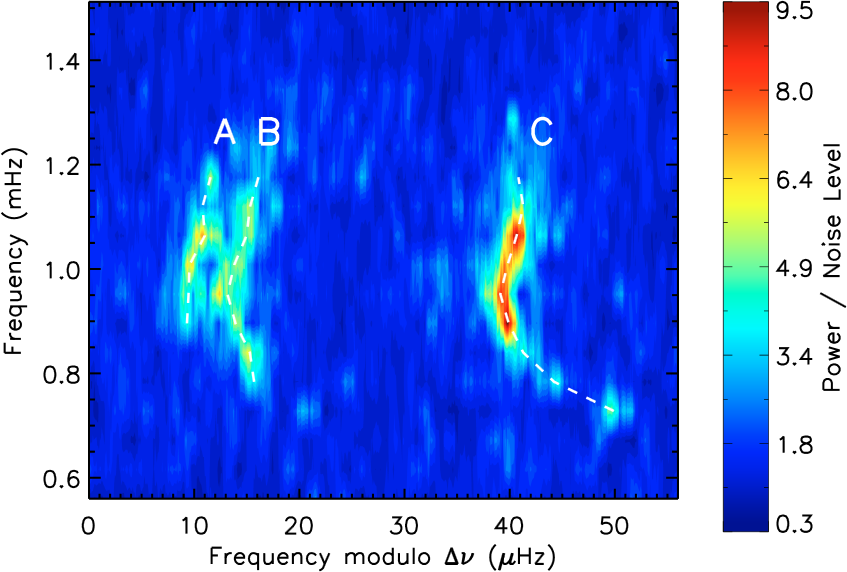}
 \legende{\'Echelle diagram with mixed modes} {\'Echelle diagram
 of the subgiant star HD 49385 observed by CoRoT. The color
 bar shows the power density level, in noise units. Radial modes correspond
 to ridge B, dipole modes to ridge A, quadrupole modes to ridge C
 \credit{2010A&A...515A..87D}.
 \labell{echelle-mixte}}
\end{figure}

%\begin{figure}[!t]
\begin{sidewaysfigure}
 \fichier{18.3}{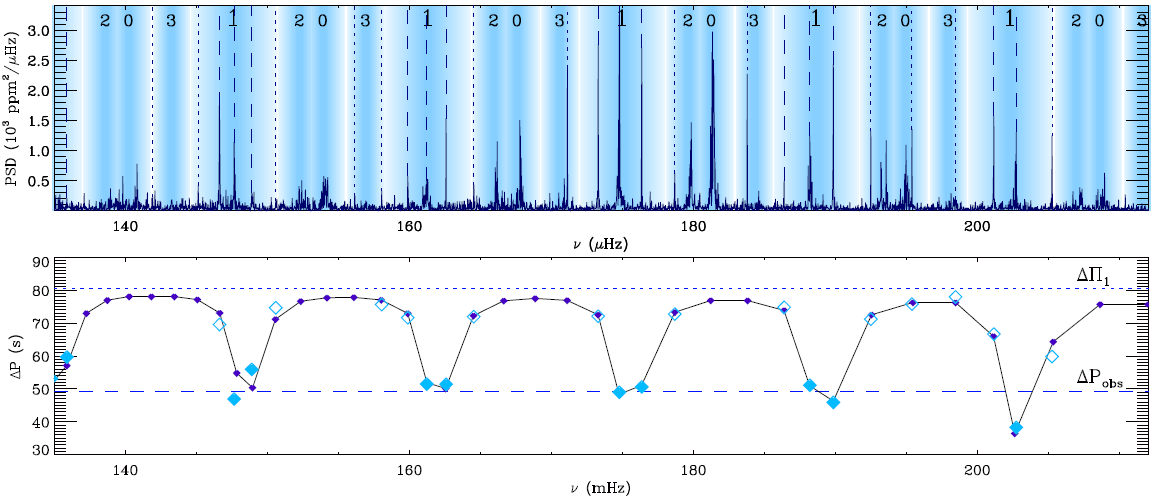}
 \legende{Mixed-mode pattern} {Mixed-mode pattern in a red giant
 observed by \Kepler. The upper diagram shows the full spectrum,
 with the indication of the angular degree of the modes. In the lower
 plot, only mixed modes were selected, as peaks with a significant
 amplitudes not corresponding neither to radial nor to quadrupoles
 modes. `Obvious' dipole mixed modes are marked with dashed lines in
 the upper diagram; dotted lines correspond to significant peaks a
 posteriori also identified as dipole mixed modes. The period spacings
 constructed with these modes (respectively full and open large diamonds
 in the lower plot) are well reproduced by the asymptotic
 expansion (small dark dots). Period spacings of gravity-dominated
 mixed modes are close (but inferior) to $\Tg$ (horizontal dotted
 line) whereas bumped period spacings of pressure-dominated mixed
 modes are much below
 \credit{2012A&A...540A.143M}.
 \labell{mixed-pattern}}
%\end{figure}
\end{sidewaysfigure}

\begin{figure}[!t]
 \fichier{8.}{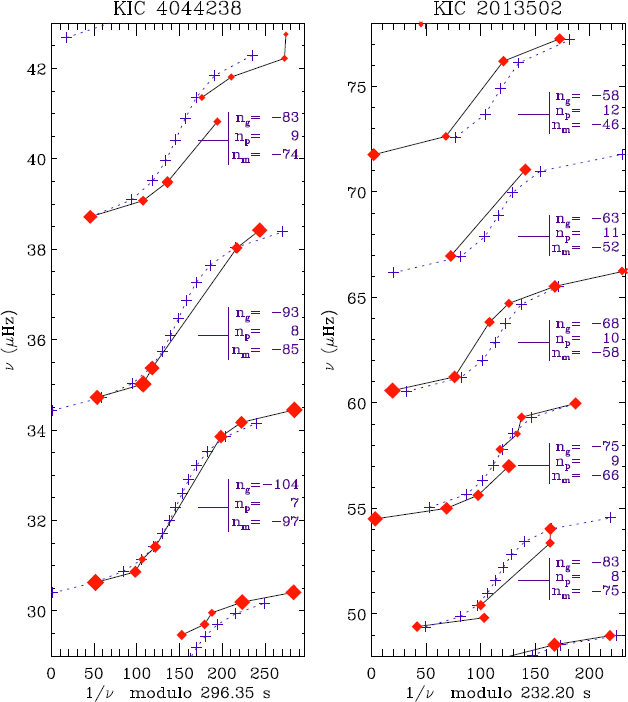}

 \fichier{8.}{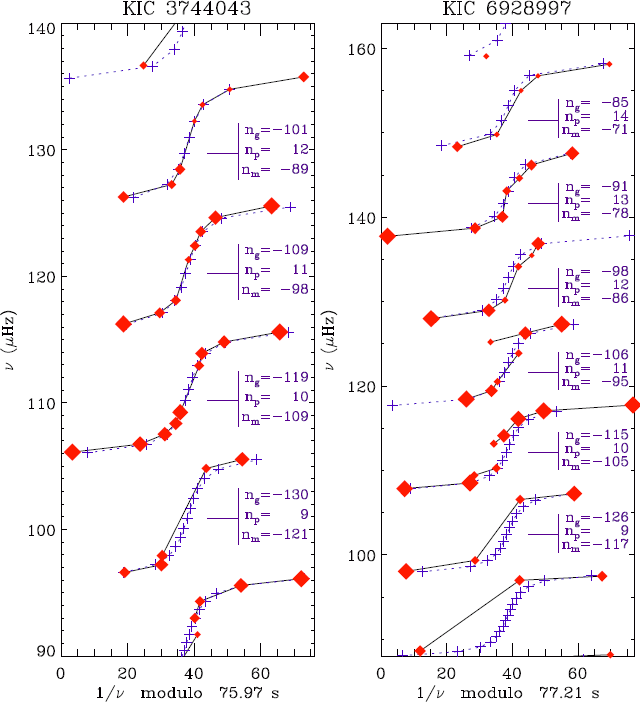}
 \legende{\'Echelle diagram of mixed modes} {Period \'echelle diagram of mixed
 modes, with the period, modulo $\Tg$, on the x-axis. Observed
 mixed modes (red symbols) are correctly fitted by the asymptotic
 expansion (blue crosses). The pressure, gravity, and mixed radial
 orders are provided for gravity-dominated mixed modes close to
 radial modes. Two spectra on the left side correspond to clump
 stars; on the right side, stars are on the red giant branch
 \credit{2012A&A...540A.143M}.
 \labell{mixed-echelle}}
\end{figure}

An asymptotic expansion for mixed modes has been derived by
\cite{1989nos..book.....U}. Following \cite{2012A&A...540A.143M},
it writes as an implicit equation
\begin{equation}
  \nu = \nu_{\np,\ell} + {\Dnu \over \pi} \arctan %
  \left[%
  \coup \tan \pi \left( {1 \over \Delta\Pi_\ell \nu} - \epsg \right) \right]
  ,
  \labell{eqt-implicite2}
\end{equation}
where $\nu_{\np,\ell}$ is the pure pressure mode frequency and $q$
is a dimensionless coupling factor. According to
\cite{1989nos..book.....U}, $q$ is by definition less than 1/4.
This is not confirmed by observations: maximum coupling, with
$q=1$, is observed in evolved subgiants; minimum coupling occurs
with $q$ or $1/q$ close to 0, for instance for quadrupole modes.
Nevertheless, Fig.~\ref{mixed-pattern} shows that the asymptotic
fit nicely fits the mixed modes that are observed.

The way used for deriving the asymptotic expansion of mixed modes
is similar to the way used for pressure modes, except that we have
to introduce the eigenfunctions of gravity waves  in the inner
cavity, and  to account for the evanescent region. Following in
\cite{1989nos..book.....U} the different ways the eigenfunctions
are expressed either for p (their Eqs. 16.23 and 16.25) or g waves
(their Eqs. 16.37 and 16.38), we see that the gravity
wavefunctions in the core region vary as
\begin{eqnarray}
  p_i   &\propto & -\sin\Phi_i ,\\
  \xi_i &\propto & \cos\Phi_i  ,
\end{eqnarray}
with
\begin{equation}\label{eqt-phii-g}
    \Phi_i = {1\over \omega \tau_i}
    \ \hbox{ and } \
    \tau_i = \left(\int_0^{r_i} \sqrt{\ell (\ell+1)}\  \NBV {\diff r\over r}\right)^{-1} .
\end{equation}
The expression of the buoyancy radius $\tau_i$ derives from the
leading term of the radial wavevector $\kr$ in the dispersion
equation (Eq.~\refer{eqt-dispersion}) developed at very low
frequency.

In the envelope, the pressure wavefunctions \citep[][and
Eqs.~\refer{A.5} and \refer{A.6}]{1989nos..book.....U} vary as
\begin{eqnarray}
  p_o   &\propto & -\sin\Phi_o , \\
  \xi_o &\propto & \cos\Phi_o  ,
\end{eqnarray}
with
\begin{equation}\label{eqt-phio-p}
    \Phi_o = \omega \tau_o
    \ \hbox{ and } \
    \tau_o = \int_{r_o}^R {\diff r\over c}.
\end{equation}
The continuity of the wavefunctions has to account for the decay
in the evanescent region, here expressed by the factors $\eta_p$
and $\eta_\xi$:
\begin{eqnarray}
  p_o   &=& \eta_p  \ p_i \\
  \xi_o &=& \eta_\xi\ \xi_i .
\end{eqnarray}
From the continuity of the wavefunctions, we get:
\begin{eqnarray}
  \sin\Phi_o  &\propto& \eta_p  \ \sin\Phi_i, \\
  \cos\Phi_o  &\propto& \eta_\xi\ \cos\Phi_i,
\end{eqnarray}
hence
\begin{equation}\label{eqt-mix-modulo}
  \tan \Phi_o\ \equiv\ q \tan \Phi_i
  .
\end{equation}
With the connection $r = r_i = r_o$ outside the \BV\ cavity, we
see that the right-hand term of Eq.~(\refer{eqt-mix-modulo})
introduces the right-hand term of Eq.~(\refer{eqt-implicite2})
since $\tau_i$ introduces $\Delta\Pi_\ell$. With $r$ at a level
low enough to ensure that $2 \tau_o$ is close to $\Dnu^{-1}$, we
retrieve also the pressure contribution, hence
Eq.~(\refer{eqt-implicite2}).\\

In the $\Dnu$-wide frequency range corresponding to the pressure
radial order $\np$ near $\nmax$, one derives $(\fac+1)$ solutions
from Eq.~(\ref{eqt-implicite2}), with $\fac \simeq
\Dnu\,\Tg^{-1}\numax^{-2}$. They correspond to the single pure p
modes and to the $\fac$ pure g modes that should be observed
without coupling.

The observations of mixed modes is crucial for measuring the
period spacing $\Tg$. Contrary to $\Dnuobs$, $\Tg$ cannot be
obtained directly from period differences, since all period
differences are smaller than $\Tg$, especially near the expected
location of pure p modes (Fig.~\refer{mixed-pattern}). The value
of $\Tg$ is then derived from a least-squares fit of the observed
mixed modes. Gravity dominated mixed modes are especially useful
for this identification (Fig.~\refer{mixed-pattern},
\refer{mixed-echelle}).

\subsection{Classification of normal modes\label{classifi}}

The radial order provides a count of radial nodes. A convenient
way to treat all modes is to assign negative values to radial
orders $n$ for gravity nodes and positive values for pressure
nodes. The absolute value of the radial order $n$ gives the number
of zeros of the radial displacement $\xir$ (a zero at $r=0$ is
counted for radial modes only); $n=1$ corresponds to the first
pressure mode with a node at the center; $n=0$ corresponds to the
fundamental mode, without any radial node.
\citep{1974A&A....36..107S}.
\\

Counting and classifying pressure modes is quite obvious in stars
as then Sun,  with an inner radiative region where the \BV\
frequency is much below the frequency domain where p modes are
stochastically excited, as illustrated by the very regular
frequency comb-like spectrum of the Sun (Fig.~\refer{fig-iphir}).
Similarly, counting gravity modes in a mostly radiative star is
not difficult (Fig.~\refer{fig-Astar}). However, classifying mixed
modes may require a thorough count. \\

For this numbering, \cite{1989nos..book.....U} propose a formalism
based on the phase of the wave function. Observationally, one can
assign two orders to a mixed mode observed in a red giant
oscillation spectrum:\\
- the mixed radial order $\nm$ helps classifying the modes,
observationally and theoretically; it allows
us to follow the change of the oscillation pattern with stellar evolution; \\
- the total number of radial nodes $\npg$ helps describing the
radial properties of the mode eigenfunctions.\\
These numbers are derived from the pressure radial order $\np\
(>0)$ and gravity radial order $\ng\ (<0) $ according to
\begin{eqnarray}\labell{eqt-class-nm}
  \npg &\simeq& \np - \ng,\\
  \nm  &\simeq& \np + \ng,
\end{eqnarray}
with
\begin{eqnarray}
  \np &=& \left\lfloor \disp{\nu \over \Dnuobs} - d_{01} - \epsobs\right\rfloor ,\\
  \ng &=& - \left\lfloor \disp{1\over \nu\Tg} \right\rfloor
  .
\end{eqnarray}
In practice:\\
- the definition of $\np$ is operating, but one has to account for
the second-order term of the asymptotic expansion to reach
precision for modes far away from $\numax$,\\
- the definition for $\ng$ works better for gravity dominated
mixed modes, far away from the mode bumping that occurs at the
avoided crossing, near the pressure-dominated mixed modes.\\

An example is provided by Table~\refer{tab-compte-mixte}, which
corresponds to the RGB oscillation spectrum plotted in
Fig.~\refer{mixed-pattern}. The observation of a sequence of mixed
modes provides a safe identification, since obvious corrections
can be made on $\nm$ in case of repeated or missing values. We
note that, near radial modes, the values of $\ng$ and $\npg$ are
repeated; only the mixed-mode order $\nm$ provides the correct
numbering of mixed modes.

\begin{table}[t]
 \centering
 \small{\begin{tabular}{lrrrrrr}
  \hline
  $\nu$ ($\mu$Hz) & $\np$ & $\ng$ & $\nm$ & $\npg$\\
  \hline
%156.14 &   9 & -79 & -70 &  88 \\
%158.05 &   9 & -78 & -69 &  87 \\
159.89         &   9 & -77 & -68 &  86 \\
161.19$_{---}$ &   9 & -76 & -67 &  85 \\
162.56         &  10 & -76 & -66 &  86 \\
164.54         &  10 & -75 & -65 &  85 \\
(...)& \\
171.12         &  10 & -72 & -62 &  82 \\
173.28         &  10 & -71 & -61 &  81 \\
174.79$_{---}$ &  10 & -70 & -60 &  80 \\
176.35         &  11 & -70 & -59 &  81 \\
178.68         &  11 & -69 & -58 &  80 \\
(...)& \\
183.77         &  11 & -67 & -56 &  78 \\
186.37         &  11 & -66 & -55 &  77 \\
188.21$_{---}$ &  11 & -65 & -54 &  76 \\
189.84         &  12 & -65 & -53 &  77 \\
192.47         &  12 & -64 & -52 &  76 \\
195.37         &  12 & -63 & -51 &  75 \\
198.49         &  12 & -62 & -50 &  74 \\
201.12$_{---}$ &  12 & -61 & -49 &  73 \\
202.70         &  13 & -61 & -48 &  74 \\
  \hline
\end{tabular}}
\legende{Classification and number of mixed modes}{Number and
classification of mixed modes in the RGB KIC 9882316 with $\Dnuobs
= 13.68\,\mu$Hz and $\Tg=80.60$\,s \citep{2012A&A...540A.143M}.
The location of the pure pressure modes is indicated with
horizontal ticks. Unobserved or unidentified mixed modes, near
radial modes, are indicated by (...).
 \labell{tab-compte-mixte}}
\end{table}

\subsection{Beyond asymptotic expansions, still asymptotics}

Departure to the second-order asymptotic expansion can be due to
rapid structure variations, called glitches
(Fig.~\refer{fig-glitch}). The discontinuity induced by the glitch
on the interior structure parameters induces a modulation in the
oscillation spectrum (Fig.~\refer{fig-observed-glitch}).\\

\begin{figure}[!t]
  \fichier{9.6}{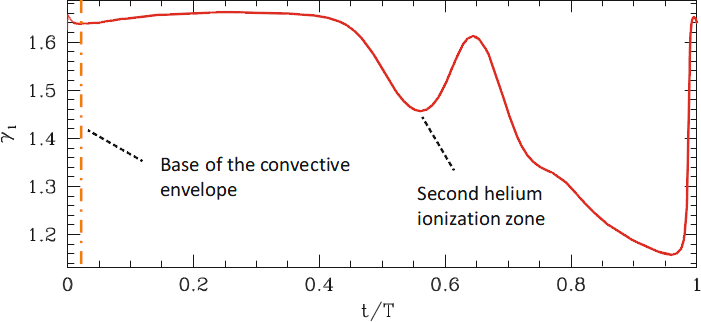}
  \legende{Glitch due to the second helium ionization zone} {Glitches in the $\Gamma_1$
  coefficient due to the second helium ionization zone in the red giant HD
  181907 (HR 7349) observed by CoRoT   \credit{2010A&A...520L...6M}.
  \labell{fig-glitch}}
\end{figure}

\begin{figure}[!t]
 \fichier{9}{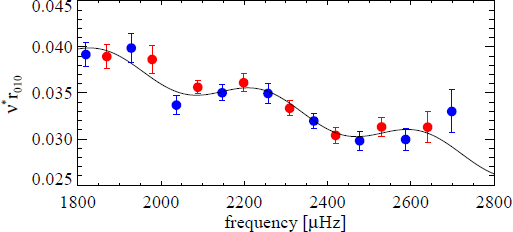}

 \fichier{9}{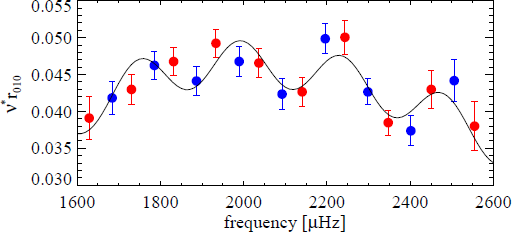}
 \legende{Observed glitches} {Glitches observed in two main-sequence stars observed with
 \Kepler\
 \credit{2014ApJ...782...18M}.
 \labell{fig-observed-glitch}}
\end{figure}

In fact, the asymptotic formalism
\cite[e.g.,][]{1993A&A...274..595P} helps investigating this
departure: the unperturbed spectrum is modulated by a cosine term.
At first order:
\begin{equation}
  {\nu\nl}^{(1)} \ =\ \left[ \ n'
    - {\eta\over \pi} \sin 2\pi
    \left( {n'\over N} -
  {\ell \over 2} \right) -{\eta^2\over \pi}
  {N-2\over   2N}
  \sin 2\pi {2n'\over N } \ \right] \ \Dnu,
  \labell{eqt-glitch1}
\end{equation}
with
\begin{eqnarray}
\left\{
\begin{array}{rcl}
  n' &=&\displaystyle{ n + {\ell\over 2} +
         \varepsilon} ,
         \\
  N &=& \displaystyle \int_0^R {\diff r\over c}
  \biggm\slash
  \displaystyle \int\ind{glitch}^R\!\! {\diff r\over c} .
  \labell{eqt-period-glitch}\\
\end{array}
\right.
\end{eqnarray}
The amplitude term $\eta$ is defined in Appendix
\refer{justi-asymp}. The definition of the period $N$ introduces
the acoustic depth $\tau\ind{glitch} = \int\ind{glitch}^R\!\!
{\diff r / c}$. A second-order form can be derived too
\citep{1993A&A...274..595P}. \\

Other developments are found in the literature. They have the same
oscillatory behavior,
\begin{equation}\label{eqt-oscill-glitch}
    \delta\nu\ind{glitch}
    \propto
    \sin \bigl( 4\pi \tau\ind{glitch} \nu\nl + \varphi \bigr)
\end{equation}
but various forms for the amplitudes, either uniform or not
\citep[e.g.,][]{2014ApJ...782...18M}. \\

What is most important is the relation between the period of the
glitch and its acoustic depth, as seen in
Eq.~(\refer{eqt-period-glitch}). Measuring this period helps
locating various structure discontinuities where the sound speed
profile is disrupted, either as the base of the convective zone,
or at the second helium ionization zone. In red giants, sound
speed glitches are mostly associated to the second helium
ionization region
\citep{2010A&A...520L...6M,2012A&A...538A..73B,2014MNRAS.440.1828B}.
\cite{2015arXiv150507280V} have observed them in a large data set
of red giants observed by \Kepler . They could derive that the
glitch properties mainly depend on the stellar evolutionary
status, so that they can be used to distinguish RBG from red clump
stars, in a complementary way to period spacings
\citep{2011Natur.471..608B,2011A&A...532A..86M}.

%%%%%%%%%%%%%%%%%%%%%%%%%%%%%%%%%%%%%%%%%%%%%%%%%%%%%%%%%%%%%%%%%%%%%%%%%%%%%%%%%%%
\section{Ensemble asteroseismology\labell{ensemble}}

\subsection{Scaling relations\labell{scaling-relations}}

Seismic scaling relations play an increasing role in stellar
physics, since they provide relevant estimates of the stellar mass
and radius. We examine how such relations work.

\subsubsection{The frequency $\numax$ of maximum oscillation signal\labell{numax}}

Oscillation modes are preferably seen around the frequency
$\numax$. Regardless of the excitation mechanism, out of the scope
of this lecture, we may investigate the consequences of this fact.

The excitation occurs in the uppermost stellar envelope, so that
is has been conjectured by
\cite{1991ApJ...368..599B,1995A&A...293...87K} that the frequency
of the maximum of the power spectrum $\numax$  scales as the
surface cutoff frequency $\nuc =\wc / 2\pi$ because the latter
corresponds to a typical time-scale of the atmosphere. This has
been justified by \cite{2011A&A...530A.142B}, who examined how
$\nuc$ and $\numax$ depend on the characteristic thermal time in
the upper convective layers. They have shown that
\begin{equation}\labell{eqt-nuc-numax-mach}
    \numax \ \propto  \ \mathcal{M}^3 \ \nuc,
\end{equation}
where $\mathcal{M}$ is the Mach number in the uppermost convective
region where modes are excited. It seems that this number has a
week dependence with the stellar evolution, since it varies
approximately as $g^{-0.012}$ \citep{2013ASPC..479...61B}, so that
Eq.~(\refer{eqt-nuc-numax-mach}) mostly reduces to
\begin{equation}\label{eqt-nuc-numax}
    \numax \ \propto\ \nuc
\end{equation}
From Eq.~(\refer{eqt-wc-g}), we get that $\nuc$, hence $\numax$,
varies as $g\,\Teff^{-1/2}$. This explains why the measurement of
$\numax$ provides measurements of the stellar gravity much more
precise than the $\log g$ parameter derived from spectrometry
\citep{2011ApJ...730...63G}.

Hence, the measurements of $\numax$ and $\Dnu$ can be scaled to
the stellar mass, radius, and effective temperature:
\begin{eqnarray}
  {\numax \over \numaxs}
  &=&
  \left({M\over\Ms}\right) \   \left({R \over\Rs}\right)^{-2} \
  \left({\Teff \over \Ts}\right)^{-1/2}
  ,
  \\
   {\Dnu \over \dnus}
   &=&
  \left({M\over\Ms}\right)^{1/2} \   \left({R \over\Rs}\right)^{-3/2}
  .
\end{eqnarray}
The scaling of $\Dnu$ with the square root of the mean density can
be derived from dimensional analysis \citep{1917Obs....40..290E}
or, more precisely, from homology
\citep[e.g.,][]{2013ASPC..479...61B}.\\

$\Dnu$ and $\numax$ jointly evolved with stellar evolution
(Fig.~\refer{fig-ensemble2}). From the previous equations, one
derives
\begin{equation}\label{eqt-scaling-evol}
  \Dnu \propto M^{-1/4} \ \Teff^{3/8} \ \numax^{3/4}
  .
\end{equation}
We may examine the meaning of this relation, related with stellar
evolution. Assuming that $M$ is constant with stellar evolution is
a reasonable assumption, except if we focus on the upper part on
the RGB and AGB regime where strong mass loss can occur.

In the main-sequence regime, the slope of the $\Dnu (\numax)$
relation is 0.80. The relation significantly depends on the
stellar mass, namely in the range [0.9 - 1.7\,$M_\odot$] for
main-sequence stars showing solar-like oscillations
\citep[e.g.,][]{2013A&A...550A.126M}. In fact, examining the slope
of relation has no deep physical meaning at these evolutionary
stages.

In the red giant regime, the slope of the $\Dnu (\numax)$ relation
is very close to 0.75 \citep[e.g.,][]{2010A&A...517A..22M}. Here,
examining the slopes has a deeper physical meaning: the $\numax$ -
$\Dnu$ diagram (Fig.~\refer{fig-ensemble2}) is composed of nearly
parallel evolution tracks of low-mass stars, according to
Eq.~(\refer{eqt-scaling-evol}). This evolution at fixed mass
explains $\Dnu \propto \numax^{3/4}$ since the $\Teff$ variation
is small.

\begin{figure}[!t]
 \fichier{11}{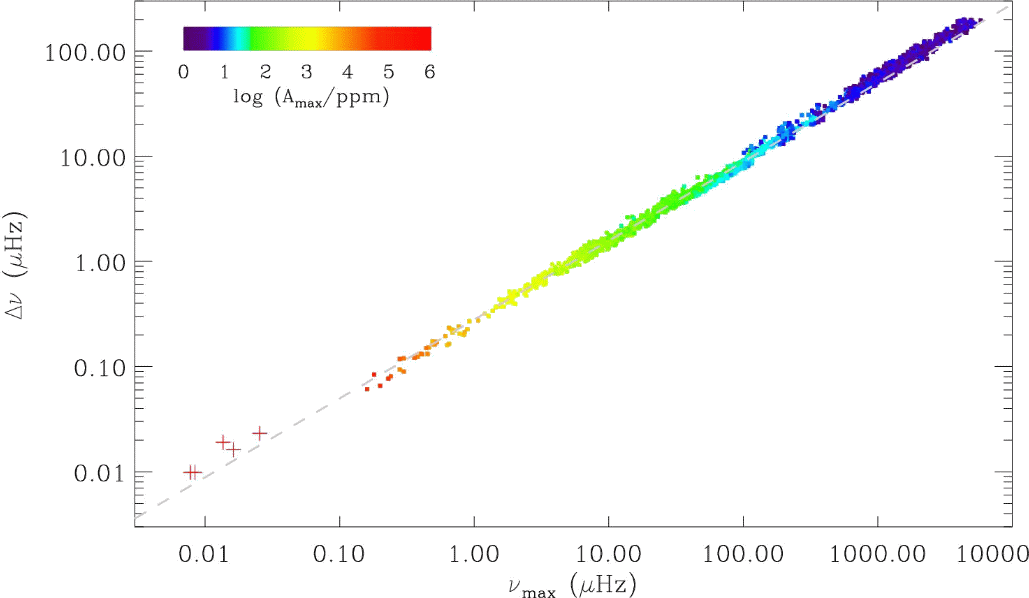}
 \legende{Ensemble asteroseismology (2)} {Stellar evolutionary tracks in the
  $\numax$ - $\Dnu$ diagram. The dashed line indicates the 3/4 slope.
 In the red giant regime, it corresponds closely to the
 evolutionary track of a 1.3-$M_\odot$ star. Tracks for other
 stellar masses are close and parallel
 \credit{2013sf2a.conf...25M}.
 \labell{fig-ensemble2}}
\end{figure}

\subsubsection{Mass and radius scaling relations}

Inverting the previous system of equations and considering that
$\Teff$ can be derived from band-photometry or from spectroscopy,
we can obtain estimates of the stellar mass and radius.
\begin{equation}
  {R \over\Rs}  = \left({\numax \over \numaxs}\right) \
     \left({\Dnu \over \dnus}\right)^{-2}
     \left({\Teff \over \Ts}\right)^{1/2}, \labell{scalingR}
\end{equation}
\begin{equation}
  {M\over\Ms} = \left({\numax \over \numaxs}\right)^{3}
     \left({\Dnu \over \dnus}\right)^{-4} \left({\Teff \over \Ts}\right)^{3/2} . \labell{scalingM}
\end{equation}
The precise calibration of these equations is a pending problem.
The solar values chosen as references are not fixed uniformly in
the literature. Usually, internal calibration is ensured by the
analysis of the solar low-degree oscillation spectrum with the
same tool used for the asteroseismic spectra. However, this does
not provide a robust calibration since this unduly supposes that
homology is ensured during stellar evolution, which is not proven.
One can find significantly different reference values for
Eqs.~(\refer{scalingR}) and (\refer{scalingM}), always close to
the solar values, e.g., $\dnus = 134.9\,\mu$Hz and $\numaxs =
3120\,\mu$Hz \citep{2010A&A...522A...1K}. We see in the next
Section that this diversity is not an issue with a coherent and
proper calibration. \\

\begin{figure}[t!]
 \fichier{10.6}{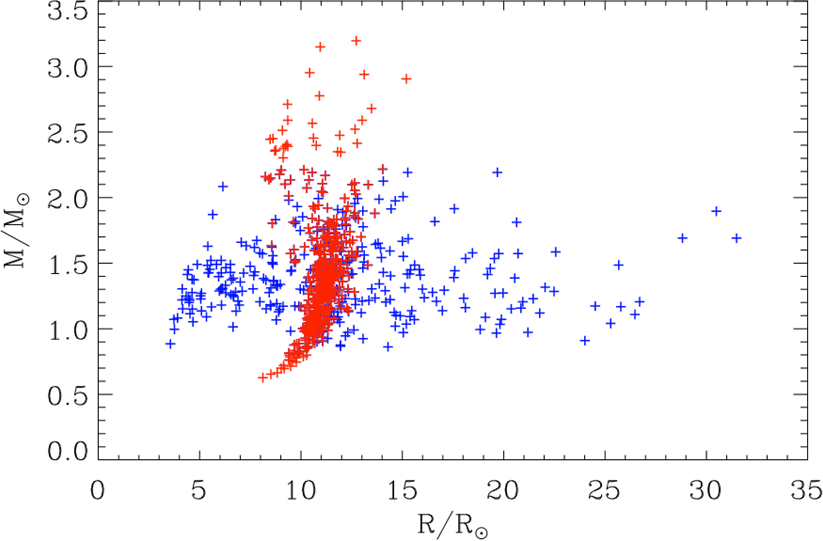}%
 \legende{Mass -- radius relation } {Mass -- radius relation for
 \emph{Kepler} red giants, with RGB stars in blue and clump stars in
  red. We note that RGB stars showing solar-like oscillations have
  typical masses in the range [1 -- 2\,$M_\odot$]: lower-mass
  stars are rare due to their too long evolution time; higher-mass
  stars are rare too since they reach the RGB at larger radius and evolve rapidly.
  The clump population
  presents a well defined mass-radius relation, due to the fact
  that clump stars start burning helium in the core in the same,
  degenerate conditions. The presence of clump stars with masses
  as low as 0.65\,$M_\odot$ indicates that strong mass loss has
  occurred on the tip of the RGB.
  Star in the secondary clump, with a mass above 1.9\,$M_\odot$,
  show a larger spread: they started helium burning in
  non-degenerate conditions.
  Figure adapted from \cite{2012A&A...537A..30M}.
  \label{fig-MR}}
\end{figure}

Independent from this calibration effort, it is clear that these
estimates of the stellar masses and radii are fully relevant (and
highly useful), as shown by the systematic comparison of modeling
and scaling relations \citep[Table 1 of][and references
therein]{2013A&A...550A.126M}. Having such precise estimates for
field stars is incredibly useful. It allows us to address stellar
physics with enriched information (Fig.~\refer{fig-MR}).

This emphasizes the importance of precise measurements of
$\Dnuobs$ and $\numax$. Very precise values of $\Dnuobs$ can be
obtained, with a relative precision better than 1\,\%. This
precision is relevant only if glitches are properly taken into
account. Deriving a precision on $\numax$ better than $\Dnu /4$ is
difficult and cannot be considered as accurate since such a value
is highly method-dependent.

\subsection{Calibration of the scaling relations}

\subsubsection{Definitions}

A proper calibration of the mass and radius scaling relations
should be based on the most relevant definition and measurements
of the seismic variables $\Dnu$ and $\numax$. The efforts in this
direction are hampered by serious problems with $\numax$. We lack
a precise definition of $\numax$, since defining it as the
frequency of maximum oscillation signal is not enough; this
maximum depends on the way oscillations are measured.  We also
lack clues for quantifying synthetic values of $\numax$: it is
indeed  possible to derive the value of $\nuc$ from a model, but
not the value of $\numax$. The estimate of $\numax$ requires an
estimate of the Mach number, according to
Eq.~(\refer{eqt-nuc-numax-mach}). Such effort is currently done
for measuring $\mathcal{M}$, but results have not yet reached the
necessary precision \citep[][and Samadi \& Belkacem's contribution
to this EES]{2012A&A...543A.120S, 2013A&A...559A..40S}.

Deriving more precise information for the large separation is
easier, but not straightforward. In fact, we have to face with
many definitions of the large separations (with a large confusion
in the literature):

- The local measurements provide local values of the frequency
spacings $\Dnu\nl = \nu_{n+1,\ell} - \nu\nl$. Following
\cite{2013A&A...550A.126M}, a global measurement is favored for
minimizing the influence of the glitch on the mean large
separation $\langle\Dnu\rangle$. The influence of the glitches can
be seen in Fig.~\ref{fig-scaling-calib}: they create extra
fluctuations. In a first approximation, global measurements can
provide a glitch-free measurement of $\Dnuobs$.

- The observed value $\Dnuobs$ of the mean large separation is, as
shown by \cite{2013A&A...550A.126M}, largely different from the
asymptotic value $\Dnuas$. According to its definition
(Eq.~\ref{eqt-def-Dnu}), $\Dnuas$ is related to the integral of
the sound speed. Except in case of perfect homology, its
measurement cannot provide directly the stellar mass and radius.

- Last but not least, for linking the seismic measurement with the
stellar mass and radius, it is mandatory to introduce the
dynamical frequency, which scales as the square root of the mean
stellar density \citep{1917Obs....40..290E}
\begin{equation}\label{eqt-def-nu0}
    \nu_0 \propto \sqrt{\G M\over R^3} .
\end{equation}
Departure from homology explains the uncalibrated scaling between
$\Dnuas$ and $\nu_0$ \citep{2013ASPC..479...61B}.

So, we should use the dynamical frequency $\nu_0$ instead of
$\Dnuobs$ or $\Dnuas$, and the acoustic frequency $\nu\ind{c}$
instead of $\numax$ in the scaling relations (Eq.~\refer{scalingR}
and ~\refer{scalingM}). As this is not the case, an intensive
calibration effort is necessary.

\subsubsection{Calibration with independent measurements}

A possible way to calibrate the relation comes from independent
information derived from complementary observations.

- An independent verification has been made for stars that have
accurate Hipparcos parallaxes, by  coupling asteroseismic analysis
with the InfraRed Flux Method \citep{2012ApJ...757...99S}. The
seismic distance determinations agree to better than 5\,\%: this
shows the relevance and the accuracy of the scaling relations in
the subgiant and main-sequence regime.

- With long-baseline interferometric measurement of the radius of
five main-sequence stars, one subgiant, and four red giant stars
for which solar-like oscillations have been detected by either
\Kepler\ or CoRoT, \cite{2012ApJ...760...32H} have shown that
scaling relations are in excellent agreement within the
observational uncertainties. They finally derive that
asteroseismic radii for main-sequence stars are accurate to better
than 4\,\%.

- Oscillations in cluster stars \citep{2011ApJ...729L..10B} were
used to compare scaling relations for red giants  in the red clump
or on the RGB. \cite{2012MNRAS.419.2077M} have found evidence for
systematic differences in the $\Dnuobs$ scaling relation between
He-burning and H-shell-burning giants. This implies that a
relative correction between RGB and clump stars must be
considered. As this correction is also related to mass loss, it is
currently not possible to measure it precisely. Independent of
this, oscillations in cluster stars provide useful constraints on
cluster membership \citep{2011ApJ...739...13S}. They also help
constraining  the relations depicting the parameters of the
pressure mode spectrum \citep{2012ApJ...757..190C}.\\

These efforts have used the different definitions and meaning of
the large separation. As a consequence, homogeneity is not
ensured. As differences between the dynamical frequency and large
separations may be as high as the magnitude of the calibration,
homogeneous definitions are mandatory for a precise calibration.

\begin{figure}[!t]
 \fichier{10.5}{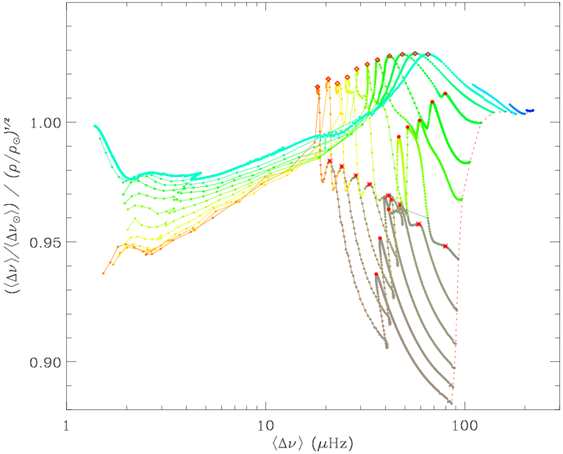}
 \legende{Scaling} {Calibration of the $\Dnu$ scaling relation
 with respect to the square root of the mean density, as a function of
 $\Dnu$ in models of near-solar metallicity ($Z = 0.017$) and mass range
 from 0.7 (green) to 2.0\,$M_\odot$ (magenta). Models that have
 effective temperatures hotter than the approximate cool edge of
 the classical instability strip are shown in gray. The zero age
 main sequence is indicated by the dotted red line
\credit{2011ApJ...742L...3W}.
 \labell{fig-scaling-calib}}
\end{figure}

\begin{figure}[t!]
 \fichier{8.4}{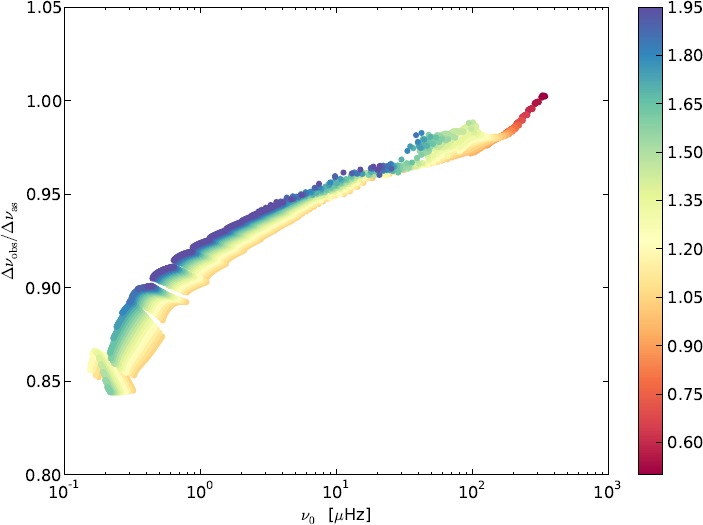}%

 \fichier{8.4}{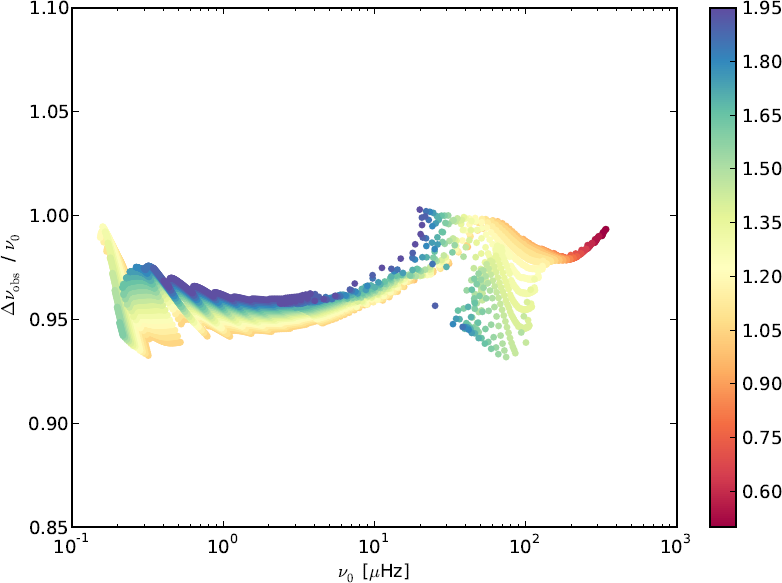}%

 \fichier{8.4}{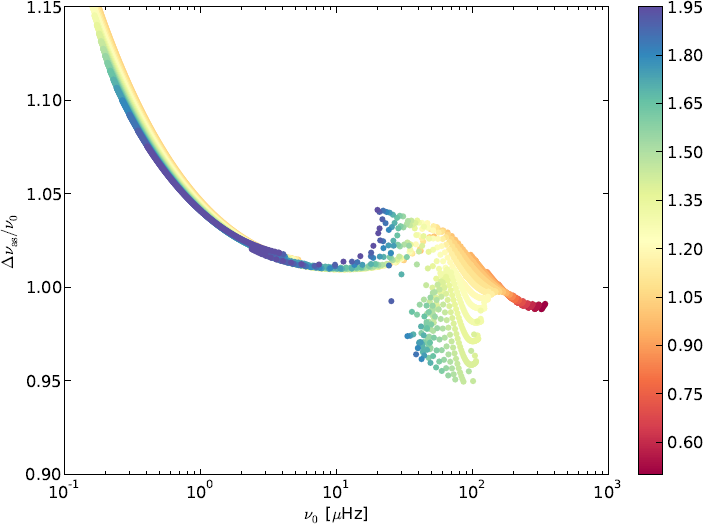}%
 \legende{Scalings between $\Dnuobs$, $\Dnuas$, $\nu_0$} {Scalings between $\Dnuobs$, $\Dnuas$,
 $\nu_0$. Surprisingly,  $\Dnuobs$ provides a better fit of $\nu_0$ than
 $\Dnuas$. There is however no other way to link $\Dnuobs$ and
 $\nu_0$ than considering $\Dnuas$ as an intermediate case
  \credit{2013ASPC..479...61B}.
  \label{fig-scaling-kevin}}
\end{figure}

\begin{table}[t]
\caption{Calibration of the mass and radius scaling
relations}\label{table-calibration}
\begin{tabular}{rrcll}
\hline
% &\multicolumn{3}{c}{Calibration} \\
%Modelling & Asymptotic&   Calibration & Asymptotic& Observation\\
%$\numaxmod$                 & $\leftarrow$           & ? & $\leftarrow$          & $ \numaxobs $\\
%\hline $\nu_0(M,R) \leftrightarrow$& $\Dnuasmod \leftrightarrow$ &? & $\leftrightarrow \Dnuasobs$&$\leftrightarrow \Dnu$\\
Observation & Asymptotic&   Calibration & Asymptotic& Modelling \\
\hline
${\numax}\ind{,obs}$             &$\rightarrow$             & ? & $\rightarrow$             &${\numax}\ind{,mod}$  \\
$\Dnuobs\leftrightarrow$&$\Dnu\ind{as,obs}\leftrightarrow$& ? & $\leftrightarrow\Dnu\ind{as,mod}$&$\leftrightarrow\nu_0(M,R)$\\
\hline
\end{tabular}

\scriptsize{Calibration process showing all formal steps for a
proper calibration of the mass and radius scaling relations,
through asymptotic values for the large separation, determined in
the modelling process from $1/(2\int_0^R \diff r / c)$ and in
observations from the asymptotically corrected  $\Dnuobs$.

}
\end{table}

\subsubsection{Calibration and modelling}

Modelling helps determining the differences between $\nu_0$,
$\Dnuobs$ and $\Dnuas$. So, modelling must be used to link those
different acceptations of the large separation.\\

However, adiabatic code cannot help for calibrating $\numax$, so
that it seems necessary to work in asymptotic conditions:
converting observed values into asymptotic values seems more
precise than the opposite, since the asymptotic regime (large
radial orders) is conceptually and practically defined, whereas
$\numax$ and $\nmax$ are not known if not observed. In other
words, it is safer to translate an observed $\Dnuobs$ into an
asymptotic value than to translate an asymptotic value into an
unknown observable value. The procedure for linking $\nu_0$,
$\Dnuas$ and $\Dnuobs$ is summarized in
Table~\refer{table-calibration}.

- With grid modelling, \cite{2011ApJ...742L...3W} have proposed a
calibration depending on the stellar temperature. They have
compared the large separation, as it should be observed, with the
mean density (Fig.~\refer{fig-scaling-calib}). So, they have
compared $\Dnuobs$ with $\nu_0$: $\Dnuobs$ is derived from the
frequency spacing in the synthetic oscillation spectrum near the
expected $\numax$ and $\nu_0$ is derived from the stellar mass and
radius.

- With grid modelling too, \cite{2013ASPC..479...61B} have
compared $\nu_0$, $\Dnuas$ and $\Dnuobs$. This work shows that
$\nu_0$ is closer to $\Dnuobs$ than $\Dnuas$
(Fig.~\refer{fig-scaling-kevin}). It defines a precise methodology
for proper calibration. It shows that the basic physical picture
is understood and that departure from the observed relation arises
from the complexity of non-adiabatic processes of convection,
which requires time-dependent 3D hydrodynamical simulations.

- With a compilation of observations and modelling,
\cite{2013A&A...550A.126M} have proposed a calibration of the
scaling. This calibration represents part of the effort, since
observed values were translated into asymptotic values: \\
1) considering asymptotic values instead of directly observed
values helps reducing the spread in the calibration process.\\
2) calibrating the relation with modelled stars at various
evolution stages ensures a more precise result than calibrating
with the Sun only.\\

The best way to use this calibration with the solar-like (but not
solar) references consists in translating the observed $\Dnuobs$
into the asymptotic value $\Dnuas$,
\begin{equation}\label{eqt-Dnuobs-Dnuas}
  \Dnuas = \Dnuobs \ (1+ \zeta)
\end{equation}
with $\zeta = 0.57 / \nmax$ in the main-sequence regime and $\zeta
= 0.038$ in the red giant regime, according to
Eqs.~(\refer{eqt-curv-MS}), (\refer{eqt-curv-RG}), and
(\refer{eqdnu}). Then,
\begin{equation}
  {R \over\Rs}  = \left({\numax \over \numaxref}\right) \
     \left({\Dnuas \over \Dnuref}\right)^{-2}
     \left({\Teff \over \Ts}\right)^{1/2}, \labell{scalingRas}
\end{equation}
\begin{equation}
  {M\over\Ms} = \left({\numax \over \numaxref}\right)^{3}
     \left({\Dnuas \over \Dnuref}\right)^{-4} \left({\Teff \over \Ts}\right)^{3/2} ,  \labell{scalingMas}
\end{equation}
with $\Dnuref=3104\,\mu$Hz and $\numaxref = 138.8\,\mu$Hz based on
the comparison with the models \citep{2013A&A...550A.126M}. The
reference $\Dnuref$ is much larger than the observed value of the
solar large separation; it is close to the asymptotic value of the
solar large separation.\\

Otherwise, corrections can be done on the $\Robs$ and $\Mobs$
values derived from Eqs.~(\refer{scalingR}) and (\refer{scalingM})
used with $\Dnuobs$. The correction and calibration provided by
Eqs.~(\refer{scalingRas}) and (\refer{scalingMas}) are equivalent
to the translations
\begin{equation}
  \Ras \simeq  \left({ 1 - 2(\zeta-\zeta_\odot)} \right) \Robs
  \hbox{ and }
  \Mas \simeq  \left({ 1 - 4 (\zeta- \zeta\odot)} \right)
  \Mobs ,
\end{equation}
with $\Robs$ and $\Mobs$ given by Eqs.~(\refer{scalingR}) and
(\refer{scalingM}). This accounts for the fact that scaling
relations providing raw estimates $\Robs$ and $\Mobs$ were
calibrated on the Sun, so that one has to deduce the solar
correction $\zeta_\odot \simeq
2.6\,$\%. \\

As a result, for subgiants and main-sequence stars, the
\emph{systematic} negative correction reaches about 5\,\% for the
seismic estimate of the mass and about 2.5\,\% for the seismic
estimate of the radius. The absolute corrections are maximum in
the red giant regime. This justifies the correcting factors early
introduced for deriving masses and radii for CoRoT red giants
\citep[Eqs. (9) and (10) of][]{2010A&A...517A..22M}, obtained by
comparison with the modeling of red giants chosen as reference.

\begin{figure}[t!]
 \fichier{11.4}{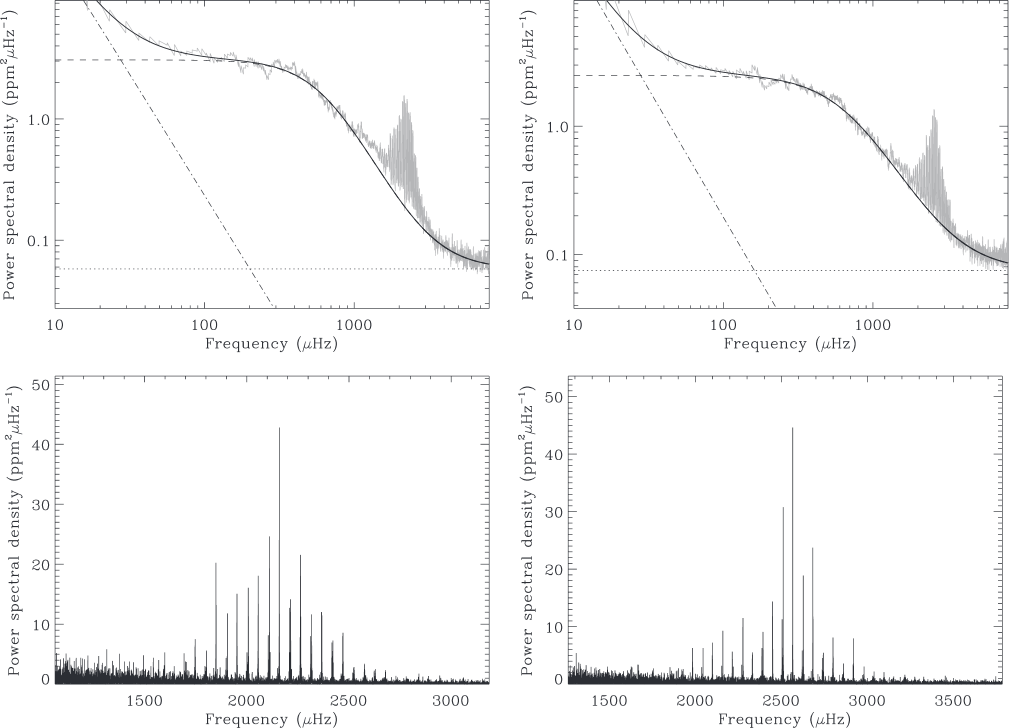}%
 \legende{Oscillation spectra of 16 Cyg A and B }{Oscillation spectra of
 16 Cyg A  (left panels) and B  (right panels)
 \credit{2012ApJ...748L..10M}.
  \label{fig-cyg}}
\end{figure}

\subsection{Seismic parameters}

Other global seismic parameters conform to a large numbers of
scaling relations. These scaling relations make that all
oscillation spectra always show similar features
(Fig.~\refer{fig-ensemble-intro}, ~\refer{fig-cyg}). These scaling
relations show even less spread in the red giant regime, as a
result of homology. For red giants, scalings with $\numax$ are
summarized in Table~\ref{table-fit}.\\

- The scaling relation between $\numax$ and $\Dnuobs$, already
discussed, provides a direct view on stellar evolution. As a by
product, one can estimate the radial order at $\numax$: $\nmax =
\numax / \Dnuobs - \epsobs$. This order $\nmax$ significantly
decreases when $\Dnuobs$ decreases. For the Sun, $\nmax\simeq 22$;
at the red clump, $\nmax\simeq 8$; and at the tip of the RGB,
$\nmax\simeq 2.5$. As already shown, observing solar-like
oscillations in red giants and even more in semi-regular variables
occur in non-asymptotic conditions.

- $\dnuenv$ is the full-width at half-maximum of the smoothed
excess power; $\nenv = \dnuenv / \Dnuobs$ provides $\dnuenv$ in
large separation unit; as $\nmax$, $\nenv$ significantly decreases
when $\Dnuobs$ decreases since $\dnuenv$ approximately scales as
$\numax$. As already shown, only a few radial orders can be
observed in semi-regular variables showing solar-like
oscillations.

- $\Hmax$ (in ppm$^2\,\mu$Hz$^{-1}$) is the mean height of the
modes at $\numax$, defined according to the description of
smoothed excess power as a Gaussian envelope
\citep[e.g.,][]{2012A&A...537A..30M}.

- $\Bmax$ (in ppm$^2\,\mu$Hz$^{-1}$) is the value of the stellar
background $B$ at $\numax$. The background is described by
Harvey-like components \citep{2008Sci...322..558M}. Each component
is a modified Lorentzian of the form $b(\nu) = a / [1 +
(2\pi\,\nu\tau)^{\alpha}]$, where $\tau$ is the characteristic
time scale. High values for the exponent $\alpha$, about 4, are
definitely preferred, with two different components in the
background \citep{2014A&A...570A..41K}. However, the local slope
of the background in the frequency range where oscillations are
observed is close to $\nu^{-2}$ \citep{2012A&A...537A..30M}.
$\Hmax / \Bmax$ is representative of the height-to-background
ratio (HBR) at $\numax$. Interestingly, this ratio shows no
significant variation all along stellar evolution.\\

\begin{table}
 \centering
 \caption{Scaling relations in red giants}\label{table-fit}
  \rotatebox{90}{%
  \begin{tabular}{lllccl }
\hline parameter &  & unit & coefficient $\alpha$ & exponent $\beta$\\
\hline
large separation & $\Dnuobs$  & $\mu$Hz           &$0.274\pm 0.004$ & $0.757\pm0.004$ & \\
                 &$\nmax=\numax/\Dnu -\epsobs$ &--&$3.26\pm0.031$   & $0.242\pm0.005$ & \\
FWHM             & $\dnuenv$  & $\mu$Hz           &$0.73 \pm 0.03 $ & $0.88\pm0.01$   & \\
                 & $\nenv= \dnuenv/\Dnu$       &--&$2.49 \pm 0.12$  & $0.13\pm0.01$ & \\
Height at $\numax$    &$\Hmax$&ppm$^2$\,$\mu$Hz$^{-1}$&$(2.03\pm 0.05)\;10^7$& $-2.38\pm0.01$ &  \\
Background at $\numax$&$\Bmax$&ppm$^2$\,$\mu$Hz$^{-1}$&$(6.37\pm 0.02)\;10^6$& $-2.41\pm0.01$ &  \\
HBR              &$\Hmax / \Bmax$              &--&$3.18\pm 0.09$   & $0.03\pm 0.03$  \\
Granulation      & $P\ind{g}$&ppm$^2$\,$\mu$Hz$^{-1}$ &                      & $-2.15\pm0.12$ &  \\
                 &$\tau\ind{g}$& s &                   & $-0.90\pm0.005$&\\
  &$P\ind{g}(\tau\ind{g})$&ppm$^2$\,$\mu$Hz$^{-1}$&        & $2.34\pm0.01$ &  \\
 \hline
radius & $R$        &   $R_\odot$                  &  $63.1\pm 1.1$  & $-0.49\pm 0.01$ \\
effective temperature& $\Teff$    & K              &  $3922\pm 50$   & $0.051\pm 0.05$ \\
\hline
\end{tabular}
}
\rotatebox{90}{%
-  Each parameter is estimated as a power law of $\numax$, with
$\alpha$ the coefficient and $\beta$ the exponent,}
\rotatebox{90}{%
except $P\ind{g}(\tau\ind{g})$.}
\rotatebox{90}{%
- All results were obtained with the COR pipeline
\citep{2012A&A...537A..30M}. }
\rotatebox{90}{%
- Granulation data are from \cite{2011ApJ...741..119M}. }
\end{table}

Some of the scaling relations were illustrated in
Fig.~\ref{fig-ensemble-intro} that shows, compared to the Sun,
oscillation spectra of red giants from the bottom to the top of
the RGB. Currently, we lack theoretical models for explaining most
of these relations. Their interpretation involves a non-adiabatic
treatment.\\

Large efforts have been devoted to explain the scaling relations
of the maximum amplitude $\Amax$.  This global parameter can be
fitted, in limited frequency range, as in
\cite{2012A&A...537A..30M}. However, the fit heavily depends on
the method \citep{2011ApJ...743..143H}, so that it is not yet
possible to provide a physically relevant result
\citep{2013MNRAS.430.2313C}. \cite{2012A&A...543A.120S} have shown
that scaling relations of mode amplitudes cannot be extended from
main-sequence to red giant stars because non-adiabatic effects for
red giant stars cannot be neglected. \cite{2013A&A...559A..40S}
have recently proposed a theoretical model of the oscillation
spectrum associated with the stellar granulation as seen in
disk-integrated intensity. With this model, they have highlighted
the role of the photospheric Mach number for controlling the
properties of the stellar granulation.

\begin{figure}[!t]
 \fichier{11}{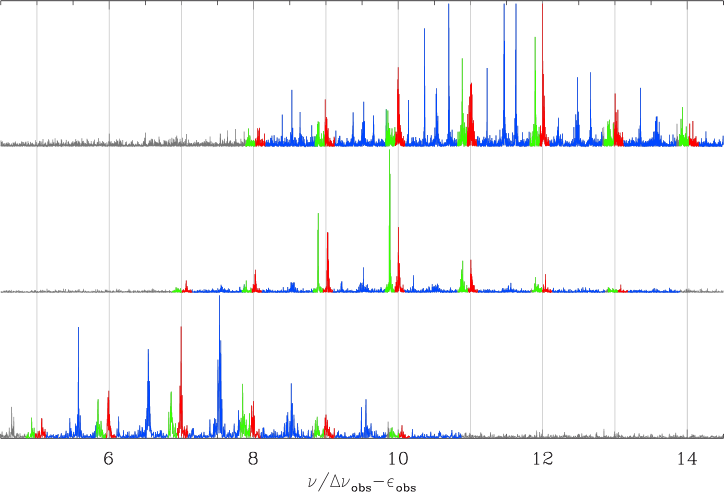}
 \legende{Depressed mixed modes}{Red giant oscillation spectra, as a function
 of the normalized frequency $\nu / \Dnuobs - \epsobs$; $\ell =
 0$,1, and 2 modes are plotted in red, blue, and green, respectively.
  \emph{Top:} typical pattern with a large number of gravity-dominated
  mixed modes.
  \emph{Middle:} dipole modes are depressed.
  \emph{Bottom:} dipole modes are mostly pressure-dominated
 \credit{2013EAS....63..137M}.
 \labell{mixed-depressed1}}
\end{figure}

\begin{figure}[!t]
 \fichier{11}{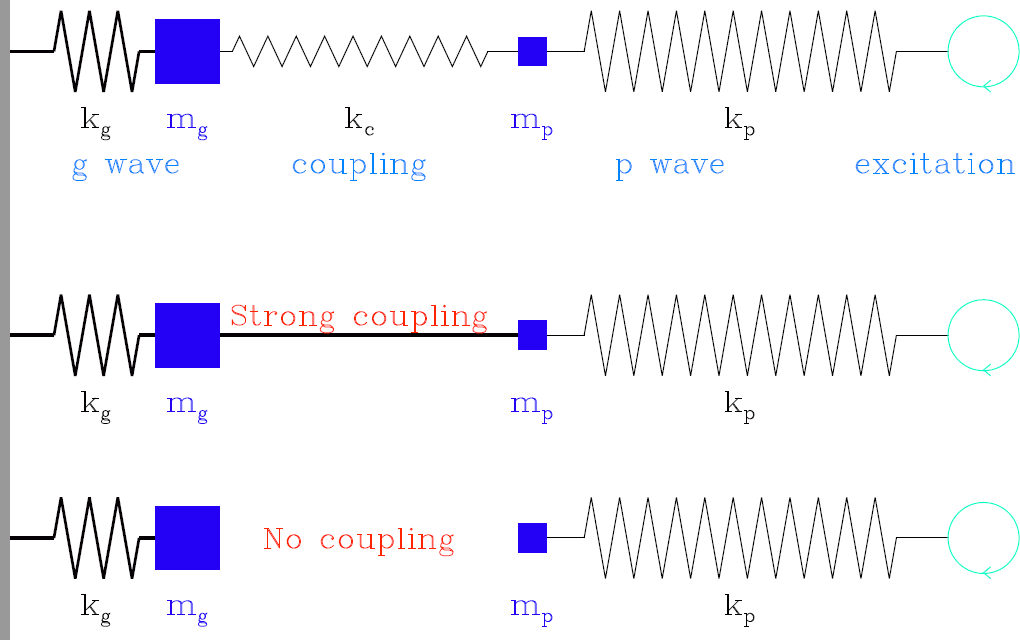}
  \legende{Toy model} {Toy model with two masses and three springs,
  representative of the coupling of gravity and pressure waves.
  Coupling conditions of the pressure and gravity waves
  contributing to a mixed mode with an angular frequency $\omega$
  imply that
  $\omega^2 \simeq k\ind{g}/m\ind{g} \simeq k\ind{p}/m\ind{p}$.
  \emph{Top:} the mean case, with an
  intermediate coupling, helps explaining the gravity-dominated
  mixed modes.
  \emph{Middle:} very strong coupling corresponds to depressed mixed modes with a huge
  mode mass; such depressed mixed modes are observed on the whole RGB.
  \emph{Bottom:} very weak coupling;
  in this case, only pressure modes are observed, as is the case at
  low frequency \citep{2014A&A...572A..11G}
  \credit{2013EAS....63..137M}.
  \labell{mixed-depressed2}}
\end{figure}

As a result, most of the global seismic parameters follow tight
scaling relations, but not all. A class of RGB stars with
depressed mixed modes has been put in evidence by
\cite{2012A&A...537A..30M}: they show very low surface amplitudes
(Fig.~\ref{mixed-depressed1}), due to a too efficient coupling
between the pressure and gravity contributions to the mixed modes
(Fig.~\ref{mixed-depressed2}). Such depressed mixed modes have a
huge mode mass, hence have no detectable surface amplitudes.

\subsection{Seismic indices}

The various seismic parameters that can be derived from solar-like
oscillation spectra can now be used as seismic indices. They
provide a complementary view to the usual fundamental constraints
used in stellar physics, as the effective temperature, luminosity,
$\log g$ value, or metallicity $Z$.\\

The effective temperature plays a crucial role, since it enters
the radius and mass scaling relations (Eq.~\refer{scalingR} and
\refer{scalingM}). As such, it carries independent information
compared to $\Dnu$ and $\numax$. On the RGB, degeneracy implies
however a close relationship between $\Teff$ and stellar
evolution, expressed by $\Teff \propto \numax^{-0.04 \to -0.06}$
\citep[e.g.,][]{2010A&A...517A..22M,2011ApJ...743..143H}.
\\

Metallicity plays a direct role in the structure equations of the
stars, through the equations of state of the heavy elements, and a
crucial role in convection since heavy elements significantly
contribute to opacity. This translates into a non-negligible role
of metallicity on the scaling relations:
\cite{2014ApJ...785L..28E} have shown that scaling relations are
not suited for metal-poor red giants, among which six stars that
are kinematically associated with the halo, from a sample observed
by both the Kepler space telescope and the Sloan Digital Sky
Survey-III APOGEE spectroscopic survey. Masses derived from the
seismic scaling relations are for such stars significantly above
astrophysical expectations. Metallicity plays also a crucial role
for comparison of asteroseismic and classical stellar information
\citep[e.g.,][]{2014A&A...564A.119M,2014ApJ...790..127B}.

As stated above (section \refer{numax}), $\numax$ mostly depends
on the stellar surface gravity $g$. As for other scaling
relations, large efforts are currently done for having the most
exact relation between $\numax$ and $\log g$. It is already clear
that the gravity information provided by $\numax$ is much more
precise than the measurement of $\log g$: $\log g$ values derived
from $\numax$ can be obtained with a precision better than
0.02\,dex, whereas this precision is limited to about 0.3\,dex for
a field star
\citep[e.g.,][]{2011ApJ...729L..10B,2014ApJS..211....2H}.\\

As $\numax$ mostly depends on the stellar surface gravity $g$, one
can see HR diagrams with the luminosity information replaced by
$1/\numax$, since from the black body relation and the $\numax$
scaling relation, the luminosity scales as
\begin{equation}\label{eqt-scaling-L}
    L \propto M\; \Teff^{7/2} \; \numax^{-1} ,
\end{equation}
where we have kept the mass dependence to enhance the relation
between $L$ and $1/\numax$. It is clear that the seismic scaling
relations can provide the full determination of the luminosity
with $\Dnu$, $\numax$ and $\Teff$, after proper calibration.

\begin{figure}[!t]
 \fichier{9.6}{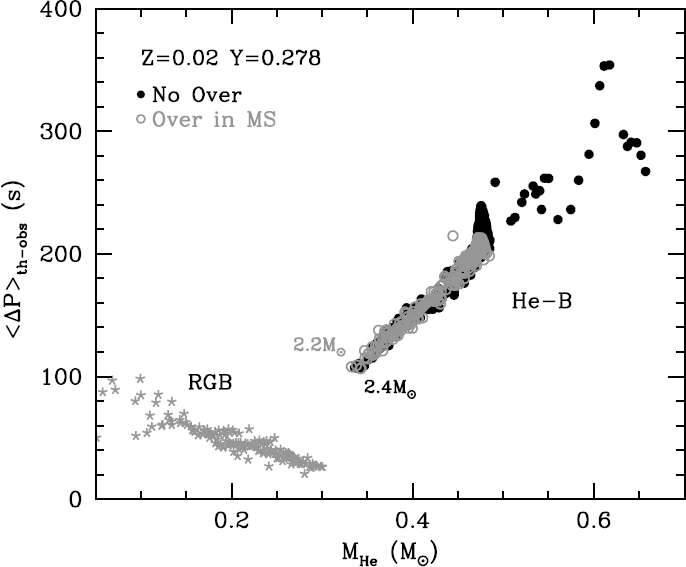}
 \legende{Helium core mass and period spacing}{Observable period spacing
 versus helium-core mass for RGB (grey stars) and clump (grey and black circles,
 with and without overshooting in the main sequence)
 \credit{2013ApJ...766..118M}.
 \labell{masse-coeur}}
\end{figure}

\begin{figure}[!t]
 \fichier{11}{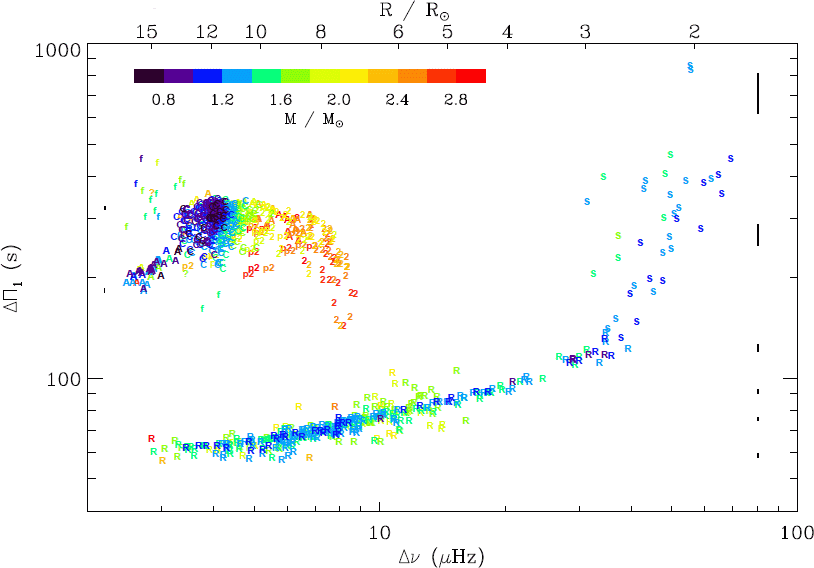}

 \fichier{10.8}{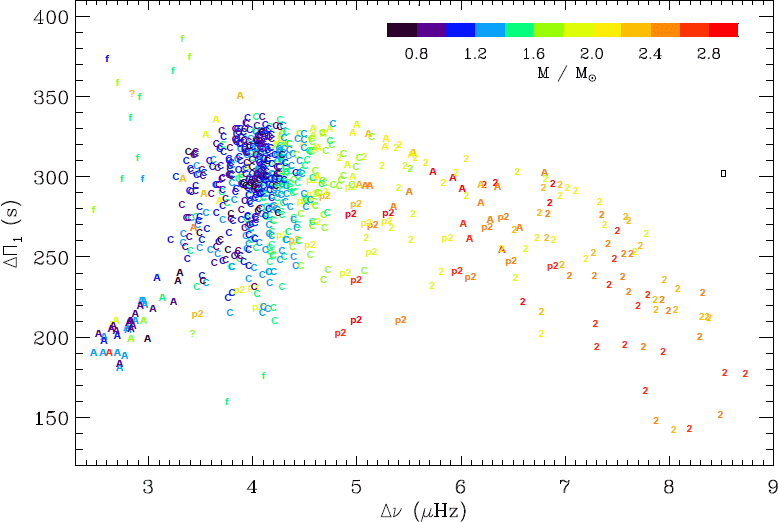}
 \legende{$\Dnu$ -- $\Tg$ diagram} {$\Dnu$ -- $\Tg$ diagram.
 The seismic proxy for the stellar mass is indicated by
the color code. The evolutionary states are indicated by S
(subgiants), R (RGB), f (helium flash stage), C (red clump), p2
(pre-secondary clump), 2 (secondary clump), and A (stars leaving
the red clump moving towards the AGB).
%The dashed line shows the limit between subgiants and red giants.
%The red clump and secondary clump in the dotted box are shown in Fig. \ref{fig_4}a.
The error boxes on the right side indicate the mean uncertainties,
as a function of $\Tg$, for stars on the RGB; for clump stars,
 uncertainties are indicated on the left side.
 \emph{Bottom:} zoom on the red clump
 \credit{2014A&A...572L...5M}.
 \labell{fig-dnudpi}}
\end{figure}

\subsection{Probing the stellar core}

We have seen that mixed modes are observed in stars evolved enough
so that the \BV\ frequency has reached values similar to the
frequency domain where pressure modes are excited. Their
observation provides the measurement of the period spacing, which
is an integral of the core properties. So, mixed modes allow us to
probe the stellar cores in subgiants and red giants and to test
their physical conditions. This was used for distinguishing red
giants on the RGB or in the red clump
\citep{2011Natur.471..608B,2011A&A...532A..86M}. We have now a
more precise view on the huge potential of mixed observed in
subgiants and red giants.

\begin{figure}[!t]
 \fichier{10.7}{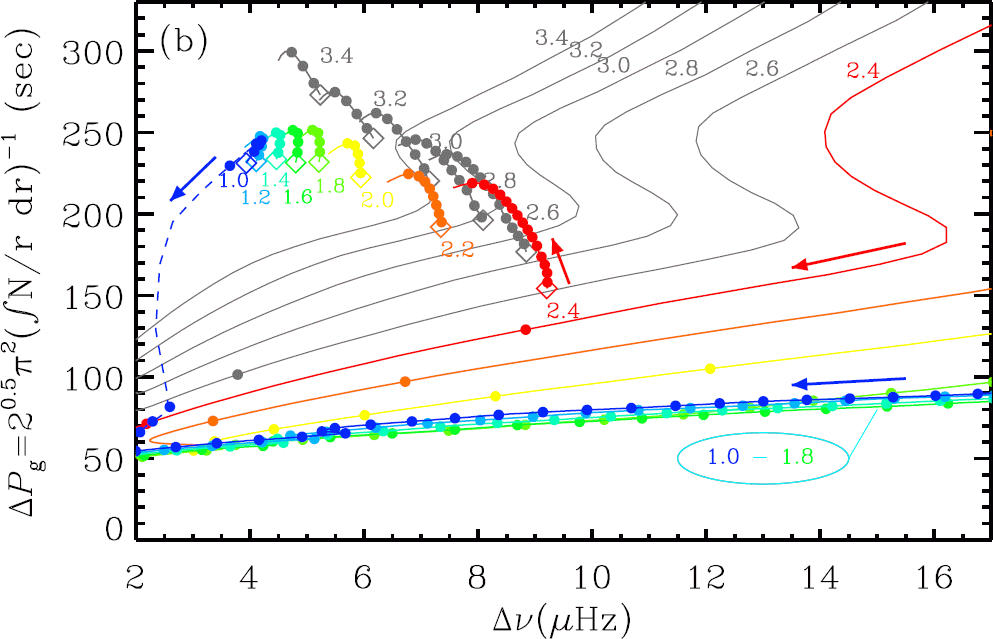}
 \legende{Theoretical $\Tg$ -- $\Dnu$ diagram} {Theoretical $\Tg$ -- $\Dnu$ diagram of grid of stellar models
 using the MESA code
 \credit{2013ApJ...765L..41S}.
 \labell{fig-dpidnu-stello}}
\end{figure}

\begin{figure}[!t]
 \fichier{11}{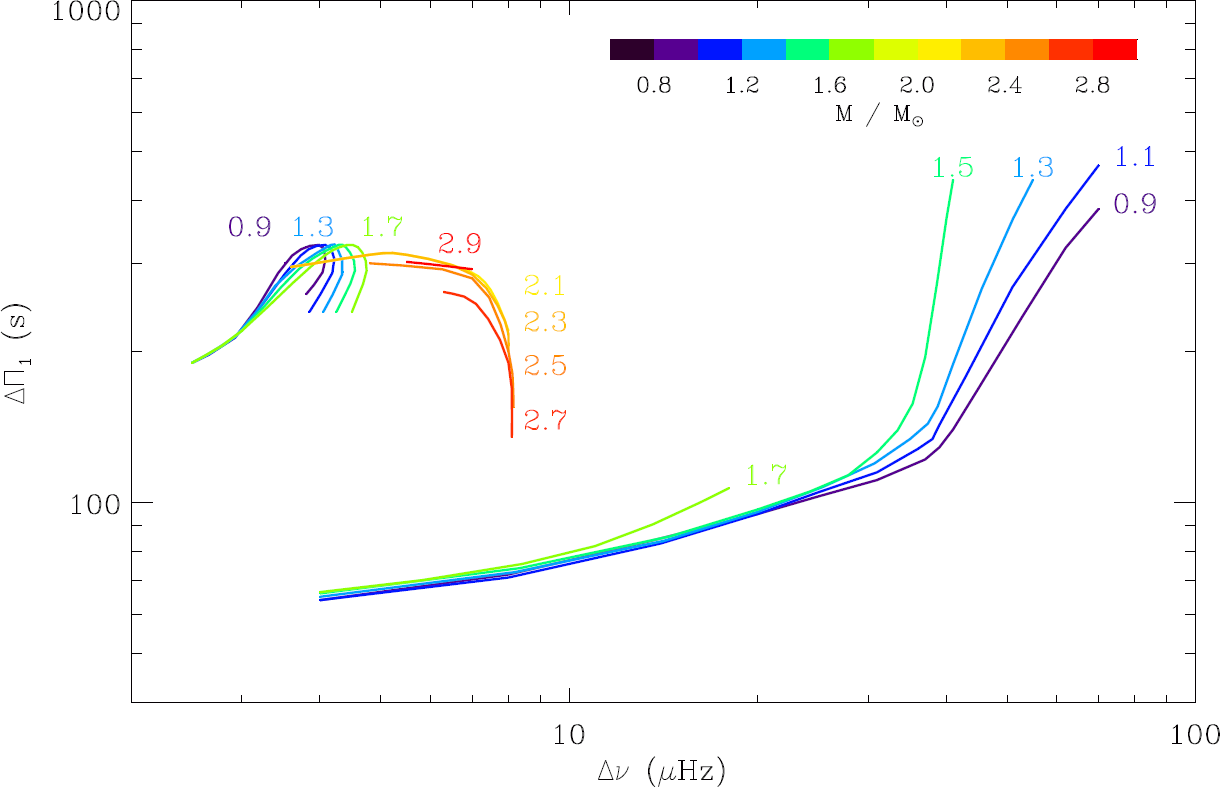}
 \legende{Stellar evolution as seen by asteroseismology}
 {Evolutionary tracks derived from the $\Tg$ -- $\Dnu$ diagram
  \credit{2014A&A...572L...5M}.
 \labell{stellar-evolution}}
\end{figure}

\subsubsection{Evolutionary tracks}

Stellar evolutionary tracks can be drawn in the $\HRsismo$ diagram
based on the observation of the large separations and period
spacings, derived from the asymptotic fit of radial and mixed
dipole modes. From \cite{2013ApJ...766..118M}, we get a
theoretical relation between $\Tg$ and the core mass
(Fig.~\refer{masse-coeur}). Ensemble asteroseismology then helps
providing a model independent clear picture of stellar evolution
(Fig.~\refer{fig-dnudpi}). In
this diagram, we benefit from all seismic information: \\
- stellar masses and radii are derived from the scaling
relations,\\
- the frequency separation provides information on the extent
of the envelope,\\
- the period spacing provides information on the size and status
of the core.\\

So, independent of modelling, we can distinguish various
evolutionary stages:

- S (subgiants): the increase of density in the core induces an
increase of the $\NBV$ frequency in the core, so that p and g
waves can couple efficiently. In this regime, the $\HRsismo$
relation still shows a significant mass dependence.

- R (RGB): the RGB regime differs from the subgiant regime. The
properties of the stellar interior become increasingly dominated
by the physical conditions of the quasi-isothermal degenerate
helium core and its surrounding hydrogen-burning shell.
Accordingly, the structural properties of the envelope are also
related to the core mass, which explains the degeneracy in the
$\HRsismo$ diagram. The change of regime is defined with an
empirical criterion: a subgiant with a mass below 1.5\,$M_\odot$
starts climbing the RGB when $(\Dnu/36.5\,\mu\hbox{Hz})^{2.5}
\,(\Tg / 126\,\hbox{s}) < 1$. The determination of this threshold
is better than 8\,\%. Translated into a stellar age, this
uncertainty represents a very short event, much less than 0.5\,\%
of the evolution time on the main sequence.

- C (red clump): red-clump stars  occupy a small region of the
$\HRsismo$ diagram, around 300\,s and 4.1\,$\mu$Hz. They have
similar core masses, hence similar luminosities, and are therefore
used as standard candles. Seismic information provides useful
constraints for improving the fine structure of the red clump,
hence for improving distance measurements. Models still have
difficulties at reproducing the period spacing in the red clump
\citep{2012ApJ...744L...6B,2013ApJ...766..118M,2013ApJ...765L..41S},
in part because they do not consider properly the extra mixing
required in the core \citep{2013ASPC..479..435N}. Now, the
accuracy of the measurements of $\Dnu$ and $\Tg$ is high enough to
track the evolution of the stars in the helium-burning phase.
Low-mass stars have lower $\Dnu$ than more massive stars, hence
lower mean density. This is in agreement with the fact that the
inner pressure is fixed by the hydrogen shell that produces the
largest part of the stellar luminosity. During the first stage of
helium burning, the core grows in mass and expands, so that the
envelope contracts: both $\Tg$ and $\Dnu$ increase. In a second
stage, both decrease. This evolution is qualitatively predicted by
models
\citep{2012A&A...543A.108L,2013ApJ...766..118M,2013ApJ...765L..41S}.

- 2 (secondary clump): in stars with masses above about
$1.9\,M_\odot$, the ignition of helium occurs gradually rather
than in a flash because the core is not fully degenerate
\citep{1999MNRAS.308..818G,2012ApJ...760...32H,2012MNRAS.419.2077M}.
%The threshold value $1.9\,M_\odot$ is, at this stage, not fully known. The uncertainty is however below 10\,\%
Therefore, these secondary-clump stars show a larger spread in the
$\HRsismo$ diagram \citep{2012ApJ...744L...6B}: $\Tg$ decreases
with increasing stellar mass up to 2.7\,$M_\odot$, as does the
mass of the helium core at ignition. Then, for masses above
2.8\,$M_\odot$, $\Tg$ increases significantly with increasing
stellar mass. This behavior is expected from stellar modelling,
which however often fails at reproducing the stellar mass
corresponding to the minimum $\Tg$ values
\citep[Fig.~\refer{fig-dpidnu-stello}, ][]{2013ApJ...765L..41S}.

-  A (stars leaving the red clump moving towards the AGB): a few
stars appear in the vicinity of the clump, but with significantly
smaller period spacings. They most probably correspond to stars in
which the core is contracting due to helium becoming exhausted,
leaving the main region of the red clump and preparing to ascend
the AGB
\citep{2012A&A...543A.108L,2012ApJ...757..190C,2013EPJWC..4303002M}.
An empirically threshold can be defined: a star leaves the red
clump and enters this stage when its large separation is 15\,\%
below the mean value observed in the clump for stars with
comparable masses. For low-mass stars, this occurs when
$(\Dnu/3.3\,\mu\hbox{Hz})^{1.5} \,(\Tg / 245\,\hbox{s}) < 1$.\\

Other evolutionary stages are identified:

- p2 (pre-secondary clump): it is possible to define, for each
mass range, the p2 status, corresponding to progenitors of
secondary-clump stars. Progenitors of secondary clump stars have a
lower $\Dnu$ (a higher luminosity) than the median stage in each
mass interval, and also a low $\Tg$ corresponding to an extended
inner radiative region.

- f (helium flash stage): a small number of stars are clearly
outside the evolutionary paths mentioned above. Such stars may
have very recently undergone the helium flash. At low $\Dnu$, we
identify stars with an unusually high period spacing,
corresponding to a small inner radiative region. This situation
matches an helium subflash \citep{2012ApJ...744L...6B}.\\

The evolutionary tracks are summarized in
Fig.~\refer{stellar-evolution}. Observations reported in
Fig.~\ref{fig-dnudpi} were used to draw these tracks, for
0.2-$M_\odot$ mass steps. Finally, stars can also be represented
in the classical HR diagram (Fig.~\ref{fig-HR-sismo}, where a few
main-sequence stars have been included).

\begin{figure}[!t]
 \fichier{11}{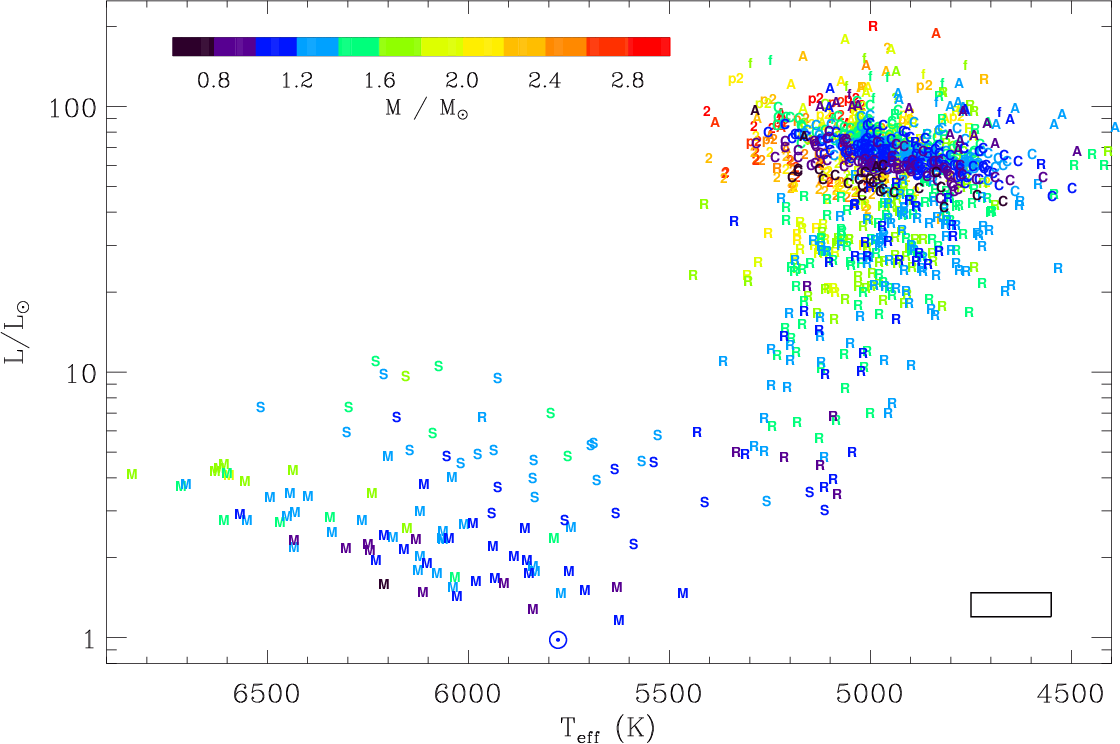}

 \fichier{11}{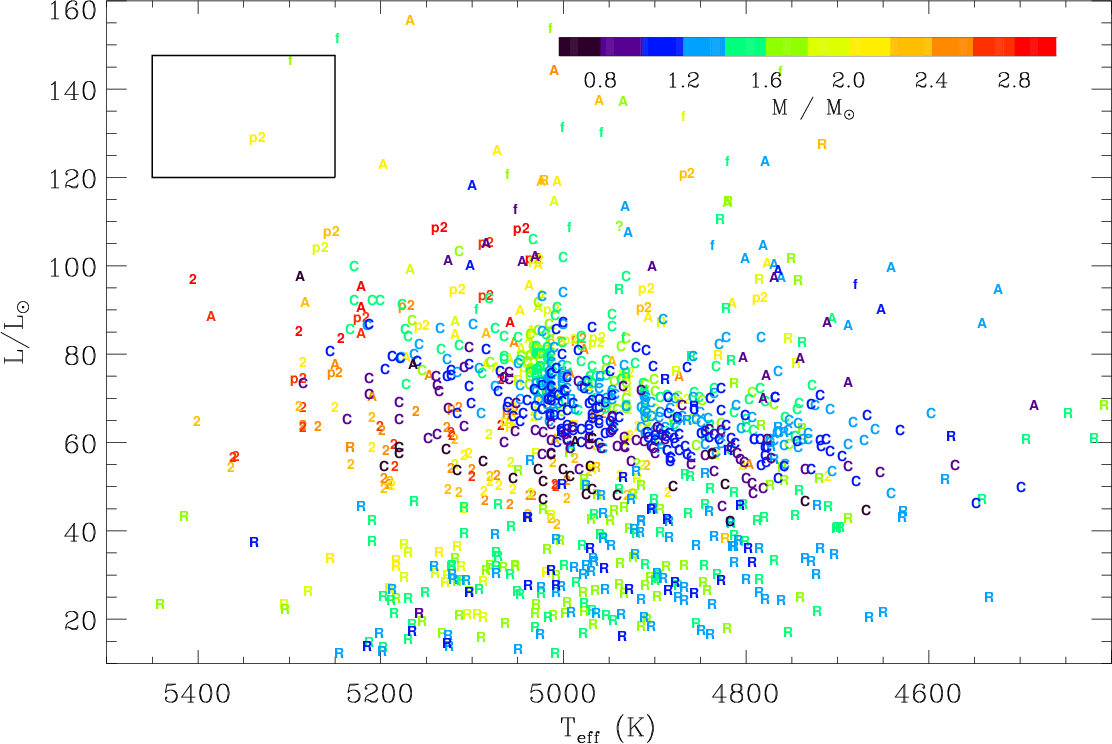}
 \legende{HR diagram with seismic information}
 {HR diagram with seismic information derived from
 Fig.~\refer{fig-dnudpi}.
 \emph{Bottom:} zoom on the red clump. Secondary clump stars are hotter and less bright than
 clump stars. The stars defined as p2 and
 A significantly more luminous than clump and secondary-clump
 stars.
  \credit{2014A&A...572L...5M}.
 \labell{fig-HR-sismo}}
\end{figure}

\begin{figure}[!t]
 \fichier{11.}{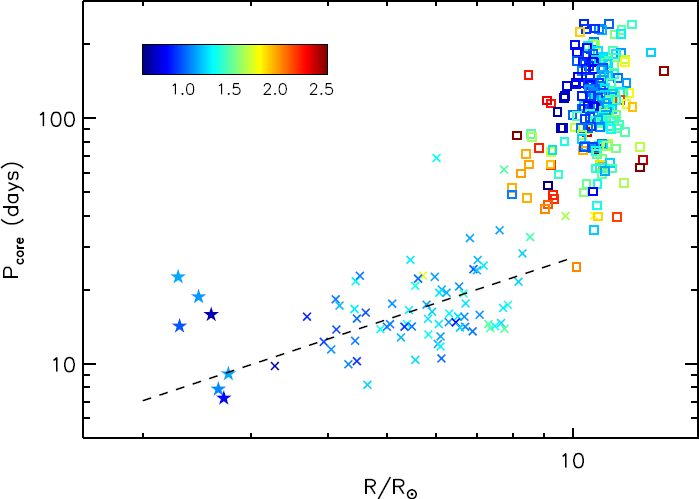}
 \legende{Core rotation in subgiants and redgiants}
 {Mean period of core rotation as a function of the
asteroseismic stellar radius, in log-log scale. Subgiants:
$\large\star$; RGB stars: $\times$; clump stars: $\small\Box$. The
dashed line indicates the fit of RGB core rotation period, varying
approximately as $R^0.5$, against $R^2$ for a solid rotation.
 \credit{2014A&A...564A..27D}.
 \labell{fig-rotation-geantes}}
\end{figure}

\subsubsection{Core rotation}

The asymptotic analysis of mixed modes provides also their
rotational splittings, from which it is possible to derive the
mean core rotation (Section \refer{rotation-mixte}). Rotational
splittings have been first observed in a handful of red giants,
putting in evidence a significant radial differential rotation
\citep{2012Natur.481...55B,2012ApJ...756...19D}. Then,
\cite{2012A&A...540A.143M} have developed a dedicated method for
automated measurements of the rotational splittings in a large
number of red giants. Their study provides the largest set of core
rotational splittings available up to now.\\

Under the assumption that a linear analysis can provide the
rotational splitting, \cite{2013A&A...549A..75G} have provided an
asymptotic description of the rotational splittings, which is
operating for both subgiants and red giants. They have proven that
the mean core rotation dominates the splittings, even for pressure
dominated mixed modes. For red giant stars with slowly rotating
cores, the variation in the rotational splittings of dipole modes
with frequency depends only on the large frequency separation, the
g-mode period spacing, and the ratio of the average envelope to
core rotation rates. Thus, they have proposed a method to infer
directly this ratio from the observations and have validated this
method using \Kepler\ data. In case of rapid rotation, rotation
cannot be considered as a perturbation any more and the linear
approach fails \citep{2013A&A...554A..80O}.\\

\cite{2012A&A...540A.143M} note a small increase of the mean core
rotation period of stars ascending the RGB
(Fig.~\refer{fig-rotation-geantes}). Alternatively, an important
spinning down is observed for red-clump stars compared to the RGB.
They also show that, at fixed stellar radius, the specific angular
momentum increases with increasing stellar mass. Similar
observations conducted in subgiants do not show the same trend
\citep{2014A&A...564A..27D}. It seems that the spinning-down
mechanism occurring on the RGB is not efficient enough in
subgiants, as recently explained
\cite{2015arXiv150505447B,2015arXiv150505452B}.

For investigating the internal transport and surface loss of the
angular momentum of oscillating solar-like stars,
\cite{2013A&A...549A..74M} have studied the evolution of
rotational splittings from the pre-main sequence to the red-giant
branch (RGB) for stochastically excited oscillation modes. They
have shown that transport by meridional circulation and shear
turbulence cannot explain the observed spin-down of the mean core
rotation. They suspect the horizontal turbulent viscosity  to be
largely underestimated.

%%%%%%%%%%%%%%%%%%%%%%%%%%%%%%%%%%%%%%%%%%%%%%%%%%%%%%%%%%%
\section{Conclusion}

After the study of adiabatic oscillations, it is necessary to
investigate non-adiabatic conditions and to enter stellar and
Galactic physics in detail, as proposed in the next chapters.

Observations with the CoRoT and \Kepler\ missions are completed,
but much more result is to come and forthcoming missions, as K2,
TESS, and Plato, will provide new data. In a near future,
calibrated scaling relations combined with spectrometric,
interferometric and astrometric measurements will boost ensemble
asteroseismology. GAIA results, when available, will add new
constraints on stellar populations. In parallel, precise stellar
modeling based on seismic data makes continuing progress.
Seismology has become a fruitful partner of stellar physics!

%%%%%%%%%%%%%%%%%%%%%%%%%%%%%%%%%%%%%%%%%%
%      ¨APPENDICE
\clearpage

\Appendix

\section{Spherical harmonics\labell{ylm}}

Spherical harmonics are the normal basis for a spherically
symmetric problem: hydrogen atom, cosmic microwave background, or
stellar interior structure... They are defined by
\begin{equation} \labell{eqt-def-ylm}
 \Ylm (\theta, \varphi)
 \egaldef
 (-1)^{(m+|m|)/2}\ \left[{2\ell
 +1\over 4\pi}{(\ell-|m|) !\over(\ell+|m|) !}\right]^{1/2}
 P_\ell^{|m|} (\cos\theta) \exp im\varphi ,
\end{equation}
where the degree $\ell$ is the number of nodal lines, and the
azimuthal order $m$  is the number of meridional nodal lines
(Figs.~\refer{fig-ylm-cartes} and \ref{fig-ylm-images}). The
number of sectors corresponds to to $2m$; $\ell-|m|$ gives the
number of nodal lines parallel to the equator.

Spherical harmonics are based on the Legendre polynomials
$P_\ell^m (\cos\theta)$, which are the generic solutions of the
differential equation
\begin{equation}
{\diff\over \diff x} \left[ (1-x^2) {\diff F (\theta)\over \diff
x} \right] + \left(L^2 - {m^2\over 1-x^2} \right) F (\theta)\ = \
0,
\end{equation}
with $x =\cos\theta$. These solutions converge only when  $L^2 =
\ell (\ell + 1)$, where $\ell$ and $m$ have integer values, and
$m$ is in the range $-\ell$ et $+\ell$.

Even if spherical harmonics form an orthogonal basis over the full
sphere,
\begin{equation}
\int\ind{sphere} \Ylm . Y_{\ell'}^{m'}\ \diff \Omega \ = \
\int\ind{sphere} \Ylm . Y_{\ell'}^{m'}\ \sin\theta \diff\theta
\diff\varphi \ = \ \delta_{\ell, \ell'}\ \delta_{m, m'},
\end{equation}
observations are never able to see the full sphere. On an
hemisphere, with limb darkening, orthogonality is not ensured:
\begin{equation}
\int\ind{hemisphere} \Ylm . Y_{\ell'}^{m'}\ \sin\theta \diff\theta
\diff\varphi \ \ne \ \delta_{\ell, \ell'}\ \delta_{m, m'},
\end{equation}
so that observations are affected by a confusion between the
$(\ell, m)$ values. The main confusion occurs between $(\ell, m)$
and $(\ell\pm 2, m)$. This may be problematic for observations
\citep[e.g.,][]{2011A&A...531A.104G}.

Spherical harmonics are introduced in the components of the wave
displacement $\depl$ projected on the spherical basis:
\begin{equation}
\depl (r, \theta, \varphi, t) = \hbox{Re} \left[ \hbox{\rm
e}^{i{\omega\nl}t}\ \left(
  \xir (r) \Ylm                     {\bf e}\ind{r}
+ \xih (r) \left[
  {\partial \Ylm \over \partial\theta} {\bf e}_\theta
+ {1\over\sin\theta}{\partial \Ylm \over \partial\varphi} {\bf
 e}_\varphi \right] \right) \right] .
  \labell{eqt-ylm-disp}
\end{equation}
The degree $\ell$ and  azimuthal order $m$ are connected with the
spherical angles $\theta$ et $\varphi$, respectively, as the
cyclic frequency $\nu$ is connected with time $t$.

\begin{figure}[!t]
 \fichier{10.7}{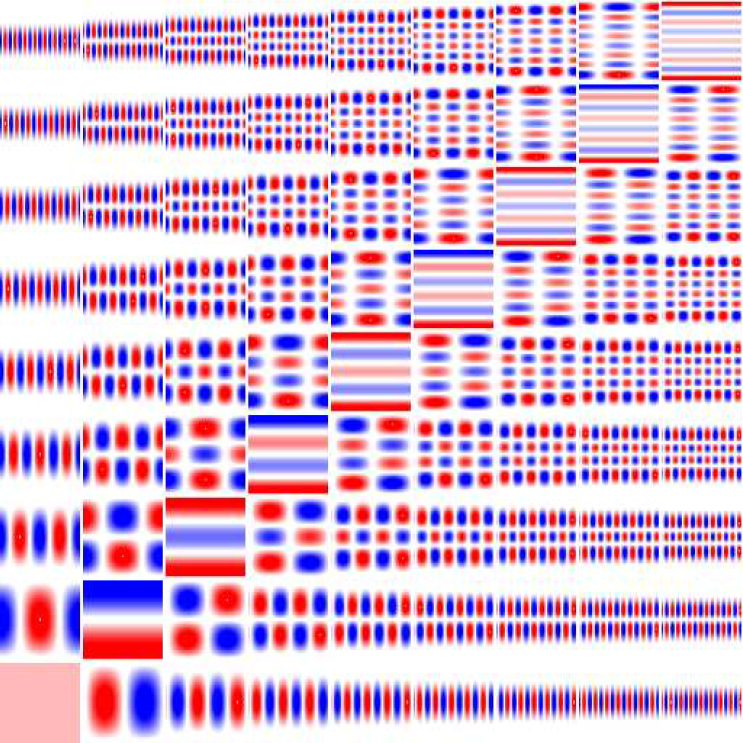}
  \legende{Spherical harmonics (1)}{Map of spherical harmonics $\Ylm (\theta, \varphi)$, for the degrees $\ell$
  from 0 to 8 (lines) and azimuthal orders $m$ from $-\ell$ to
  $\ell$. The $2\ell+1$ components of the multiplets of degree
  $\ell$ are located on the upper and right sides of a square with side
  $\ell$:
  $m=0$ spherical harmonics are on the first diagonal; they show bands parallel to the stellar
  equator;
  $|m|=\ell$ spherical harmonics are on the left and lower sides;
  their structure is made of meridian segments.
  The $\Ylm$ are confined near equatorial regions when $|m|/\ell$ increases.
  \labell{fig-ylm-cartes}}
\end{figure}

\begin{figure}[!t]
  \fichier{10.7}{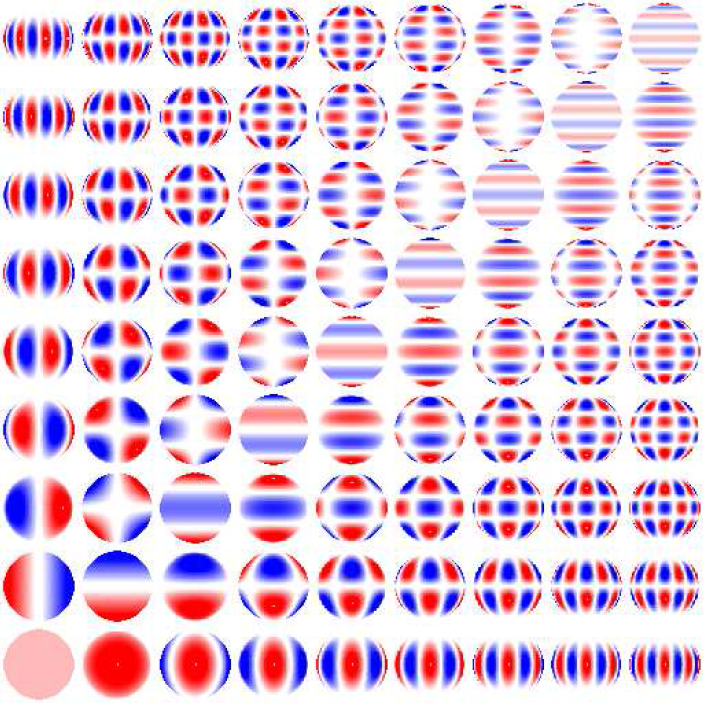}
  \legende{Spherical harmonics (2)}{Spherical harmonics $\Ylm$ of Fig. \refer{fig-ylm-cartes}
  projected onto a sphere. \labell{fig-ylm-images}}
\end{figure}

The quantization expressed by the degree $\ell$ comes from the
relation
\begin{equation}
 \nabla\ind{h}^2\ \Ylm \ =\ -{\ell (\ell +1) \over r^2} \ \Ylm ,
 \labell{eqt-ylm-ell}
\end{equation}
where $\nabla\ind{h}$ is the horizontal gradient. This relation,
crucial for the seismic study, derives from the generic properties
of the functions $f$ with separate angular variables $\theta$ et
$\varphi$ ($f(\theta, \varphi) = f_1 (\theta) f_2 (\varphi)$),
which obey the Laplace relation
\begin{equation}
 \nabla^2 \ind{h}\ f \ = \ -{1\over r^2} \ L^2 f
\end{equation}
where $L$ is a constant term. This equation can be developed as:
\begin{equation}
{1\over \sin\theta} {\partial \over \partial\theta} \left(
\sin\theta {\partial f\over \partial\theta} \right) + {1\over
\sin^2\theta} {\partial^2 f\over \partial\varphi^2} \ = \ -L^2 \ f
\end{equation}
and provides the basis of the characteristic differential equation
defining the Legendre polynomials.\\

The quantization expressed by Eq.~(\refer{eqt-ylm-ell})
corresponds to a key relation for the horizontal wavenumber:
\begin{equation}\labell{eqt-kh}
  \kh \  = \ {\sqrt{ \ell (\ell + 1)  } \over r}.
\end{equation}

Finally, an obvious property of the angular derivative with
respect to $\varphi$ is
\begin{equation}\labell{eqt-ylm-phi}
 {\partial \Ylm \over \partial \varphi} = m \ \Ylm .
\end{equation}
This explains the main term of the rotational splitting
(Eq.~\ref{eqt-rot-kernel}).

%\clearpage
\section{Derivation of the asymptotic expression\labell{justi-asymp}}

In the section, we follow the analysis of
\cite{1980ApJS...43..469T}, hereafter T80.

\subsection{Eigenfunctions and eigenfrequencies }

Following the JWKB analysis, the linearized equations governing
the low-degree adiabatic oscillations are developed, as functions
of a large parameter, namely the frequency. Near the internal and
external turning points, respectively the center and the upper
troposphere, the pressure perturbation and the radial
eigendisplacement are expressed in terms of Bessel functions.

The Bessel functions are characteristic solutions of the
differential equation
\begin{equation}\label{eqt-def-bessel}
    x^2 {\diff^2 y \over \diff x^2} +  x {\diff y \over \diff x}
    + (x^2 - \alpha^2)\; y
    = 0
    .
\end{equation}
In a spherical problem, the order $\alpha$ has half-integer value
$(\alpha = \ell +1/2)$. The Bessel functions are developed to the
second order in the argument $\omega$, where $\omega=2\pi \nu$ and
$\tau$ is some characteristic time, in the following way:
\begin{equation}\labell{A.1}
 J_\alpha (\omega \tau)\ =\ \sqrt{ 2\over \pi \omega \tau}\ \left[ \cos
 \chi - {4\alpha^2 - 1\over 8 \omega \tau} \sin \chi \right] ,
\end{equation}
with $\chi\ =\ \omega \tau -\alpha\pi /2 - \pi /4$. The
development is valid as soon as the argument $\omega \tau$
satisfies the relation
\begin{equation}\labell{A.2}
  \omega \tau\ \gg\ (4\alpha^2 - 1)(4\alpha^2 - 9) / 128 .
\end{equation}

For a pressure mode of degree $\ell$, the expressions of the
radial eigendisplacements and the Lagrangian pressure
perturbations, developed up to the second order in frequency, are
combinations of the Bessel functions $J_{\ell+1/2}$ and
$J_{\ell+3/2}$ in the central region and $J_{n_o}$ and of $J_{n_o
+1}$ in the envelope. We use Lagrangian perturbations in order to
treat possible discontinuities, as for instance in the presence of
a dense core, or in the region of second ionization of helium.\\

When the condition given by Eq.~(\refer{A.2}) is fulfilled,  one
writes for pressure modes \citep[T80,][]{1993A&A...274..595P}:
\begin{eqnarray}
  (\delta p)_i
 &\propto&
  {\sqrt{\rho c} \over r} \left[
  \cos\phi_i - \left\{ F_i + {\ell(\ell+1) \over 2\tau_i(r)} -{g\over c} \right\} {\sin\phi_i \over \omega} \right]
  \labell{A.3} \\
  \xi_i
&\propto&
  - {\omega^{-1} \over r \sqrt{\rho c} } \left[ \sin\phi_i + \left\{ H_i + {\ell(\ell+1) \over 2\tau_i(r)} \right\}
  {\cos\phi_i \over \omega} \right]
  \labell{A.4}\\
     (\delta p)_o
 &\propto&
  - {\sqrt{\rho c} \over r}  \left[ \sin\phi_o \vphantom{{H_o \tau_o(r)\over 2}} \right.
 + \left. {\left\{ {H_o \tau_o(r)\over 2} {4 n_o ^2-5 \over 8\tau_o(r)}
 + {g\over c} \right\} } {\cos\phi_o \over \omega} \right]
 \labell{A.5}\\
  \xi_o
 &\propto&
  {\omega^{-1} \over r \sqrt{\rho c}} \left[ \cos\phi_o - \left\{ {2F_o\over \tau_o(r)}  + {4 n_o ^2-1 \over
 8\tau_o(r)} \right\} {\sin \phi_o \over \omega} \right]
 \labell{A.6}
\end{eqnarray}
The amplitude factors have been omitted. The index $i$ and $o$
refer, respectively, to quantities in the central region and in
the envelope. All the symbols which are here not defined have the
same signification as in T80. The expressions derive from the
developments of Bessel functions (in the inner region,
$J_{\ell+1/2}$ for the pressure perturbation, $J_{\ell+3/2}$ for
the displacement).\\

The two phases $\phi_i$ and $\phi_o$, respectively related to
propagation in the central region and in the envelope, are
\begin{eqnarray}
  \phi_i &=& \omega \tau_i - (\ell+1/2)\ {\pi\over 2} -
{\pi\over 4} , \labell{A.7}\\
  \phi_o &=& \omega \tau_o - n_o \,{\pi\over 2}
- {\pi\over 4} , \labell{A.8}
\end{eqnarray}
with
\begin{equation}\labell{A.9}
  \tau_i(r)\, =\int_0^{r_{io}}\! {\diff r\over c} \hbox{ \  and \ }
  \tau_o(r)\, =\int_{r_{io}}^R\! {\diff r\over c}.
\end{equation}
$\tau_i$ and $\tau_o$ are the values of $\tau_i (r_{io})$ and
$\tau_o (r_{io})$ at the boundary $r_{io}$ where solutions will be
connected. The expression of the eigenfrequency is obtained by
assuming the continuity of $\delta p$ and $\xi$ at the boundary.\\

In the general case without structure discontinuity, the exact
location $r_{io}$ of the boundary plays no role so that the
validity of the development ensures that the value of $r_{io}$
finally disappears. In case of a density or sound-speed
discontinuity, $r_{io}$ is chosen at the frontier. Then:
\begin{equation}
     \varrho\varsigma\ = {(\rho c)_i \over (\rho c)_o}
\end{equation}
is different from 1. The index $i$ and $o$ refer now,
respectively, to quantities estimated at the boundary. For an
intensive variable $x$, $x_i$ is necessarily equal to $x_o$, but
extensive parameters may have
different values on the inner and outer side of the boundary.\\

The implicit calculation of the eigenfrequencies is then given by
\begin{eqnarray}\labell{A.10}
\nonumber
  &
  \varrho\varsigma\
  &  \left[ \cos\phi_i - \omega^{-1}
  \left\{ F_i + \disp{\ell(\ell+1) \over 2\tau_i} - \disp{g\over c_i} \right\}
  \sin\phi_i \right] \\
\nonumber
  & & \left[ \cos\phi_o - \omega^{-1} \left\{
  F_o \disp{2\over \tau_o}  + \disp{4n_o ^2-1 \over 8\tau_o} \right\} \sin \phi_o
  \right] \\
\nonumber
  &
 =
  & \left[ \sin\phi_i + \omega^{-1} \left\{ H_i + \disp{\ell(\ell+1)
    \over 2\tau_i} \right\} \cos\phi_i \right] \\
  &
  & \left[ \sin\phi_o + \omega^{-1} \left\{ H_o \disp{\tau_o\over 2} + \disp{4 n_o ^2-5 \over 8\tau_o}
  + \disp{g\over c_o} \right\} \cos\phi_o \right] .
\end{eqnarray}
When the terms varying as $\omega^{-1}$ are supposed to be small
(which is not a priori satisfied if the inner turning point is
deep), this implicit relation leads to the following explicit form
of the dispersion equation:
\begin{equation}\labell{A.11}
\cos(\theta_i + \theta_o) \ =\ - \discon \ \cos(\theta_i
-\theta_o)
\end{equation}
where $\discon$ is the ratio $(\roc -1)/(\roc+1)$. The values of
$\theta_i$ and $\theta_o$ can be derived from Eqs.~(\refer{A.3})
to (\refer{A.6}) at the first and second orders in $\omega^{-1}$.
Let us note that in the continuous case $\discon=0$, we recover
Tassoul's solution for the eigenfrequency, up to the second order
in frequency.

\subsection{First-order terms}

To the first order in frequency, ${\theta_i}^{(1)}$ and
${\theta_o}^{(1)}$ identify to
\begin{eqnarray}
  {\theta_i}^{(1)}
  &=& \phi_i  \labell{A.12}\\
  {\theta_o}^{(1)}
  &=&
  \phi_o \labell{A.13}
\end{eqnarray}
and the phase combinations of Eq.~(\refer{A.11}) are
\begin{eqnarray}
  {\theta_i}^{(1)}+\theta_o^{(1)}
 &=&
  \omega (\tau_i+\tau_o)-\ell{\pi\over 2} - n_o  {\pi\over 2}-{3\pi\over
  4}\labell{A.14}\\
  {\theta_i}^{(1)}-\theta_o^{(1)}
 &=&
  \omega  (\tau_i-\tau_o)-\ell{\pi\over 2} + n_o  {\pi\over 2}-{\pi\over 4}
  \labell{A.15}.
\end{eqnarray}
The solution in the continuous case ($\discon=0$) writes
\begin{equation}
  \theta_i+\theta_o \
  =\ {\pi\over 2} + p \pi\quad (p\in \mathrm{I\!N}). \labell{A.16}
\end{equation}
Since $\discon$ is small we search the solution of
Eq.~(\refer{A.11}) at the first order in frequency as a
perturbation $\delta\omega$ of the solution of Eq.~(\refer{A.16}).
The development of Eq.~(\refer{A.11}) gives the first-order
expression of the eigenfrequency and leads to the development of
the second-order solution in $\discon$.

\subsection{Second-order terms }

Obtaining symmetric phases after the expansion of the condition of
continuity introduces a correction to the second-order
coefficients of T80. The new phases $\theta_i$ and $\theta_o$ that
appear in Eq.~(\refer{A.11}) are then, including the second-order
corrections $\varphi_i$ and $\varphi_o$:
\begin{eqnarray}
  \nonumber
  \theta_i{}^{(2)}
&=&
  \phi_i - \omega^{-1} \Biggl\{ {1 \over 2} \Bigl( F_i + H_i - {\ell(\ell+1) \over \tau_i} \\
&-&
  {g\over c_i} + \discon T_i + {2\roc\over\roc+1}{T_i-T_o\over
  \roc-1} \Bigr) \Biggr\}
  \labell{A.17}
  \\
  \nonumber
&=&
  \phi_i - \omega^{-1}\varphi_i
  \\
  \nonumber
  \theta_o{}^{(2)}
&=&
  \phi_o - \omega^{-1} \Biggl\{ {1 \over 2} \Bigl( {2F_o\over \tau_o} + {H_o
  \tau_o\over 2} + {4 n_o ^2-3\over 4\tau_o} \\
&+& {g\over c_o} + \discon T_o -
   {2\roc\over\roc+1}{T_i-T_o\over \roc-1}
   \Bigr) \Biggr\}
 \labell{A.18}
  \\
  \nonumber
 &=&
 \phi_o - \omega^{-1}\varphi_o
\end{eqnarray}
The functions $F$ and $H$ are expressed in T80. The terms $T_i$
and $T_o$ are respectively
\begin{equation}
  T_{i,o}\ =\ {c_{i,o} \over 2} \left[\ {1\over \rho}{ \diff\rho\over
  \diff r} + {1\over c   }{\diff c  \over \diff r} - {2 \over r} \
  \right]_{i,o}.
 \labell{A.19}
\end{equation}
First, the function $T$ differs from its expression in T80 because
of taking account for the Lagrangian perturbation. Second, the
values of $T_i$ and $T_o$ may differ at the frontier because of
the discontinuity. This difference induces then discontinuous
terms related to the density and sound-speed jumps at the
boundary. The functions $F$ and $H$, principally given by the
integration of the term $\Omega_2$ of T80, give the second-order
coefficient in term of density and sound speed scale heights. If
we introduce two parameters, $\psi_i$ and $\psi_o$, which
represent the two integrals of Eq.~(66) in T80, the family of the
second-order constants $V$ expresses by
\begin{eqnarray}
   L^2 V_1\ + \ V_2
&=&
   2\, (\varphi_i+\varphi_o)
    \labell{A.20}
    \\
    \nonumber
&=&
  \left( {\psi}_i + {\psi}_o\right)
   + 3g \left( {1\over c_i} - {1\over c_o} \right)
   + 2\ (T_i - T_o)
   - \discon\ (T_i + T_o )
   \\
   L^2 V_3\ + \ V_4
&=&
   2\ \left[{\tau_o\over \tau_o+\tau_i} \varphi_i - {\tau_i\over \tau_o+\tau_i}
  \varphi_o\right]
   \labell{A.21}
  \\
  \nonumber
&=&
   {\tau_o\over \tau_o+\tau_i}
   \left[ \psi_i + {3g\over c_i}
   + (2-\discon) T_i  - {2\roc\over \roc+1}{T_i-T_o\over\roc -1}
   \right] \cr
&-&
   {\tau_i\over \tau_o+\tau_i}
   \left[ \psi_o - {3g\over c_o}
   - (2+\discon) T_o + {2\roc\over \roc+1}{T_i-T_o\over\roc -1}
   \right].
\end{eqnarray}
The values of $\varphi_i$ and $\varphi_o$ are derived from
Eq.~(\refer{A.21}) and are used to calculate the implicit
asymptotic values, with Eqs.~(\refer{A.11}), (\refer{A.17}) and
(\refer{A.18}). If we further assume that the interior is a fully
convective polytrope, characterized by the adiabatic coefficients
$\Gamma_i$ in the central region and $\Gamma_o$ in the outer
region, we may calculate the functions $\psi_i$ and $\psi_o$,
derived from the function $\Omega_2$ of T80:
\begin{eqnarray}
   \psi_i
&=&
   L^2 \left[ \int_0^{r_i}\!\! {\diff c\over r} - {c_i\over r} \right]
   +  {1\over 4} \left( {\Gamma_i-3 \over
   \Gamma_i-1} \right)^2 \int_0^{r_i}\!\!{ 1\over M}{\diff M \over
   \diff r} \diff c
    \labell{A.22}
   \\
\nonumber &+& {1\over 2} \left[ {4 \over\Gamma_i-1} -
\left({\Gamma_i
   -3\over \Gamma_i-1}\right)^2 \right] \int_0^{r_i}\!\! {\diff c
   \over r} + {(\Gamma_i-3)(3\Gamma_i-5) \over 8(\Gamma_i-1) }
   {g\over c_i}
   \\
   \psi_o
&=&
    L^2 \left[ \int_{r_o}^R\!\! {\diff c\over r} + {c_o\over r}\right]
    +{1\over 4} \left( {\Gamma_o-3 \over \Gamma_o-1}\right)^2 \int_{r_o}^R\!\!{ 1\over M}{\diff M \over
    \diff r} \diff c
    \labell{A.23}
    \\
 \nonumber
 &+&
    {1\over 2} \left[ {4 \over\Gamma_o-1} - \left({\Gamma_o -3 \over \Gamma_o-1}\right)^2 \right] \int_{r_o}^R\!\! {\diff c \over
    r} - {(\Gamma_o-3)(3\Gamma_o-5) \over 8 (\Gamma_o-1) } {g\over
    c_o}
\end{eqnarray}
where $M$ is the mass inside the sphere of radius $r$. Continuity
of the solution implies $r_i =r_o$ at the boundary $r_b$. The term
$V_1$ becomes simply:
\begin{equation}
  V_1\ = \ \left[ \int_0^{r_b}\! {\diff c \over r}
             + \int_{r_b}^R\! {\diff c \over r} \right]
  \ + \ \left( {\Delta c \over r} \right)_{\mathrm{\small{boundary}}}
  \ =\ \int_0^{R}\! {\diff c \over r}
  \labell{A.24}
\end{equation}
In case of discontinuity of the sound-speed profile at $r_b$, the
term $(\Delta c /r)_{\mathrm{\small{boundary}}}$ may correct
severely the negative contribution of the integral. In case of
non-adiabaticity, all previous equations which lead to the
expression of the 2$^{nd}$ order terms $V$ remain valid, except
the ones which give $\psi_i$ and $\psi_o$, which have to be
changed according to Eq.~(68) of T80.

\subsection{Implicit asymptotic expression}

The implicit expression is obtained when denying the factorization
of the sinus and cosinus in Eqs.~(\refer{A.3}) to (\refer{A.7}),
which supposes that the terms in $\omega^{-1}$ are small. Then,
with $\psi_i'=\psi_i +3g/c_i$ and  $\psi_o'=\psi_o - 3g/c_o$, the
continuity of the eigenfunctions at the boundary expresses by
\begin{eqnarray}
\nonumber &\varrho\varsigma&
   \left[\cos\phi_i +{1\over 2\omega}(\psi_i'+\hphantom{3}T_i) \sin\phi_i\right]
   \\
   \nonumber
& &
   \left[\cos\phi_o +{1\over 2\omega}(\psi_o'-3T_o) \sin\phi_o\right]
   \\
   \nonumber
&=&
   \left[\sin\phi_i -{1\over 2\omega}(\psi_i'+3T_i)
   \cos\phi_i\right]
   \\
& &
   \left[\sin\phi_o -{1\over
   2\omega}(\psi_o'-\hphantom{3}T_o) \cos\phi_o\right].
   \labell{A.25}
\end{eqnarray}
The second-order coefficients are increasing functions of $\ell$,
as indicated by Eqs.~(\refer{A.22}) and (\refer{A.23}). The
difference with the continuous case lies in the terms $g/c_i$ and
$g/c_o$, $T_i$ and $T_o$, as well as in the terms $L^2\thinspace
c_i/r$ and $L^2\thinspace c_o/r$ implying the $\ell$ dependence.

%\clearpage
\section{Variational principle\labell{variationalprinciple}}

Adiabatic oscillations can be seen as an eigenvalue problem in an
Hilbert space. Hilbert spaces can be considered as extensions of
the Euclidean space, in the sense they offer many mathematical
properties derived from the usual Euclidean space to any spaces
with any dimensions. Spectral methods in an Hilbert space are used
to study the behavior of eigenvalues and eigenfunctions of
differential equations.\\

The equation of motion (Eq.~\refer{eqt-moment}) can be rewritten
with an operator ${\cal L }$
\begin{eqnarray*}
{\cal L }(\depl )  &=&
 {1\over \rho_0^2}\nabla p_0 \nabla . (\rho_0 \depl )
-{1\over \rho_0}\nabla \bigl( \nabla  p_0 . \depl \bigr) -{1\over
\rho_0}\nabla \bigl( c^2 \rho_0 \nabla  \depl \bigr)
\\
& &  \ + \nabla\left[ \G\int{\nabla . (\rho_0\depl)\ \diff³ {\bf
x}\over
\Bigl| {\bf x} - {\bf r} \Bigr|}\right]\\
&=& \ \omega^2 \depl ,
\end{eqnarray*}
where $\omega$ is introduced by the time dependence $\partial
/\partial t \equiv i\omega$ of  the spectral method. \\

The Hilbert multidimensional space possesses the structure of an
inner product defined by
\begin{equation}\label{eqt-inner-product}
\langle x, y \rangle = \int \rho x^\star\cdot y \ \diff³ {\bf r} ,
\end{equation}
where $^\star$ is the complex conjugate. The operator is
symmetric:
\begin{equation}\label{eqt-op-sym}
\langle x, {\cal L }(y) \rangle =\langle {\cal L } (x), y \rangle
.
\end{equation}
Eigenvalues of the operator ${\cal L }$ are given by this inner
product \begin{equation}
 \omega\nl^2 \ = \ {{\disp\int  \depl\nl^\star \cdot {\cal L}
 (\depl\nl )\ \rho_0\ \diff³ {\bf r}} \over {\disp\int
 \depl\nl^\star \cdot \depl\nl \ \rho_0\ \diff³ {\bf r}}} ,
\end{equation}
which is the spirit of the Hilbert space. The variational
principal shows that a small change in the
interior model implies:\\
- a change  $\delta {\cal L}$ of the Hilbertian operator,\\
- a change  $\delta \omega^2\nl$ of the eigenvalues,\\
- but no change, at first order,  of the eigenfunctions  $\depl\nl$.\\

A model can be used to infer the eigenvalues $\omega\nl$ and
eigenfunctions $\depl\nl$. The comparison of the observed and
modelled frequencies helps determining\\
- the frequency differences  $\delta \omega^2\nl$,\\
- the difference $\delta {\cal L}$, since $\depl\nl$ are known
from the synthetic model, \\
according to
\begin{equation}
  \delta \omega\nl^2
  \ = \
  {\langle\depl\nl, \delta {\cal L} (\depl\nl) \rangle_\ell\over
 \langle\depl\nl,           \depl\nl  \rangle_\ell}
  \ = \
  {{\disp\int  \depl\nl^\star \cdot \delta {\cal L}
  (\depl\nl )\ \rho_0\ \diff³ {\bf r}} \over {\disp\int
  \depl\nl^\star \cdot \depl\nl\ \rho_0\ \diff³ {\bf r}}} .
\end{equation}
So,
\begin{equation}
  {\delta \omega\nl \over \omega\nl}
  \ = \
  {{\ifmmode{1\over 2}\else
  {$1\over 2$}\fi}} \ {\delta \omega\nl^2 \over \omega\nl^2}
  \ = \
  {1\over 2\omega\nl^2}\ {\langle\depl\nl, \delta {\cal L}
  (\depl\nl) \rangle_\ell\over
  \langle\depl\nl,\depl\nl  \rangle_\ell} .
\end{equation}

\begin{figure}[!t]
 \fichier{9.231}{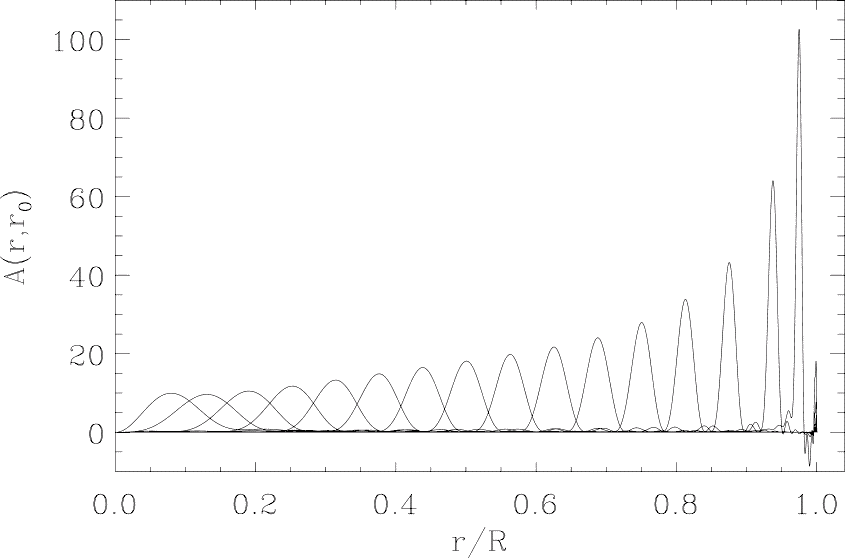}
 \legende{Local averaging kernels}{Local averaging kernels for the squared sound-speed estimates
 \credit{1997SoPh..175..287R}.
  \labell{fig-kernel-av}}
\end{figure}

We infer the linear variation of the operator with the kernels
$K^{n, \ell}\ind{X}$ that measure  the influence of the variable
$Z$ on the eigenvalue $\omega\nl$
\begin{equation}
 {\delta \omega\nl \over \omega\nl} \ = \ \int_0^R \left[
   K\ind{c^2}^{n, \ell} (r) {\delta c^2\over c^2}
 + K\ind{\rho}^{n, \ell} (r) {\delta \rho\over \rho}
 + K\ind{X}^{n, \ell} (r) {\delta Z\over Z} \right]
 \ \diff r .
 \labell{eqt-kernel}
\end{equation}
This relations is of great use in asteroseismology since it allows
improving the match of the model to the observed data. One often
uses weighted kernels, able to probe a given region of the stellar
interior (Fig.~\refer{fig-kernel-av}).

\newcommand\Omegarot{\langle\Omega\rangle}

%\clearpage
\section{Rotation\label{perturbationrotation}}

Rotation was always neglected in what precedes. We may investigate
its action as a perturbation of the non-rotating case. In case of
rapid rotation, a perturbative approach is not convenient and one
has to consider a direct approach that takes the centrifugal
distortion into account \citep[e.g.,][]{2013A&A...554A..80O}. This
is beyond the scope of this lecture.

The introduction of rotation into the equation of motion
introduces the non-galilean terms expressing the Coriolis and
centrifugal contributions. It also introduces the difference
between the rotation frame where the star does not rotate and the
frame of the observer.

The eigensolutions are perturbed in the following way
\citep{2010aste.book.....A}:
\begin{equation}\label{eqt-rot1}
    \delta\omega  = -i\
    {
    \disp{\int \rho_0 \depl^\star (\mathbf{v_0} . \mathbf{\nabla})
    \depl \ \diff³ {\bf r}}
    \over
    \disp{\int \rho_0 | \depl |^2 \ \diff³ {\bf r}}
    }.
\end{equation}
If the rotation profile is smooth enough so that a linear approach
makes sense, it is possible define a mean rotation value by
\begin{equation}\labell{eqt-mean-rot}
    \langle \Omega \rangle =
    { \int \Omega \ | \depl |^2  \ \rho_0 \diff³ {\bf r}
    \over
    \int | \depl |^2  \ \rho_0 \diff³ {\bf r}
    }.
\end{equation}
For mixed modes, we have seen that the displacement associated
with a mode strongly depends on the density and sound speed, since
the density of kinetic energy is approximately conserved
(Eq.~\refer{eqt-densi}). So the previous equation is equivalent to
\begin{equation}\label{eqt-mean-rot-p}
    \langle \Omega \rangle_p
    \simeq
    { \int \Omega \  \diff \tau
    \over
    \int  \diff \tau }
    \simeq
    2\Dnu \int \Omega  {\diff r \over c}
    ,
\end{equation}
if we assume that the variation in $\Omega$ are dominated by
radial gradients. This means that the mean value of the rotation
profile for pressure modes is heavily weighted by the inverse of
the sound speed. In other words, stellar rotation as seen by the
pressure modes is mostly probed in the outer stellar envelope only.\\

For gravity modes, the situation can be inferred by analogy.
Instead of having the mean rotation weighted by the acoustic
radius, the integration takes the buoyancy radius into account
\begin{equation}\label{eqt-mean-rot-g}
    \langle \Omega \rangle_g \simeq
    { \disp\int \Omega \ \NBV\ \disp{ \diff r \over r}
    \over
    \disp\int  \NBV\ \disp{ \diff r \over r}
    }
    .
\end{equation}

\subsection{Axisymmetrical case}

With the reasonable hypothesis of axisymmetry, we suppose that
$\Omega= \Omega (r, \theta)$. We can then perturb the normal mode
approach, with a new specification: the symmetry axis of the
spherical harmonics is defined by the stellar rotation axis. This
introduces the azimuthal order $m$. In the general case, one uses
the eigenfunctions associated with the unperturbed solutions
$\omega\nl$. This helps defining the eigenfunctions associated
with $\omega\nlm$, with the $m$ dependence derived from the
definition of the spherical harmonics (Eq.~\refer{eqt-def-ylm}).\\

Classical computations give a generic form where all the
rotational information is provided by the rotational kernels
$K\nlm$ \citep{2010aste.book.....A}:
\begin{equation}\label{eqt-rot-kernel}
  \omega\nlm \ = \ \omega\nl + m\int\!\!\!\!\int K_{n,
 \ell, m} (r, \theta)\ \Omega (r, \theta)\ r \diff r \diff \theta
 .
\end{equation}
Having $\omega_{n,\ell,0} = \omega\nl$ derives from the assumption
of slow rotation. In case rotation is rapid enough for rendering
an
oblate shape, this is no more the case \citep{1990Icar...87..198M}.\\

\begin{sidewaysfigure}
%\begin{figure}[!t]
% \fichier{10.}{kernel_rotation_soleil2.png}
% \fichier{13.}{kernel_rotation_soleil.png}
 \fichier{15.}{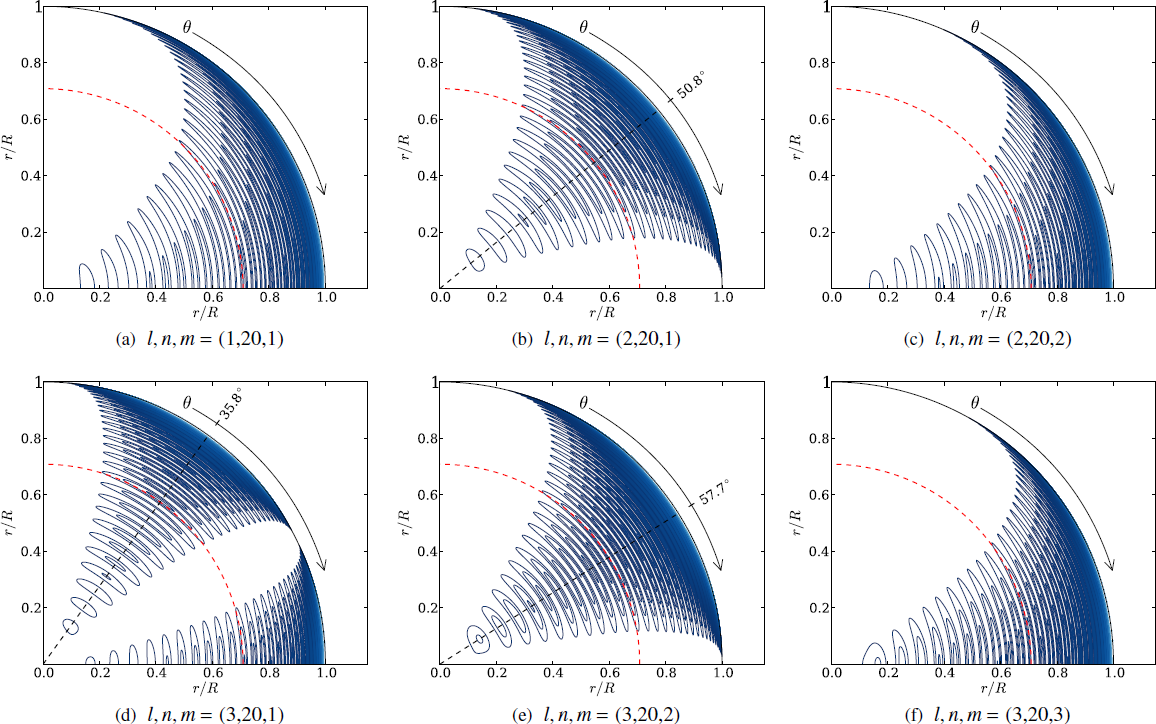}
 \legende{Rotation kernels}{Contour plots of rotation kernels for modes with $n=10$ and
 $\ell =$ 1, 2, and 3. The dashed red circle indicates the base of the convection zone.
 The
 rapid decrease of the kernels with depth shows that rotation is
 essentially probed in the outer layers. The different $\Ylm$
 provide different mappings of the stellar interior
 \credit{2014ApJ...790..121L}.
  \label{kernel_rotation_soleil}}
%\end{figure}
\end{sidewaysfigure}

\begin{figure}[!t]
 \fichier{11.}{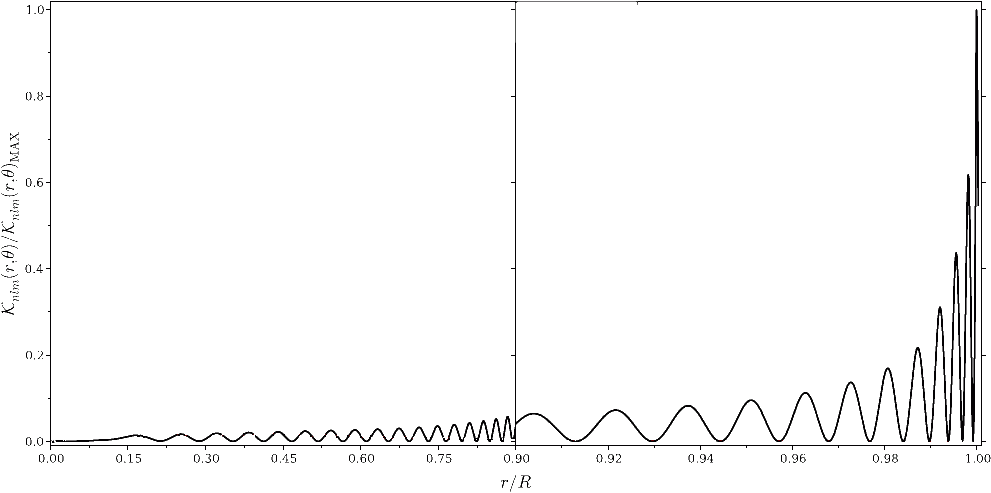}
 \legende{Normalized rotation kernels} {Normalized rotation
 kernels
 \credit{2014ApJ...790..121L}.
  \label{kernel_rotation}}
\end{figure}

Rotational kernels are useful to map the regions from where
rotational information can be inferred
(Fig.~\ref{kernel_rotation_soleil}). Since the information is
weighted by the travel time of the wave, rotational kernels are
mostly sensitive to the external region
(Fig.~\refer{kernel_rotation}).

\subsection{Spherically symmetric case}

Even if differential rotation is largely observed, we may examine
the simple case where $\Omega = \Omega (r)$ for two reasons:\\
- in such a case mathematical manipulations are considerably
simplified, so it is possible to emphasize the physics of the
phenomenon;\\
- radial differential rotation is known to be huge in evolved
stars \citep{2012Natur.481...55B}, whereas azimuthal differential
rotation is often soft and limited to a few tens percent
\citep[e.g.,][]{2009A&A...506..245M}.

Such a rotation profile is called shellular rotation
\citep{1992A&A...265..115Z}.\\

According to the definition of the mean rotation
(Eq.~\ref{eqt-mean-rot}) derived from the perturbation approach,
one can write $\delta\omega\nlm = \omega\nlm - \omega\nl$ in the
form
\begin{equation}\label{eqt-rot-xi}
    \delta\omega\nlm =  m\
    {
    \disp\int_0^R \left( \xir^2 + \ell (\ell+1) \xih^2 - 2 \xir\xih -
    \xih^2 \right) \ \rho_0\, r^2\, \Omega \, \diff r
    \over
    \disp\int_0^R \left( \xir^2 + \ell (\ell+1) \xih^2 \right) \ \rho_0\, r^2\, \diff r
    }
    ,
\end{equation}
when the rotation depends on $r$ only. This is often abbreviated
as
\begin{equation}\label{eqt-rot-def-dwnlm}
    \delta\omega\nlm =  m\ \beta\nl\ \langle \Omega \rangle
    ,
\end{equation}
or, equivalently, with the Ledoux coefficients,
\begin{equation}\label{eqt-rot-def-ledoux}
    \delta\omega\nlm =  m\ (1-C\nl)\ \langle \Omega \rangle
    .
\end{equation}
For dipole modes, the dominating terms in Eq.~(\refer{eqt-rot-xi})
are $\xir^2$ and $\ell (\ell +1) \xih^2$, so that
\begin{equation}\labell{eq-ledoux-p}
    \beta\nl \simeq 1 \ \hbox{ and } \  C\nl \simeq 0
    .
\end{equation}
The splitting is mainly related to the rotating frame, in
agreement with the fact that terms introduced by the Coriolis
force, in fact measured by $C\nlm$, are neglected.\\

For gravity modes, $\xih$ dominates in Eq.~(\refer{eqt-rot-xi}) so
that
\begin{equation}\labell{eqt-ledoux-g}
    \beta\nl \simeq 1 - {1\over \ell (\ell+1)} \ \hbox{ and } \  C\nl \simeq {1\over \ell
    (\ell+1)} .
\end{equation}
For low degree $\ell$, and especially for dipole modes with
$\ell=1$, the Coriolis force plays a non-negligible role.

\begin{figure}[!t]
 \fichier{10.6}{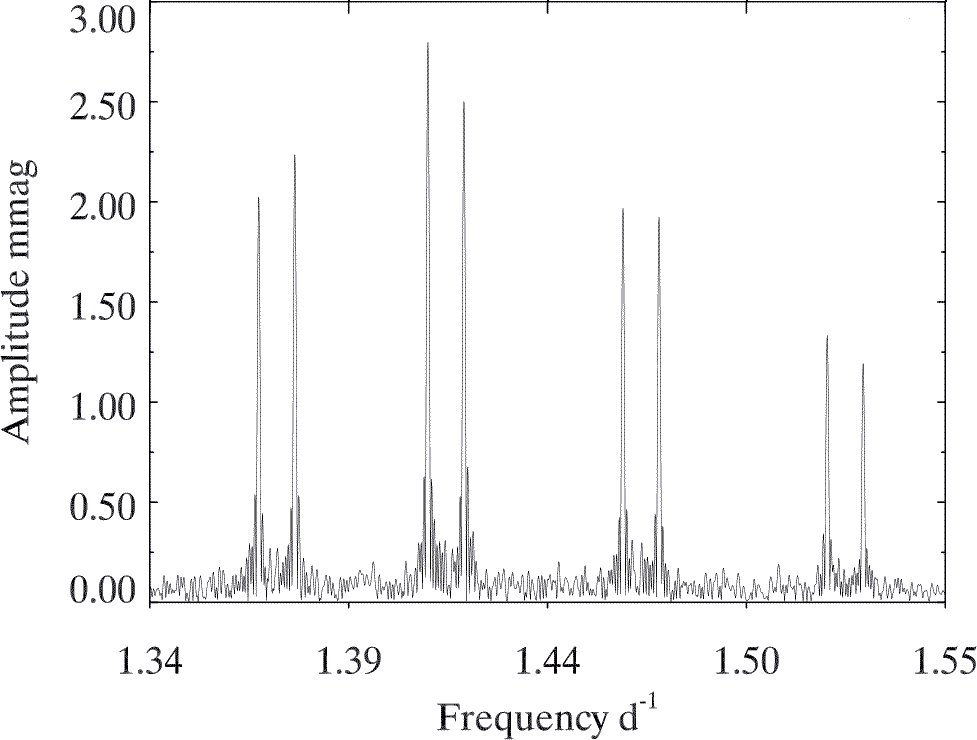}
 \legende{Rotational splittings in an A star} {Rotational splittings
 observed in an A star showing both $\delta$ Scuti pressure-mode pulsations
 and $\gamma$ Dor gravity-mode pulsations, observed by \Kepler\
 \credit{2014MNRAS.444..102K}.
 \labell{fig-gammador-rot}}
\end{figure}

\begin{figure}[!t]
 \fichier{10.6}{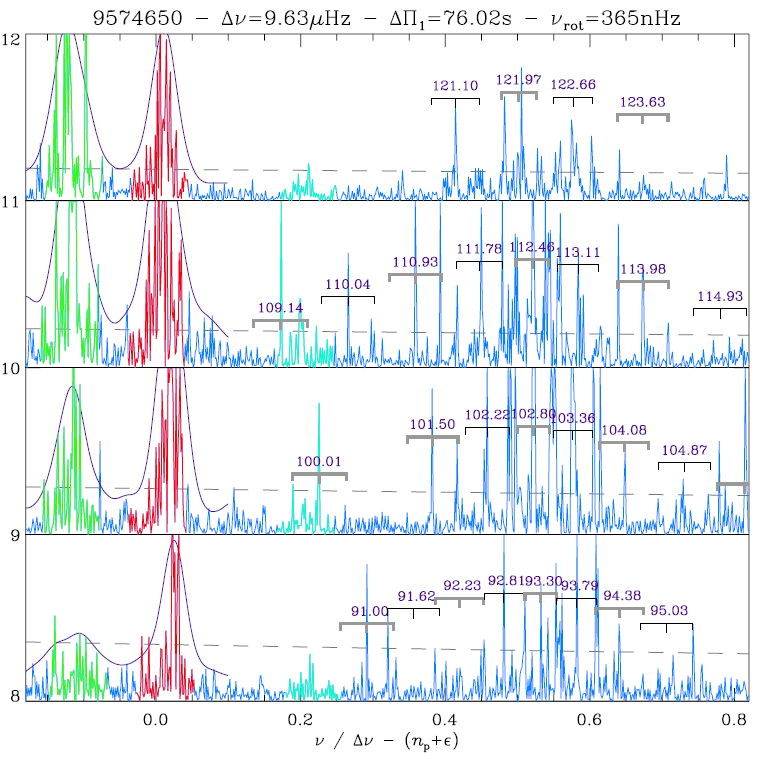}
 \legende{Rotational splittings in a red giant}{Zoom on the rotational splittings of the mixed modes
  corresponding to the radial orders $\np = 8 \to 11$ in the giant
  KIC 9574650 observed by \Kepler, in an \'echelle diagram as a function of the reduced
  frequency $\nu/\Dnu - (\np+\varepsilon)$. At low frequency, multiplets are
  overlapping
  \credit{2012A&A...548A..10M}.
  \labell{fig-kepler-rotg}}
\end{figure}

\subsection{Mixed modes\labell{rotation-mixte}}

Measuring rotational splittings of mixed modes is useful for
inferring the inner rotation profile in the radiative region. This
can be done in $\gamma$ Dor-type stars for instance
(Fig.~\refer{fig-gammador-rot}). The rotational splittings of
mixed modes observed in subgiants and giant stars are described by
\cite{2013A&A...549A..75G}. As mixed modes share properties of
pressure and gravity modes, their rotational splittings are more
or less dominated by the pressure and gravity terms, depending on
the mode frequency. As a result, the splittings undergo the mixed
influence of the mean core rotation and of the mean envelope
rotation (Fig.~\ref{fig-kepler-rotg}). Of course, one has to
suppose that this mean rotation can be defined, which seems
confirmed by observations, which show significant spin down with
stellar
evolution \citep{2012A&A...548A..10M}.\\

These mean rotations are respectively defined by
\begin{equation}\label{eqt-rot-moy-core}
    \langle \Omega \rangle\ind{core}
    =
    \int\ind{core} \Omega (x) {\NBV \over x} \diff x
    \Bigm/
    \int\ind{core} {\NBV \over x} \diff x
\end{equation}
and
\begin{equation}\label{eqt-rot-moy-env}
    \langle \Omega \rangle\ind{env}
    =
    \int\ind{env} \Omega (x) {\diff x\over c}
    \Bigm/
    \int\ind{env} {\diff x\over c}
    ,
\end{equation}
where $x=r/R$ is a normalized radius.\\

The rotational splitting writes
\begin{equation}\labell{eqt-rot-split}
    \delta\nu
    =
    \beta\ind{core}
    {\langle \Omega  \rangle\ind{core} \over 2\pi}
    +
    \beta\ind{env}
    {\langle \Omega  \rangle\ind{env}  \over 2\pi}
    ,
\end{equation}
with
\begin{equation}\labell{eqt-rot-beta}
    \beta\ind{env,core}
    =
    \int\ind{env, core}
    K(x) \diff x
    .
\end{equation}

\begin{figure}[!t]
 \fichier{9.2}{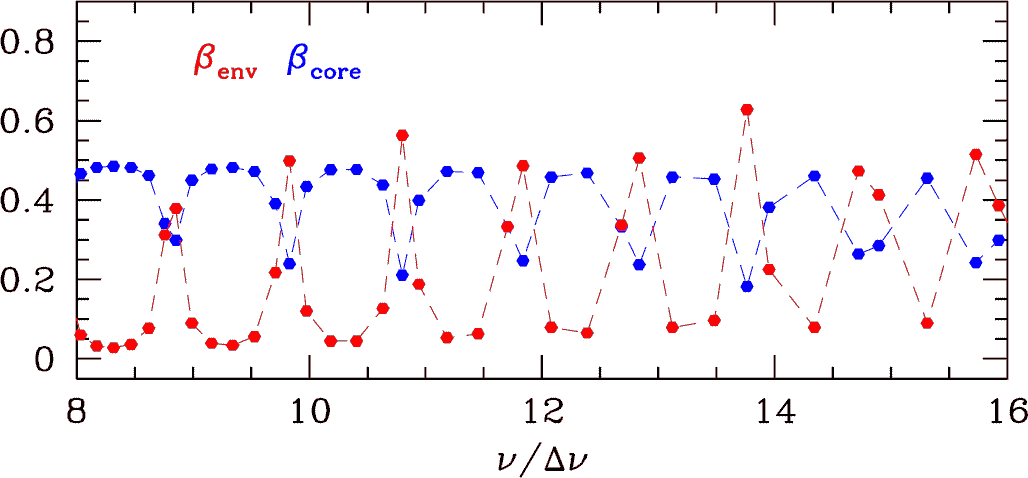}
 \legende{Rotational splittings in red giants (1)} {Coefficients
 $\beta\ind{core}$ and $\beta\ind{env}$ introduced for an estimate
 of the rotational splittings  (Eq.~\refer{eqt-rot-beta}). We note
 that, according to the Ledoux term, $\beta\ind{core} \simeq 1/2$
 for gravity-dominated modes;  $\beta\ind{env} > 1/2$ for pressure-dominated
 modes, due to the influence of the pressure contribution
 \credit{2013A&A...549A..75G}.
 \labell{fig-rot-beta}}
\end{figure}

The coefficients $\beta\ind{core}$ and $\beta\ind{env}$
(Fig.~\ref{fig-rot-beta}) result from the expression of the
inertia, assuming shellular rotation, with new variables
\begin{eqnarray}\labell{eqt-rot-z1z2}
  \inertia
  &=&
  4\pi\, r^3 \ \int_0^1 (\xi
  r^2 + \ell(\ell+1) \xih^2) \rho_0 x^2
    \diff x \\
   &=&
   \int_0^1 (z_1^2 + z_2^2) {\diff x \over x}
   ,
\end{eqnarray}
with
\begin{equation}\labell{eqt-rot-z1}
   z_1^2
    =  {3\rho_0\over \langle  \rho_0\rangle}
    \ \left({\xir \over R}\right)^2 \ x^3
\end{equation}
and
\begin{equation}\labell{eqt-rot-z2}
   z_2^2
    =  \ell (\ell+1) \
    {3\rho_0\over \langle  \rho_0\rangle} \
    \left({\xih \over R}\right)^2 \ x^3
    .
\end{equation}

\begin{figure}[!t]
 \fichier{10.6}{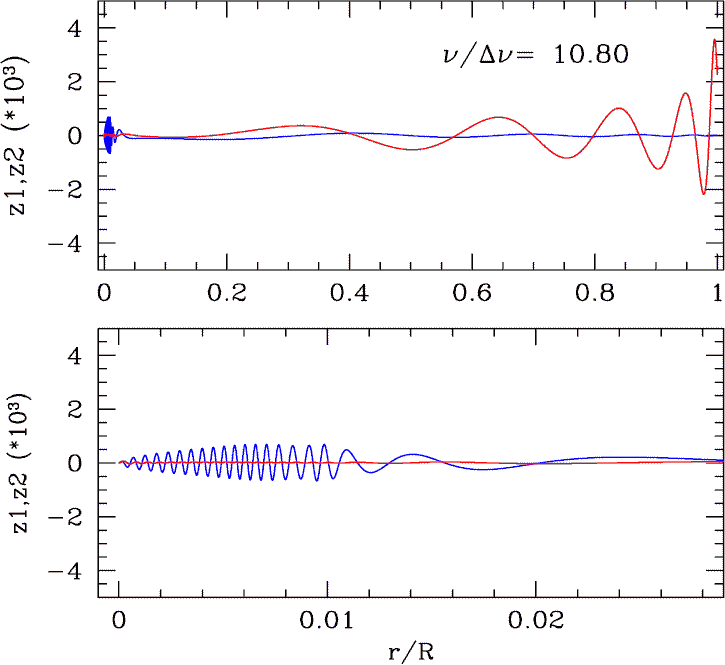}
 \legende{Rotational splittings in red giants (2)} {Coefficients
 $z_1$ and $z_2$ introduced for an estimate
 of the rotational splittings
 (Eq.~\refer{eqt-rot-beta}): vertical displacement of $z_1$ in red, horizontal displacement of
 $z_2$ in blue
 \credit{2013A&A...549A..75G}.
 \labell{fig-rot-z1z2}}
\end{figure}

\begin{figure}[!t]
 \fichier{10.6}{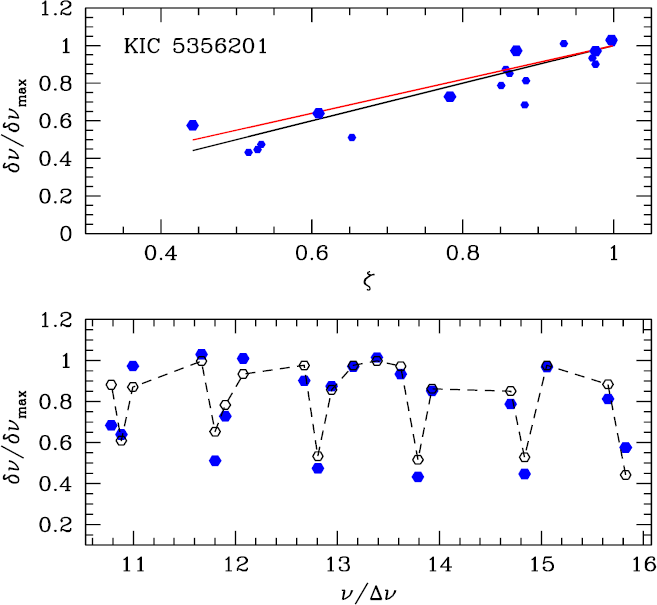}
 \legende{Rotational splittings in red giants(3)} {Rotational splittings as a
 function of $\zeta$
 (Eq.~\refer{eqt-rot-zeta})
 \credit{2013A&A...549A..75G}.
 \labell{fig-giant-splitting}}
\end{figure}
The kernel then writes
\begin{equation}\labell{eqt-rot-kernz}
    K
    =
    {1\over \inertia}
    \left(
    z_1^2
    - {2\over \sqrt{\ell (\ell+1)}} \, z_1 z_2
    + \left(1-{1 \over \ell (\ell+1)}\right) z_2^2
    \right)\
    {1\over x}
    .
\end{equation}
The variations of $z_1$ and $z_2$ as a function of $x$ explain
that splittings show variations  between the mean envelope and
core  rotations
(Fig.~\refer{fig-rot-z1z2}).\\

\cite{2013A&A...549A..75G} have asymptotically developed the
function $z_1^2$ and $z_2^2$ to obtain an explicit expression of
the rotational splitting. The formalism is quite similar to the
development used in Section \refer{asymp-mixte} for obtaining the
asymptotic expansion of mixed modes, so that results are expressed
in similar forms. Rotational splittings are expressed by the
coefficient
\begin{equation}\label{eqt-rot-zeta}
    \zeta = {1 \over 1 + \Dnu \Tg\ \chi^2}
    ,
\end{equation}
with
\begin{equation}\label{eqt-rot-chi}
    \chi \simeq 2{\nu \over \Dnu} \
    \cos \left({\pi \over \nu \Tg}\right)
    .
\end{equation}
The splitting writes
\begin{equation}\label{eqt-rot-splitting-mixte}
    {\delta\nu\over \delta\nu\ind{max}}
    =
    {1 - 2 \mathcal{R} \over 1 + \Dnu \Tg\ \chi^2}
    + 2 \mathcal{R}
    ,
\end{equation}
where $\mathcal{R} = \langle \Omega \rangle\ind{env} / \langle
\Omega \rangle\ind{core}$ is the ratio of the mean rotations in
the envelope and in the core, and $\delta\nu\ind{max}$ is very
close to $\langle \Omega \rangle\ind{core} / 4\pi$. The extra 1/2
factor coming in this relation is the Ledoux coefficient of dipole
gravity modes (Eq.~\refer{eqt-ledoux-g}). \\

Figure~\refer{fig-giant-splitting} compares the splitting observed
in an RGB star with the modelling. The determination of the global
seismic parameter helps explaining the affine relation with the
parameter $\zeta$. This provides a precise fit of the rotational
splittings, where discrepancies are mostly explained by the
resolution of the observed peaks.

The factor $\zeta$ in Eq.~(\refer{eqt-rot-zeta}) can be used to
analyze the asymmetry in the splitting. In fact, a multiplet near
a pressure-dominated mode cannot be symmetric: the splitting of
the component $m=\pm 1$ close to the pressure-dominated mode is
smaller than the opposite component $m=\mp 1$.

%\clearpage
\section{Surface term\labell{surface}}

\subsection{Pressure modes}

As shown by its definition (Eq.~\refer{eqt-rad-as}), the
asymptotic frequency spacing $\Dnuas$ is heavily weighted by the
external region where the sound speed has low values
(Fig.~\refer{fig-intc-sol}). However, this region cannot be
depicted with the same precision as the inner layers since
hypothesis valid in the inner regions are no longer valid. We note : \\
- the inadequacy of the adiabatic approximation,\\
- pressure and density scale heights with low values, so that one
can no more consider that they are much larger than the
wavelength...\\

We also have seen that the low-frequency waves are reflected in
deeper region. Their frequencies are not perturbed by an improper
treatment of the uppermost layers, contrary to higher frequencies
(Fig.~\refer{fig-coupure-jup}). This is a general problem, for
instance studied for Jovian seismology
\citep{1994A&A...291.1019M,1995A&A...293..586M}, which requires
the introduction of a surface term for correcting the modelled
frequencies before comparison with observed frequencies. In stars,
a generic form was proposed by \cite{2008ApJ...683L.175K}, varying
as
\begin{equation}\label{eqt-surface-cor}
  \nu\ind{obs} = \nu\ind{model} + a \left({\nu \over
  \nu\ind{cor}}\right)^b
  ,
\end{equation}
where the coefficient $a$ and the exponent $b$ have negative
value: high-frequency waves spend more time in the upper
atmospheric regions than low-frequency waves and require a larger
correction in absolute value. This measure how frequencies
computed with a stellar model have to be corrected to match
observed frequencies (Fig.~\ref{fig-surface-cor}).

\subsection{Mixed modes\labell{surface-mixte}}

As we deal with solar-like oscillations, we are interested by
stars with a convective envelope. Gravity modes are trapped well
below the surface and are not affected by this correction. Mixed
modes however are affected, but in a way that depends on their
characteristics: as one can imagine, a gravity-dominated mixed
mode is less affected than a pressure-dominated mixed mode.

The correction on mixed modes can be inferred from the correction
on pressure modes (and no correction for gravity modes) and
involves the mode inertia. So, the correction is
\begin{equation}\label{eqt-cor-surf-mixed}
  (\delta\nu\ind{mixed})\nl = (\delta\nu\ind{p})_{n,0} \ {Q_{n,0}
  \over Q\nl},
\end{equation}
where $(\delta\nu\ind{p})_{n,0}$ is the correction of the closest
radial mode \citep{2010aste.book.....A}. Inertia are defined by
Eq.~\refer{eqt-def-inertia}.

\begin{figure}[!t]
 \fichier{8.5}{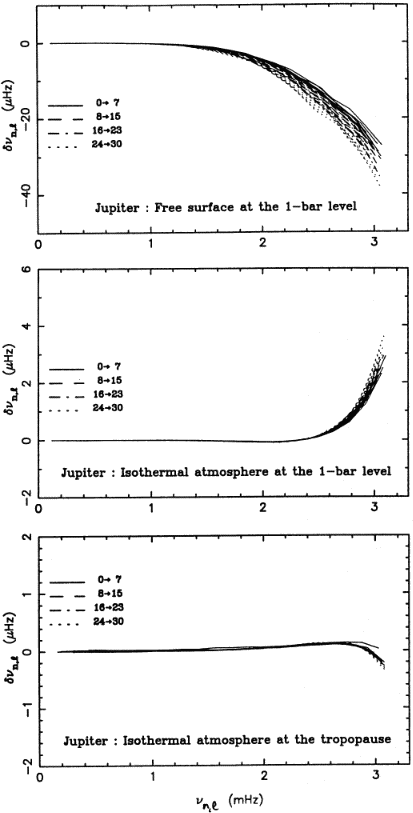}
 \legende{Surface correction}{Surface correction in the Jovian
 upper envelope. An isothermal atmosphere at the 1-bar level,
 corresponding to the external boundary of interior models, helps
 reducing the correction; a much better solution is found when the
 isothermal atmosphere is considered at the tropopause, namely the region
 where the temperature reaches its minimum value, similarly to the
 stellar photosphere
 \credit{1994A&A...291.1019M}.
  \labell{fig-surface-cor}}
\end{figure}

%%-----------------------------
%%      your bibliography
%%-----------------------------
%\begin{thebibliography}{} % (do not forget {})
\bibliographystyle{aa} % style aa.bst
\baselineskip = 10pt
\bibliography{biblio_r}

\end{document}